\documentclass[twocolumn,numberedappendix, twocolappendix, apj]{openjournal}

\usepackage[english,english]{babel}
\usepackage{amsmath}
\usepackage[usenames]{color}
\usepackage{soul}
\usepackage{graphicx}
\usepackage{float}
\usepackage{bm}
\usepackage{comment}
\usepackage{xcolor}
\usepackage{enumitem}
\usepackage{physics}

\usepackage{hyperref}

\newcommand{\mP}{{\cal P}}
\newcommand{\ii}{\hbox{i}}
\newcommand{\niton}{\not\owns}
\newcommand{\hrho}{{\hat\rho}}

\newcommand{\bkap}{{\tilde \kappa}}

\newcommand{\hypgeo}[1]{{}_0F_1\left( #1 \right)}

\definecolor{darkerred}{rgb}{0.8,0,0}
\def\change#1{\noindent{#1}}

\begin{document}

\title{It takes two to know one: Computing accurate one-point PDF covariances\\ from effective two-point PDF models }

\author[0000-0001-7831-1579]{Cora Uhlemann$^{1,\star}$}
\author{Oliver Friedrich$^{2}$}
\author[0000-0002-7537-6921]{Aoife Boyle$^{3}$}
\author[0000-0002-1524-6949]{Alex Gough$^{1}$}
\author[0000-0003-1060-3959]{Alexandre Barthelemy$^{2}$}
\author{\\ Francis Bernardeau$^{4,5}$}
\author{Sandrine Codis$^{3}$}
\thanks{$^\star$\href{mailto:cora.uhlemann@newcastle.ac.uk}{cora.uhlemann@newcastle.ac.uk}}

\affiliation{$^{1}$ School of Mathematics, Statistics and Physics, Newcastle University, Herschel Building, NE1 7RU Newcastle-upon-Tyne, U.K.}
\affiliation{$^{2}$ Universitäts-Sternwarte, Fakultät für Physik, Ludwig-Maximilians Universität München, Scheinerstr. 1, 81679 München, Germany}
\affiliation{$^{3}$ AIM, CEA, CNRS, Université Paris-Saclay, Université Paris Diderot, Sorbonne Paris Cité, 91191 Gif-sur-Yvette, France}
\affiliation{$^{4}$ Université Paris-Saclay, CNRS, CEA, Institut de physique théorique, 91191, Gif-sur-Yvette, France }
\affiliation{$^{5}$ CNRS \& Sorbonne Universit\'e, UMR 7095, Institut d'Astrophysique de Paris, 75014, Paris, France}

\begin{abstract}
One-point probability distribution functions (PDFs) of the cosmic matter density are powerful cosmological probes that extract non-Gaussian properties of the matter distribution and complement two-point statistics. Computing the covariance of one-point PDFs is key for building a robust galaxy survey analysis for upcoming surveys like Euclid and the Rubin Observatory LSST and requires good models for the two-point PDFs characterising spatial correlations. In this work, we obtain accurate PDF covariances using effective shifted lognormal two-point PDF models for the mildly non-Gaussian weak lensing convergence and validate our predictions against large sets of Gaussian and non-Gaussian maps. We show how the dominant effects in the covariance matrix capturing super-sample covariance arise from a large-separation expansion of the two-point PDF and discuss differences between the covariances obtained from small patches and full sky maps.
Finally, we describe how our formalism can be extended to characterise the PDF covariance for 3D-dimensional spectroscopic fields using the 3D matter PDF as an example. We describe how  covariances from simulated boxes with fixed overall density can be supplemented with the missing super-sample covariance effect by relying on theoretical predictions validated against separate-universe style simulations.
 \end{abstract}
 
\maketitle

\section{Introduction}
\label{sec:intro}
The recent rise of non-Gaussian statistics for galaxy clustering and weak lensing has highlighted the need for accurate models of the likelihood function and, in particular, the covariance matrices of such statistics.  Aside from the challenge to model non-Gaussian statistics and capture their response to changing cosmological parameters in the nonlinear regime, getting accurate covariance matrices can prove difficult and require thousands of numerical simulations to ensure convergence \citep{TaylorJoachimi2014CovariancesEstimation,Colavincenzo2019MockCovariancesBispectrum}. For statistics relying on higher-order correlation functions involving $N$ points, modelling the covariance generally requires assumptions for the shape of $2N$-point correlations. Even for Gaussian fields, this task has only been achieved recently at all orders $N$ \citep{Hou2021}. To go beyond the assumption of Gaussianity, log-normal models of the cosmic density field have become a tool of choice in higher-order analyses such as studies of the bispectrum \citep{Martin2012, Halder2021}, moments of cosmic random fields \citep{Gatti19}, density split statistics \citep{Friedrich18, Gruen18}, the full probability distribution function (PDF) of cosmic random fields \citep{Uhlemann2020Fisher, Friedrich2020, Boyle2021}, and even map-based inference approaches (\citealt{Sarma_Boruah2022}). The reason why (shifted) log-normal simulations \citep{Xavier2016} have become so popular is that they obey a similar hierarchy between variance and skewness as the physical cosmic density field. In particular, they can be tuned to match both a desired power spectrum at all scales and a desired skewness at one given scale (c.f.\ Section 5.2 and Figure 4 of \citealt{Friedrich2020}). This however also highlights a limitation of those simulations: they potentially fail to realistically capture the hierarchy of moments beyond the skewness \citep[for an extension see][]{Baratta2020} as well as the full shape of higher order N-point functions.

With the large data vectors that future multi-probe analyses of upcoming surveys like Euclid \citep{Euclid, Euclid16} and the Rubin Observatory LSST \citep{LSST} are aiming for, estimated covariances - whether from log-normal simulations or other mock data - can lead to a huge degradation of cosmological constraining power \citep{DS2013, Friedrich2017precisionmatrixexp, Percival2022}. This makes it desirable to have an analytic understanding of measurement uncertainties. One-point statistics such as the PDF of the weak lensing convergence have the advantage that both their dependence on cosmological parameters and their covariance, being given in terms of the two-point PDF, can be modelled theoretically \citep{Codis16a,Uhlemann17Kaiser}. In this work, we will focus on the weak lensing convergence smoothed on mildly nonlinear scales (of order $10'$ at source redshifts of $z\simeq 1-2$), where the bulk of the PDF and its cosmology dependence can be predicted at the percent level using large deviation statistics \citep{BernardeauValageas00,LDPinLSS,Barthelemy2020,Boyle2021} and its covariance is well-approximated by a shifted lognormal distribution \citep{Boyle2021}. In the opposite regime for small smoothing scales, the high-convergence tail of the weak lensing convergence PDF and its covariance have been modelled using a halo-model formalism in \cite{Thiele2020WLconvergence}, finding  qualitative agreement with simulations but significant quantitative discrepancies attributed to strong sensitivity to small-scale effects.

For galaxy clustering, the error on and correlation among the factorial moments of galaxy counts in cells $F_k= \langle N(N-1)\ldots (N-k)\rangle$ have been computed for hierarchical models in \cite{SzapudiColombi96} and \cite{Szapudi99}. Those studies included  cosmic errors due to a finite volume and discreteness errors due to a finite number of tracers relevant for clustering but not the weak lensing convergence, which is instead affected by shape noise due to shear measurements. In \cite{Repp2021varcovar}, the variance and covariance among galaxy counts in cells was obtained for the case of \change{an
averaged} cell correlation, which has previously been considered in \cite{Codis16a}. As we explain here, and derive in Appendix~\ref{app:large_sep_refine}, the  large-separation expansion in terms of bias functions adopted in \cite{Codis16a,Uhlemann17Kaiser,Repp2021varcovar,Repp2021indicator} can be generalised to the physically more realistic case of a separation-dependent correlation. The large-separation two-point PDF and the resulting super-sample covariance effect can also be related to the idea of the position-dependent matter PDF studied in \cite{Jamieson2020}.
Recent work of \cite{Bernardeau2022} determined covariances of density PDFs in hierarchical models, which can be used to understand the dominant terms in the covariance matrix of cosmological density fields in the strongly non-Gaussian regime. In this work, we build on this foundation but focus on the weakly non-Gaussian weak lensing PDF for which hierarchical models are less suitable and better results can be achieved using shifted lognormal models.

\subsection*{Structure}
In Section~\ref{sec:covPDF}, we discuss the general covariance modelling for the two-dimensional case of the weak lensing (or photometric galaxy clustering) PDF based on a model for the joint two-point PDF. In Section~\ref{sec:covPDF_largesep}, we use a large-separation expansion for this two-point PDF and link it to dominant terms in an eigendecomposition of the covariance matrix. In Section~\ref{sec:quant_tests}, we validate our covariance models using several quantitative tests. In Section~\ref{sec:covPDF_3D}, we present the basics for generalising our findings to the case of the three-dimensional matter (or spectroscopic galaxy clustering) PDF. We conclude in Section~\ref{sec:conclusion} and provide an outlook for how our results can be applied to related one-point observables, include tomography and information from two-point statistics.

\subsection*{Executive Summary}
\begin{figure}
    \centering
    \includegraphics[width=\columnwidth]{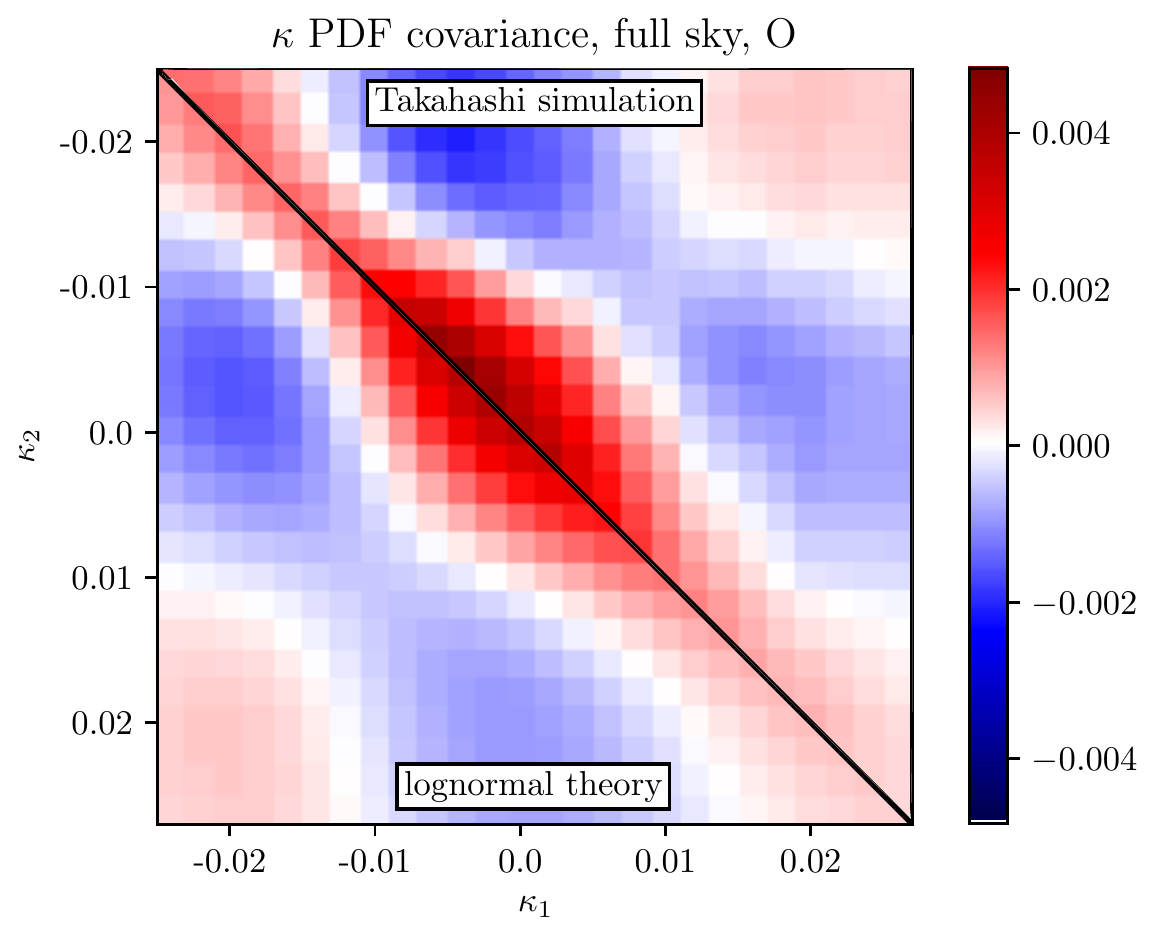}
    \caption{PDF covariance matrix for circular apertures of radius $\theta_s=7.5'$ at source redshift $z_s=2$ comparing the measurement from 108 full sky Takahashi lensing maps (upper triangle) to the shifted lognormal prediction (lower triangle).}
    \label{fig:PDFcovariance_Takahashi}
\end{figure}
\begin{figure}
    \centering
    \includegraphics[width=0.99\columnwidth]{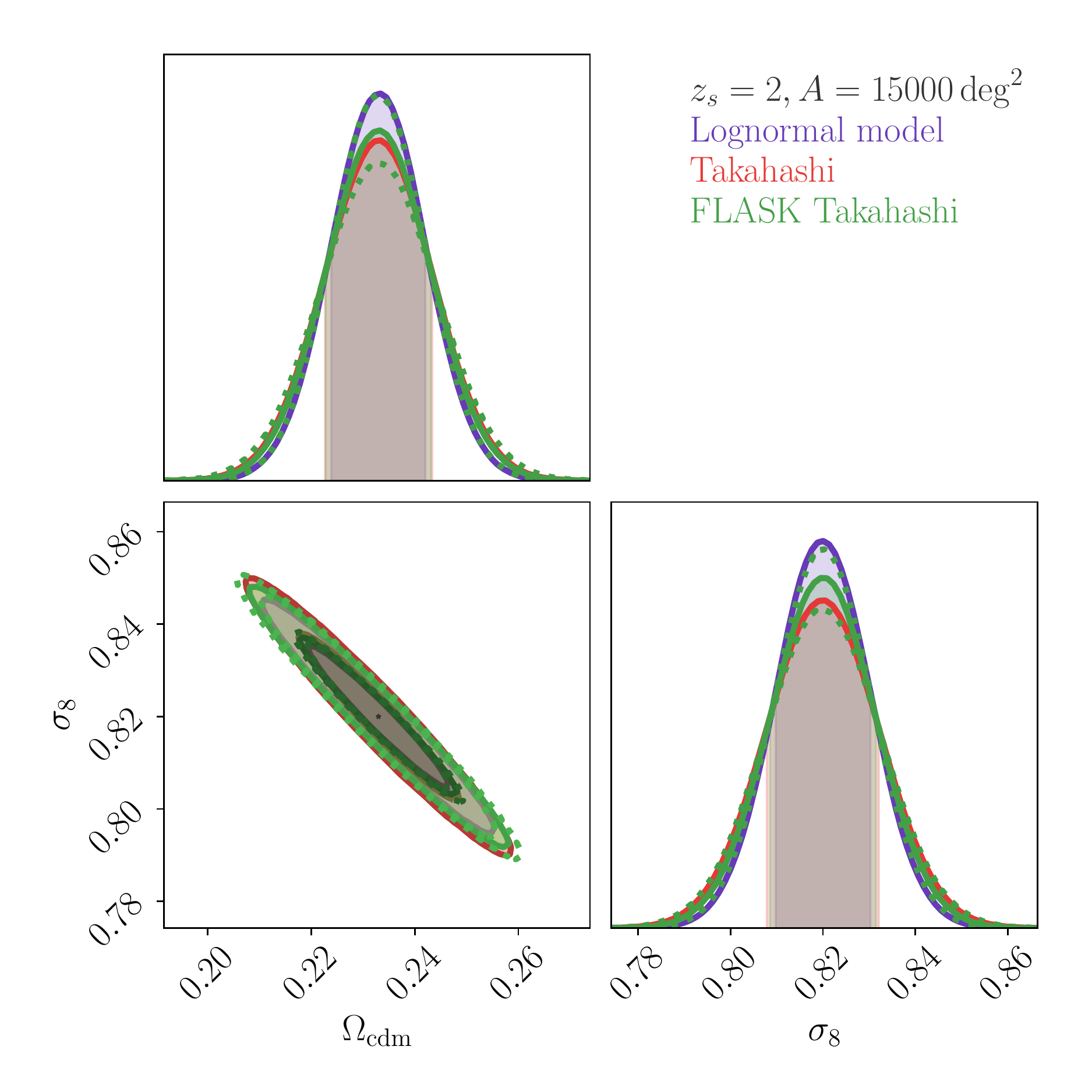}
    \caption{Fisher forecasts validating the predicted shifted lognormal PDF
    covariance (purple) against measured covariances from 108 realistic full sky simulated maps (red) and 500 associated mock FLASK maps (green). The parameter errors obtained from the raw measured covariances have additionally been corrected by the Hartlap factor~\eqref{eq:hartlap}. To highlight the variation that can result from small samples, we also show contours for two subsamples of 100 of the 500 mock FLASK maps (green dotted).}
    \label{fig:fisher_Takahashi_lognormal_FLASK}
\end{figure}
We develop a theoretical model for the covariance of measurements of the one-point weak lensing convergence PDF \change{on mildly nonlinear scales} in finite bins. Our model is based on an integration over the two-point PDF~\eqref{eq:covfromjointPDFmodel} at different separations and we rely on a joint shifted lognormal model for that two-point PDF~\eqref{eq:joint_LN} tuned to match the PDF skewness and shown to capture the density-dependence of the two-point correlation~\eqref{eq:jointPDFlargesep} in simulations, as illustrated in Figure~\ref{fig:kappa_b1_LN}. We  establish expressions for the {\it finite sampling}~\eqref{eq:covFS} and {\it super-sample covariance}~\eqref{eq:cov_SSC} contributions to the PDF covariance and their scaling with the inverse of the survey area. We validate our predictions against large sets of simulated convergence maps both from N-body simulations \citep{Takahashi17} and using the FLASK tool for generating log-normal random fields \citep{Xavier2016}. To verify the quality of our covariance model we employ a range of tests including visual comparisons, an eigendecomposition of the covariance, a band-decomposition of the precision matrix, $\chi^2$ tests and Fisher forecasts, with Figures~\ref{fig:PDFcovariance_Takahashi} and \ref{fig:fisher_Takahashi_lognormal_FLASK} summarising two main results. In Figure~\ref{fig:PDFcovariance_Takahashi} we show the measured covariance for the central region of the binned $\kappa$-PDF for a \change{mildly nonlinear} smoothing radius of $\theta_s=7.5'$ at source redshift $z_s=2$ from \change{all the} $N_{\rm sim}=108$ simulated full sky $\kappa$ maps from \cite{Takahashi17} (upper triangle) along with the shifted lognormal prediction (lower triangle, using equations~\eqref{eq:covfromjointPDFmodel}~and~\eqref{eq:joint_LN}). For the predictions, we used the cosmological parameters of the simulation along with the shift parameter $s\simeq 0.085$ obtained from the second and third moments measured from the PDF. In Figure~\ref{fig:kappa_b1_LN} in the main text we show that the environment-dependent clustering that enters the eigendecomposition of the covariance is well captured by the shifted lognormal model. 

Aside from the good visual agreement, we quantify their consistency using two tests. First, we test the compatibility of the diagonal elements of the two covariances shown in Figure~\ref{fig:PDFcovariance_Takahashi} with a $\chi^2$ test, taking into account their uncertainty and correlation among each other\footnote{Note that typically only the diagonal elements of estimated covariances are well described with multi-variate Gaussian noise.}
(covariance of the covariance). We use the covariance between the numerically estimated diagonal covariance elements $\hat C_{ii}$ and $\hat C_{jj}$ as given in equation~\eqref{eq:covariance_diagonal_error} following \cite{Taylor2013},
where $N_{\rm sim} = 108$ is the number of simulations used to estimate and the covariance elements $C_{ij}$ are evaluated with our theoretical model. Using the above expression to compute a $\chi^2$ between the diagonal of the estimated and predicted covariance we find $\chi^2=36.8$ for a number of $26$ bins. This corresponds to a marginal $\sim 1.\change{75}\sigma$ deviation between the two matrix diagonals. 

Finally, we assess the quality of the off-diagonal elements of our covariance model by performing a Fisher forecast. Figure~\ref{fig:fisher_Takahashi_lognormal_FLASK} compares Fisher contours on the parameters $\sigma_8$ and $\Omega_{\mathrm{cdm}}$ derived from a number of different covariance matrices: our covariance model \change{(purple)}, the covariance estimated from the \change{108} N-body simulations of \citet{Takahashi17} \change{(red)} and the covariance estimated from \change{500} log-normal simulations \change{(green)}. \change{For the simulations we use the maximum number of full sky maps available ($N_{\rm sim}=108$) which is not ideal for a PDF data vector of length 26. To mitigate this limitation, we include the Hartlap correction factor~\eqref{eq:hartlap} for the Fisher forecasts. Additionally, we estimated the covariance from 500 mock full sky maps generated from FLASK and checked compatibility of the results.} The differences in the width and alignment of the constraints obtained from these matrices is negligible compared to the overall parameter uncertainties. Hence, our covariance model is sufficiently accurate for application to observational data. We demonstrate the flexibility of our covariance model by adapting it to predict super sample covariance effects for the 3-dimensional matter PDF paving the way to spectroscopic galaxy clustering.

\section{Weak lensing PDF covariance}
\label{sec:covPDF}

The weak lensing convergence PDF is a well-suited complementary summary statistic that can extract non-Gaussian information from cosmic shear observations, thus complementing two-point statistics. In this work, we focus on mildly nonlinear scales, where  weak lensing convergence PDF and its response to fundamental cosmological parameters can be predicted at percent-level accuracy from large-deviation statistics as shown in \cite{Barthelemy2020,Boyle2021}. In the same work it was shown how one can use shifted lognormal convergence fields generated by FLASK \citep{Xavier2016} to match the convergence PDF and
obtain precise covariance estimates in good agreement with maps from N-body simulations, of which often only a very limited number is available. 
Following this work, we will study the covariance of the central region of the one-point PDF of the weak lensing convergence smoothed with a top-hat filter of radius $\theta_s=7.3'$ at a fixed source redshift $z_s=2$ using 26 linearly spaced bins in the range $-0.024<\kappa<0.026$. In this Section we will use both Gaussian and shifted lognormal fields generated with FLASK to test our theoretical framework for predicting the weak lensing PDF covariance based on different models for the two-point PDF of the weak lensing convergence in cells of a given separation. While we focus on the two-dimensional case of the weak lensing PDF here, the formalism can be adapted to describe the three-dimensional clustering PDF as we lay out in Section~\ref{sec:covPDF_3D}.

\subsection{Covariance prediction from joint 2-point PDF}
Before calculating the covariance matrix of a PDF measurement, let us first specify how this measurement is conducted. Let us assume that we have observed the gravitational lensing convergence field on a part of the sky with the area $A_{\mathrm{survey}}$~. Now let us smooth this convergence field with a circular aperture of some angular radius $\theta_s$  (potentially cutting away parts of the edges of the survey, so that each aperture falls fully inside the survey area), and let us pixelise the resulting map. Each pixel of that map contains the value of the average convergence within a radius $\theta_s$ around the pixel center.

We would now like to estimate the probability density of a certain value $\kappa$ for the smoothed convergence field. In practice one will need to compute this PDF within a set of finite bins, say around a set of central values $\kappa_i$ and with constant bin width $\Delta \kappa$, which results in the bins $\mathrm{bin}_i = [\kappa_i - \Delta\kappa/2, \kappa_i + \Delta\kappa/2]$~. The PDF at each $\kappa_i$ can then be approximated as
\begin{equation}
    \mP(\kappa_i) \approx \frac{P(\kappa \in [\kappa_i - \Delta\kappa/2, \kappa_i + \Delta\kappa/2])}{\Delta\kappa}\ ,
\end{equation}
where $\mP(\kappa)$ is the PDF (obtained in the limit $\Delta\kappa\rightarrow 0$) and $P(A)$ is the probability of statement $A$. With our pixelised map we can estimate the right-hand side of this to obtain the measurement
\begin{align}
\hat{\mP}(\kappa_i) &\equiv \frac{\# \lbrace \mathrm{pix\ with}\ |\kappa - \kappa_i| \leq \Delta\kappa/2 \rbrace}{\# \lbrace \mathrm{pix}\rbrace\Delta\kappa} \ .
\end{align}
Following \cite{Bernardeau2022}, we introduce the weight function
\begin{equation}
    \chi_i(\kappa) = \begin{cases} 1 & \mathrm{if}\ \kappa \in [\kappa_i - \Delta\kappa/2, \kappa_i + \Delta\kappa/2] \\ 0 & \mathrm{else}  \end{cases}\,,
\end{equation}
such that the estimator $\hat{\mP}(\kappa_i)$ is given by
\begin{equation}
\hat{\mP}(\kappa_i)
=\frac{\sum_P \chi_i(\kappa_{\rm P})}{N_{\rm P} \Delta\kappa}\approx \frac{1}{A_{\mathrm{survey}}} \underset{\mathrm{survey}}{\int} d^2 \bm{\theta}\ \frac{\chi_i(\kappa(\bm{\theta}))}{\Delta\kappa}\ ,
\end{equation}
where \rm P labels the $N_{\rm P}=A_{\rm survey}/A_{\rm pix}$ pixels of area $A_{\mathrm{pix}}$, $\bm{\theta}$ are angular coordinates on the sky, and $\kappa(\bm{\theta})$ is the smoothed, continuous convergence field at that location. The second moment of our above estimator is given by
\begin{align}
\label{eq:2nd_moment_without_Pd}
    & \langle \hat{\mP}(\kappa_i) \hat{\mP}(\kappa_j) \rangle\nonumber \\
    &\approx \frac{1}{A_{\mathrm{surv.}}^2} \underset{\mathrm{surv.}}{\int}d^2 \theta_1  \underset{\mathrm{surv.}}{\int} d^2 \theta_2\  \frac{\langle\chi_i(\kappa(\bm{\theta}_1)) \chi_j(\kappa(\bm{\theta}_2))\rangle}{(\Delta\kappa)^2}\nonumber \\
    &\approx \frac{1}{ A_{\mathrm{surv.}}^2} \underset{\mathrm{surv.}}{\int}d^2 \theta_1  \underset{\mathrm{surv.}}{\int} d^2 \theta_2\ \mP(\kappa_i, \kappa_j ; |\bm{\theta_1} - \bm{\theta_2}|)\ ,
\end{align}
where $\mP(\kappa_i,\kappa_j;\theta)$ is the joint PDF of two points of the smoothed convergence field at distance $\theta$. This equation can be simplified to
\begin{subequations}
\label{eq:covfromjointPDFmodel}
\begin{equation}
\langle\mP(\kappa_i)\mP(\kappa_j)\rangle = \int d\theta\, P_d(\theta) \mP(\kappa_i,\kappa_j;\theta)\ ,
\end{equation}
with $P_d(\theta)$ indicating the distribution of angular distances in a given survey area. The covariance of the PDF estimator can then be computed from this moment via
\begin{equation}
    \rm{cov}(\mP(\kappa_i),\mP(\kappa_j)) =\langle \hat\mP(\kappa_i)\hat\mP(\kappa_j)\rangle-\bar\mP(\kappa_i)\bar\mP(\kappa_j)\ ,
\end{equation}
\end{subequations}
where $\bar\mP\equiv\langle\hat\mP \rangle \approx \mP$~. 
We see that covariance predictions rely on two key ingredients: the distance distribution of angular separations and the two-point PDF of the lensing convergence at those separations, which will be discussed in more detail in Section~\ref{subsec:jointPDFmodels}.

{\it Distance distributions.} 
\begin{figure}
\centering
\includegraphics[width=\columnwidth]{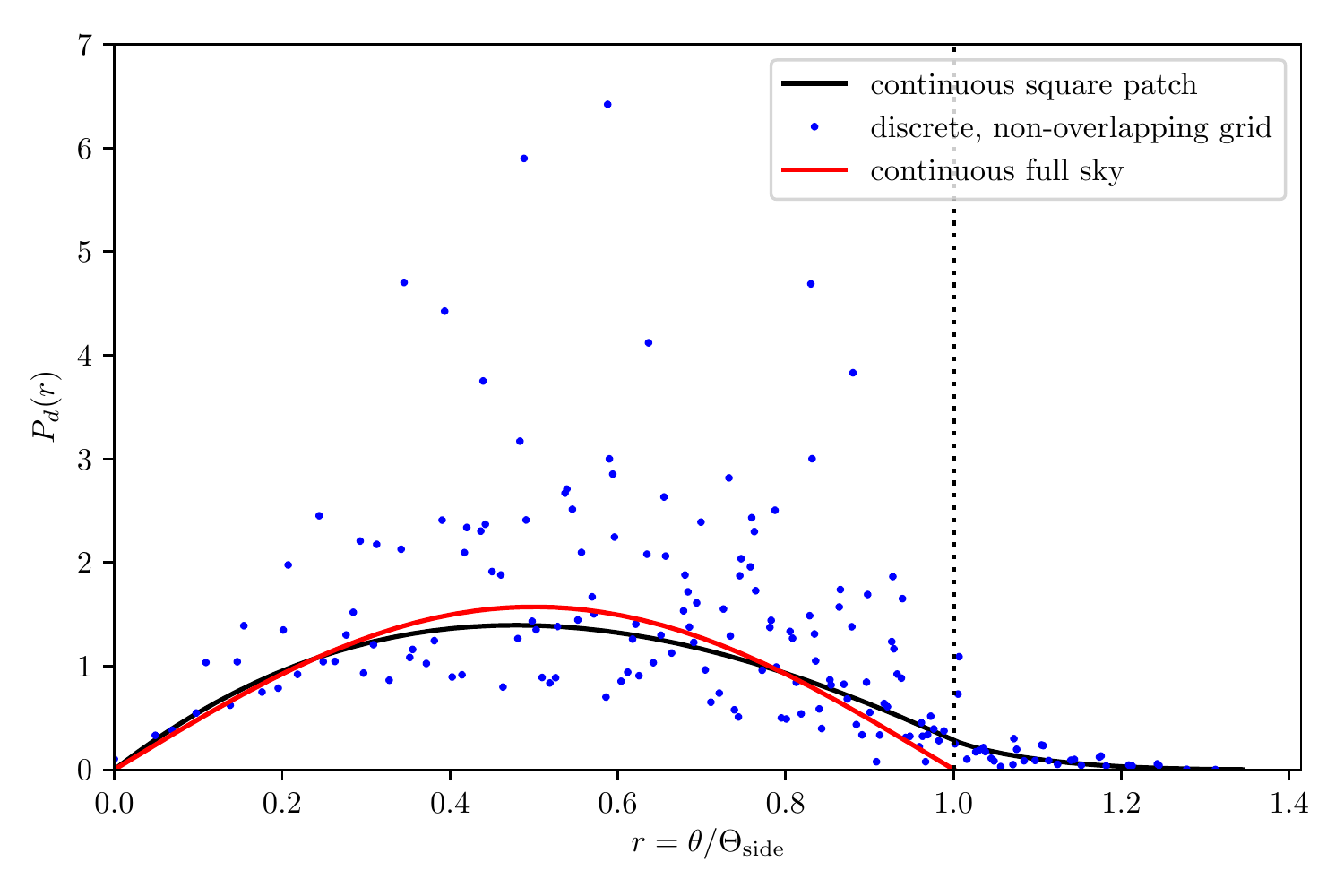}
\caption{Comparison of the continuous distribution of relative distances in a square Cartesian survey patch of side length $\Theta_{\rm side}=5$deg$=300'$ (black) and the discrete version for a coarse grid corresponding to non-overlapping circular cells of radius $7.3'$ (blue). We also show the distribution of relative distances on the full sky with maximum $\Theta_{\rm side}=\pi=180$deg (red).}
\label{fig:distance_dist_square_patch}
\end{figure}
The relevant distance distributions for weak lensing  describe a two-dimensional field. For the full sky, the uniform distribution of angular separations (measured in radians) reads
\begin{equation}
P_{d,\text{full sky}}(\theta)=\frac{1}{2}\sin\theta\,,
\end{equation}
with a maximum at $\theta_{\rm max}=\pi/2$. This distribution when written in terms of distances relative to the maximal distance $\Theta_{\rm side}=\pi$ is illustrated as red line in Figure~\ref{fig:distance_dist_square_patch}.

For a square survey patch, the distribution of angular distances relative to the size of the survey patch, $\hat\theta=\theta/\Theta_{\rm side}$, is
\begin{align}
    \label{eq:distancedist_square}
&P_{d,\text{square}}(\hat\theta)= \\
\notag &\begin{cases} 2\hat\theta[\pi+(\hat\theta-4)\hat\theta] & \hat\theta\in[0,1[\,, \\
2\hat\theta[\pi-2+4\sqrt{\hat\theta^2-1}-\hat\theta^2-4\rm{sec}^{-1}(\hat\theta)]
& \hat\theta\in [1,\sqrt{2}]\,.\end{cases}
\end{align}
This distribution is illustrated as black line  in Figure~\ref{fig:distance_dist_square_patch}, with a maximum at roughly half the patch size where $\hat\theta_{\rm max}\simeq 0.479$ and a linear growth for small distances, $P_{d,\text{square}}(\hat\theta)\simeq 2\pi \hat\theta$. 

Before looking at PDF covariance matrix examples in Section~\ref{subsec:cov_measurements} and introducing models for the two-point PDF in Section~\ref{subsec:jointPDFmodels}, let us discuss the main universal effects affecting the PDF covariance: i) the shot-noise effect related to the finite sampling of cells and ii) the super-sample covariance effect related to a finite survey area. 

\subsubsection{Finite sampling (shot noise) effects}
\label{subsec:SN}
Previously we discussed the covariance of the PDF in the continuum limit. In the more realistic discrete case, we estimate the PDF in a bin centred at $\kappa_i$ with width $\Delta_i$ (in our case linearly spaced bins $\Delta_i=\Delta$), so $\kappa\in [\kappa_i-\Delta_i/2,\kappa_i+\Delta_i/2]$ using a finite sample of $N_T$ cells that are placed on a regular grid in the survey area. The histogram-value of the PDF in a bin is obtained from averaging the continuous PDF $\mP(\kappa)$ over that bin
\begin{equation}
\label{eq:PDFbin}
\mP_i=\int_{-\Delta_i/2}^{\Delta_i/2} \hspace{-0.75cm} d\delta\kappa_i \frac{\mP(\kappa_i+\delta\kappa_i)}{\Delta_i} \simeq\mP(\kappa_i)\,,
\end{equation}
and for small bin sizes close to the value of the continuous PDF at the bin center $\bar\kappa_i$. To compute the covariance for the discrete case, we can discretise equation~\eqref{eq:covfromjointPDFmodel} by averaging the continuous joint PDF over the $\kappa$-bins and replacing the integration over the distance distribution by a sum-based average over the set of $N_d=N_T^2=\sum_{\theta\in\Theta} N_d(\theta)$ distances $\Theta$ of all grid cells with multiplicities $N_d(\theta)$ that are part of the survey patch
\begin{align}
\notag\langle\mP_i\mP_j\rangle &= \sum_{\theta \in \Theta} \frac{N_d(\theta)}{N_d}\!\!\! \int_{-\Delta_i/2}^{\Delta_i/2} \hspace{-0.75cm} d\delta\kappa_i \int_{-\Delta_j/2}^{\Delta_j/2} \hspace{-0.75cm} d\delta\kappa_j \frac{\mP(\kappa_i+\delta\kappa_i,\kappa_j+\delta\kappa_j;\theta)}{\Delta_i\Delta_j}\\
\label{eq:covfromjointPDFmodel_discrete}
&= \sum_{\theta \in \Theta} \frac{N_d(\theta)}{N_d} \mP_{ij}(\theta)\,,
\end{align}
where for small enough bins $\mP_{ij}(\theta)\simeq\mP(\kappa_i,\kappa_j;\theta)$.
In Figure~\ref{fig:distance_dist_square_patch} we show a comparison between the continuous distribution of distances in a square patch and the discrete case for a coarser grid of non-overlapping cells. We now split this sum over cell separations showing that the contribution from zero separation leads to a finite sampling or shot-noise term, and the remainder resembles the super-sample covariance contribution that dominates for the case of well-separated cells. Since the one-point PDF can be obtained from the two-point PDF by marginalisation, $\mP(\kappa_i)=\int d\kappa_j \mP(\kappa_i,\kappa_j;\theta)$, for zero separation we require that  $\mP(\kappa_i,\kappa_j;\theta=0)=\mP(\kappa_i)\delta_D(\kappa_i-\kappa_j)$. In the discrete case the joint two-point PDF at zero separation is given by $\mP_{ij}(\theta=0)=\bar\mP_i \delta_{ij}/\Delta_i$, required by the relation $\bar\mP_i = \sum_j \Delta_j \mP_{ij}(\theta=0)$.
The zero separation multiplicity equals the cell number $N_d(\theta=0)=N_T$, such that the zero-separation term in equation~\eqref{eq:covfromjointPDFmodel_discrete} produces a finite sampling (or shot-noise) covariance contribution\footnote{This shot-noise term was already described in \citep{Codis16a}.
Even if there was no spatial correlation in the sample, in the Poisson limit the covariance of finding $N_i$ cells in convergence bin 
$\text{cov}_{\rm SN}(N_i,N_j) = \bar N_i \delta_{ij}$,
where $\bar N_i$ is the mean number of cells expected to be in the convergence bin centred on $\bar\kappa_i$. The number of cells, $N_i$ in a given convergence bin of width $\Delta_i$ centered on $\kappa_i$ is related to the histogram-value of the binned PDF from equation~\eqref{eq:PDFbin}
such that $N_i=N_T\Delta_i \mP_i$ and similarly $\bar N_i=N_T\Delta_i\bar \mP_i$.}
\begin{equation}
\label{eq:covFS}
\text{cov}_{\rm FS}(\mP_i,\mP_j) =\frac{N_T}{N_d} \mP_{ij}(\theta=0)= \frac{\bar \mP_i}{\Delta_i N_T} \delta_{ij}\,.
\end{equation}
Unsurprisingly, this finite sampling term is inversely proportional to the number of cells. If the cells are placed at a fixed separation, an increase of the area corresponds to a  suppression of the shot-noise covariance term as $\sigma^2_{\rm FS}(\mP)\propto N_T^{-1}\propto \Theta_{\rm side}^{-2}\propto A_{\rm survey}^{-1}$, which is inversely proportional to the area.

While in principle the finite sampling term can be rendered arbitrarily small by using an increasing number of cells, this comes at the price of heavy cell overlaps at small separations. In this regime, a small-scale expansion with small $\Delta_\xi^{1/2}(\theta)=\sqrt{\sigma^2-\xi(\theta)}\ll\xi(\theta)\ll 1$ can be used to capture effects of heavy cell overlaps as discussed in \cite{Bernardeau2022}. In the closed-form two-point PDF models discussed in Section~\ref{subsec:jointPDFmodels} those overlap effects will be automatically included.

\subsubsection{Super-sample covariance effects}
\label{subsec:SSC}
\begin{figure}
    \centering
    \includegraphics[width=\columnwidth]{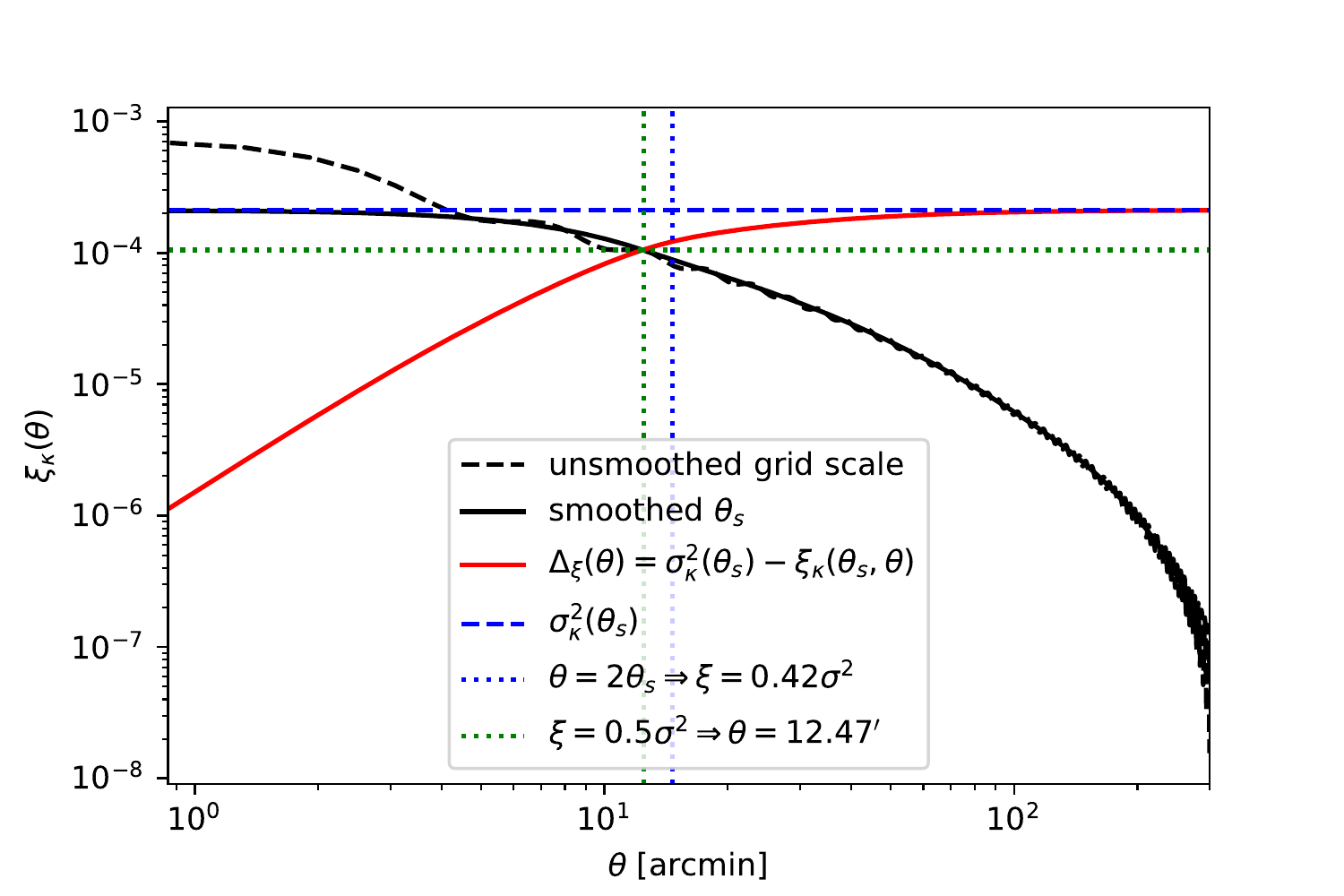}
    \caption{Two-point correlation of circular cells of radius $\theta_s\simeq 7.3'$ (solid black) in comparison to the matter correlation (dashed black), the variance in cells (dashed blue) and the difference $\Delta_\xi(\theta)$ (red line) that vanishes in the limit of small separations. The blue dotted lines indicate the point at which the cell separation equals twice the smoothing scale, while the green dotted lines indicate the point where the correlation is half of the variance, corresponding to slightly smaller scales.}
    \label{fig:expansion_correlation_variancediff}
\end{figure}

Beyond assuming specific (log-)normal models for the bivariate PDF, one can obtain accurate estimates of the two-point PDF in the large separation regime illustrated in Figure~\ref{fig:expansion_correlation_variancediff}. A large-scale expansion with small correlation $\xi(\theta)\ll\Delta_\xi(\theta)\ll 1$ is suitable for well-separated cells where the joint PDF at leading order can be approximated as \cite{Codis16a}
\begin{align}
\label{eq:jointPDFlargesep}
\mP(\kappa_i,\kappa_j;\theta\gtrsim 2\theta_{\rm s}) &\simeq  \mP(\kappa_i)\mP(\kappa_j)[1+\xi(\theta)b_1(\kappa_i)b_1(\kappa_j) ]\,,
\end{align}
with some bias function $b_1$, which describes how the two-point correlation at a given separation is modulated by the local weak-lensing convergences. This functional has been found to be very robust and the previous result extends what is expected in cosmological density fields \citep{1996A&A...312...11B}. When cells are at sufficiently large separations and the initial conditions are Gaussian, this modulation is independent of separation. We will introduce a discretised estimator in equation~\eqref{eq:b1_estimator} and derive the functional form of those bias functions for specific forms of the two-point PDF discussed in Section~\ref{subsec:jointPDFmodels} by an expansion in the correlation $\xi(\theta)$. For the related case of densities in spheres and cylinders, this bias function can be accurately predicted by spherical or cylindrical collapse, respectively. The large-deviation statistics techniques used to predict the one-point weak lensing PDF \citep{Barthelemy2020,Boyle2021} based on cylindrical collapse for thin redshift slices could be extended to derive the functional form of this bias, although this is beyond the scope of this work. This large-separation expansion predicts a leading order super-sample covariance term and further sub-leading contributions that we will describe in Section~\ref{sec:covPDF_largesep}. 
When all cells are at sufficiently wide separations $\theta\gtrsim 2\theta_s$, 
the covariance is obtained from integrating the large-separation limit of the two-point PDF~\eqref{eq:jointPDFlargesep} to obtain a super-sample covariance (SSC) covariance term
\begin{subequations}
\label{eq:cov_SSC}
\begin{equation}
\text{cov}_{\rm SSC}(\mP(\kappa_i),\mP(\kappa_j)) = \bar\xi\, (b_1\mP)(\kappa_i) (b_1\mP)(\kappa_j)  \,,
\end{equation}
where we defined the mean correlation
\begin{equation}
\label{eq:meanxi}
\bar\xi =\int^{\theta_{\rm max}} \hspace{-0.5cm} d\theta\, P_d(\theta) \xi(\theta)\,,
\end{equation}
\end{subequations}
with $P_d(\theta)$ being the probability distribution of angular distances and $\theta_{\rm \max}$ being the maximum distance set by the size of the survey patch, respectively. This mean correlation agrees with the variance of the mean of $\kappa$ measured across different patches, so $\bar\xi=\sigma_{\bar\kappa}^2$.

For a square patch of side length $\Theta_{\rm side}$, the maximal distance in the patch is $\theta_{\rm max}\simeq \sqrt{2} \Theta_{\rm side}$ where the normalised distance distribution $P_d(\hat\theta)$ is given in Equation~\eqref{eq:distancedist_square}, by conservation of probability we have 
\begin{equation}
P_d(\theta)=P_d(\hat\theta) \frac{d\hat\theta}{d\theta}= P_d\left(\hat\theta=\frac{\theta}{\Theta_{\rm side}}\right) \frac{1}{\Theta_{\rm side}} \stackrel{\hat\theta\ll 1}{\propto}  \Theta_{\rm side}^{-2} \,.
\end{equation}
The main contribution in the mean correlation integral comes from small angles $\theta$, where the correlation function is largest and $P_d(\hat\theta)\propto \hat\theta$. Hence, the SSC covariance term from equation~\eqref{eq:cov_SSC} scales as $\text{cov}_{\rm SSC}(\mP_i,\mP_j)\propto \Theta_{\rm side}^{-2} \propto A_{\rm survey}^{-1}$, which is inversely proportional to the survey area.

In the discrete case, the covariance can be expressed as a sum of the previously discussed finite sampling term~\eqref{eq:covFS} and a super-sample covariance like term
\begin{subequations}
\label{eq:covfromexp}
\begin{align}
\rm{cov}(\mP_i,\mP_j) &=\langle \mP_j\mP_j\rangle-\bar\mP_i\bar\mP_j \\
\label{eq:covcontributions}
&= \text{cov}_{\rm FS}(\mP_i,\mP_j) \\
\notag &+\frac{1}{N_d} \sum_{\theta\in \Theta\setminus \{0\}} \Big[
\mP_{ij}(\theta)-\bar\mP_i\bar\mP_j \underbrace{\frac{N_d}{N_T(N_T-1)}}_{\simeq 1}\Big]\
\,.
\end{align}
\end{subequations}
If cells at separation $\theta$ were completely independent, such that $\mP_{ij}(\theta)=\bar\mP_i\bar\mP_j$, then only the shot-noise term from the finite sampling of cells would remain. The second term in equation~\eqref{eq:covfromexp} includes significant contributions from cells at wide separation $\theta\gtrsim 2\theta_s$, where we can use the large-separation PDF expansion~\eqref{eq:jointPDFlargesep}. This leads to a super-sample covariance term analogous to equation~\eqref{eq:cov_SSC}
\begin{subequations}
\label{eq:cov_SSC_discrete}
\begin{equation}
{\rm cov}_{\rm SSC}(\mP_i,\mP_j)=\bar{\xi}(b_1\mP)_i(b_1\mP)_j\,,
\end{equation}
where $(b_1\mP)_i$ are obtained from a bin-average of the involved functions resembling equation~\eqref{eq:PDFbin} and the mean correlation can be obtained as
\begin{equation}
\label{eq:meanxi_discrete}
\bar\xi =\sum_{\theta\in\Theta\setminus\{0\}} \frac{N_d(\theta)}{N_d} \xi(\theta) 
 \,.
\end{equation}
Using the relationship $\left\langle\kappa'_j|\kappa_i\right\rangle_{\kappa'_j}=b_1(\kappa_i)/\bar\xi$, the value of the bias function in the bin $\kappa_i$ can be estimated by comparing the mean of the convergence at separation $\theta=2\theta_s$, $\kappa'_j$, from this given convergence $\kappa_i$ to the average correlation $\bar\xi$ using
a sum over the 6 nearest neighbour (NN) pixels at that separation 
\begin{equation}
    \label{eq:b1_estimator}
    \hat b_1(\kappa_i) = \frac{1}{\bar\xi}
     \frac{\sum_{\rm P} \sum_{\rm P'\in NN(2\theta_s)}\chi_i(\kappa_{\rm P})\kappa_{\rm P'}}{6\sum_{\rm P}\chi_i(\kappa_{\rm P})}\,.
\end{equation}
This definition is the analogue of the estimator for densities in spheres and cylinders \citep{Codis16a,Uhlemann17Kaiser,Uhlemann18cyl}.

\end{subequations}
Since the correlation function decreases with increasing angular separation and the probability distribution of cell distances peaks at roughly half the patch size, this super-sample covariance term is the most relevant for small patches while it is heavily suppressed for full sky maps. In some cases, like for simulation-based analysis or estimation of covariances, it is desirable to construct small patches which closely resemble the full sky behaviour. The main contribution from the SSC covariance term can be reduced by subtracting the mean $\kappa$ in every small patch, thus reducing the mean correlation $\bar\xi$ and modifying the shape of the bias function $b$ as we will discuss in Section~\ref{sec:cov_meansub}. 

\subsection{Illustrative covariance measurements}
\label{subsec:cov_measurements}

\begin{figure}
\centering
\includegraphics[width=\columnwidth]{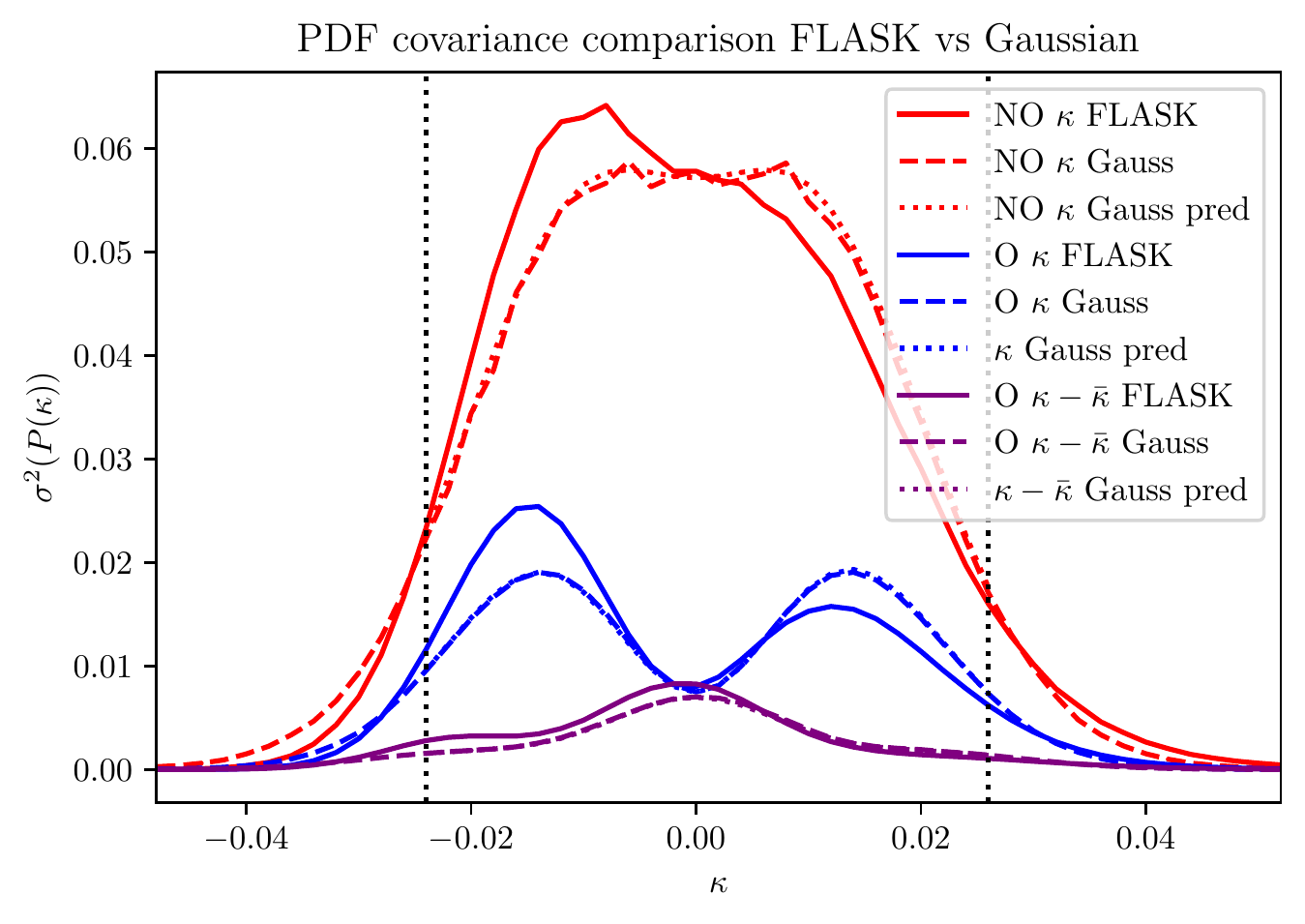}
\caption{Diagonal of the covariance matrix for the non-Gaussian FLASK PDFs (solid lines) and the Gaussian version (dashed lines) measured from small patches with non-overlapping cells (red), overlapping cells (blue) and overlapping cells with mean-subtraction in every patch (purple). The Gaussian case shows a symmetric covariance contribution as expected. The corresponding correlation matrices are shown in Figure~\ref{fig:kappa_covariance_correlation}. Additionally, we show our theoretical predictions for the Gaussian case (dotted lines, partially obscured by the dashed lines). The dotted vertical lines indicate the range of $\kappa$ values considered for the covariance matrices and Fisher forecasts.}
\label{fig:kappa_covariance_diag_FLASK_Gauss}
\end{figure}

\begin{figure*}
    \centering
    \includegraphics[width=\textwidth]{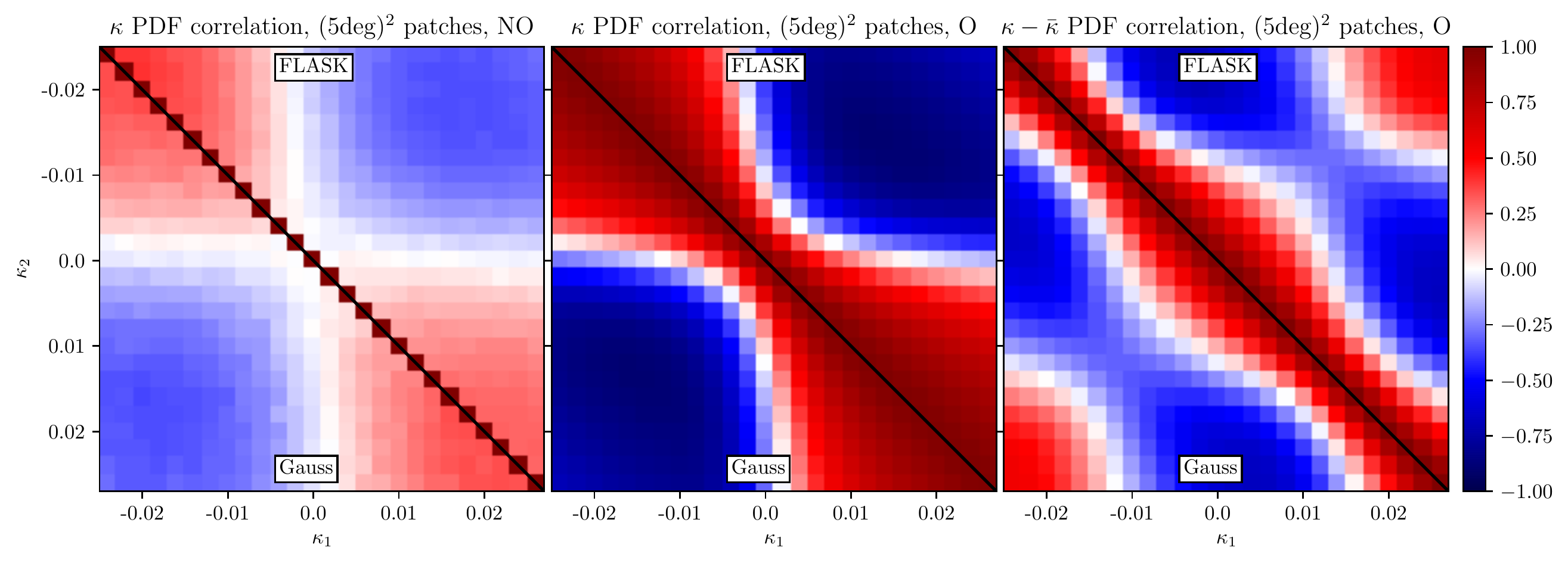}
    \caption{
    Correlation matrix between the bins of the weak lensing convergence $\kappa$ PDF measured at a smoothing scale $\simeq 7.3'$ 
    \change{measured from 10$\times$1200} (5 deg)$^2$ patches of \change{10 full sky FLASK maps} of $\kappa$ including the nonzero mean in the patches (left and middle panel) with non-overlapping (NO) cells (left) and overlapping cells (middle) and $\kappa-\bar\kappa$ after subtracting the mean with overlapping cells (right panel). The upper triangles labelled by FLASK show results from shifted-lognormal realisations, while the lower triangles those of Gaussian realisations. The diagonal of the covariance matrix is shown in Figure~\ref{fig:kappa_covariance_diag_FLASK_Gauss}.}
    \label{fig:kappa_covariance_correlation}
\end{figure*}

To validate our covariance models, we generate a large set of Gaussian as well as shifted lognormal (FLASK) $\kappa$ maps (with the shift parameter tuned to fit the $\kappa$-PDF skewness). We start from \change{10} full-sky convergence maps (healpix $N_{\rm side}=4096$ corresponding to $0.85'$ resolution) that we \change{each} divide into 1200 non-overlapping square patches with side-length of 5deg. \change{We chose the patch size to resemble the typical size of simulated convergence maps for which differing cosmologies are available \citep{Liu2018,Giocoli18,HarnoisDeraps2018}.} 
We build a data vector $\vec{S}$ from the values of the PDF histogram in the central region $-0.024<\kappa<0.026$ measured from the top-hat smoothed $\kappa$ map by considering the regular grid of all cells (overlapping) or only cells separated by at least twice the smoothing radius (non-overlapping). The covariance matrix of the binned $\kappa$-PDF across those patches is obtained using the formula
\begin{equation}
\label{eq:covariance}
C_{ij} = \langle (S_i-\bar{S}_i)(S_j - \bar{S}_j) \rangle\,,\quad \bar{S}_i = \langle S_i \rangle~\,,
\end{equation}
where $\langle\cdot\rangle$ indicates an ensemble average over different map realisations for the full sky or patches, respectively. We rescale all covariances to mimic a total Euclid-like survey area $A_{\rm survey}=15,000$ deg$^2$ using the rescaling factor $A_{\rm patch}/A_{\rm survey}$ for the patches and $A_{\rm fullsky}/A_{\rm survey}$ for the full-sky maps. We verified the validity of this rescaling using covariance measurements from full sky maps in addition to the theoretical arguments in Section~\ref{subsec:SSC}. For visualisation purposes, it is often useful to normalise the covariance matrix, with components $C_{ij}$, by its diagonal components, which defines the correlation matrix $R$ such that
\begin{equation}
R_{ij} = \frac{C_{ij}}{\sqrt{C_{ii}C_{jj}}}\,.
\end{equation}
In Figures~\ref{fig:kappa_covariance_diag_FLASK_Gauss}~and~\ref{fig:kappa_covariance_correlation} we illustrate two effects that impact the overall size and structure of the PDF covariance and the derived correlation matrix, \textit{finite sampling (shot noise)} and \textit{super-sample covariance}. In Figure~\ref{fig:kappa_covariance_diag_FLASK_Gauss} we show the diagonal of the covariance matrix for FLASK patches (solid lines) and Gaussian patches (dashed lines). In Figure~\ref{fig:kappa_covariance_correlation} we show the corresponding correlation matrices ordered by decreasing covariance, with the upper triangle showing the FLASK patches and the lower triangle showing the Gaussian patches. 

We consider three cases for the covariance in small patches of side length $5$deg:
\begin{itemize}
    \item raw $\kappa$ measured from non-overlapping cells (red lines in Figure~\ref{fig:kappa_covariance_diag_FLASK_Gauss}, left panel in Figure~\ref{fig:kappa_covariance_correlation}),
    \item raw $\kappa$ measured from overlapping cells (blue lines in Figure~\ref{fig:kappa_covariance_diag_FLASK_Gauss}, middle panel in Figure~\ref{fig:kappa_covariance_correlation}), and
    \item mean-subtracted $\tilde\kappa=\kappa-\bar\kappa$ from overlapping cells (purple lines in Figure~\ref{fig:kappa_covariance_diag_FLASK_Gauss}, right panel in Figure~\ref{fig:kappa_covariance_correlation})\,.
\end{itemize}
We see that while the lognormal maps have a visible asymmetry in the covariance diagonal compared to the Gaussian maps, both lead to very similar correlation matrices. Comparing the red and blue lines shows that given a fixed total area, increasing the number of cells for the PDF measurement decreases the finite sampling (or shot noise term) that determines the covariance matrix diagonal, as we explained in Section~\ref{subsec:SN}. The suppression of the diagonal for the case of an increased number of cells comes at the price of making cells overlap which creates a strong positive correlation band around the diagonal, as can be seen from comparing the left and middle panel in Figure~\ref{fig:kappa_covariance_correlation}. Comparing the blue and purple lines in Figure~\ref{fig:kappa_covariance_diag_FLASK_Gauss} illustrates that subtracting the mean in small patches significantly reduces the impact of the super-sample covariance term discussed in Section~\ref{subsec:SSC} and causes the 4-tiled pattern in the middle panel of Figure~\ref{fig:kappa_covariance_correlation} to transform to a 9-tiled pattern in the right panel. Additionally, the removal of this dominant term causes the overlap contribution in the band along the diagonal to become more apparent. As we will see later, the mean-subtracted convergence in small patches provides a  proxy for the convergence on the full sky, as evident from \change{comparing the right panel of Figure~\ref{fig:kappa_covariance_correlation} to Figures~\ref{fig:PDFcovariance_Takahashi}~and} ~\ref{fig:kappa_covariance_FLASK_theory}.
While most of our discussion is focused on the underlying weak lensing convergence field, we discuss the inclusion of shape noise arising from cosmic shear measurements in Section~\ref{subsec:shapenoise}.
\bigskip

\subsection{Specific models for the joint two-point PDF}
\label{subsec:jointPDFmodels}
As evident from equation~\eqref{eq:covfromjointPDFmodel}, the PDF covariance can be computed directly if the two-point PDF $\mP(\kappa_i,\kappa_j;\theta)$ is known a priori for all separations. In the following we will discuss three possible cases relying on closed-form expressions for that joint PDF, a bivariate Gaussian (normal) PDF, a bivariate shifted lognormal PDF and a bivariate PDF built from a marginal one-point PDF and a correlation structure, for example specified by an expansion for large separations or heavy overlaps.
    
\subsubsection{Bivariate Gaussian}
A Gaussian joint PDF is parametrised by the one-point PDF variance $\sigma^2$ and the two-point correlation between cells $\xi=\xi_{12}(\theta)$ such that 
    \begin{align}
    \label{eq:2ptPDFGauss}
    \mP_{\rm G}(\kappa_1,\kappa_2;\theta)&=\mP_G(\kappa_1,\kappa_2;\sigma^2,\xi_{12}(\theta))\\
    &=  \frac{\exp\left[-\frac{\sigma^2}{2d}[{\kappa_1}^2+{\kappa_2}^2]+\frac{\xi_{12}(\theta)}{d}\kappa_1\kappa_2\right]}{2\pi\sqrt{d}}\,,
    \end{align} 
    with the determinant $d=\sigma^4-{\xi_{12}}^2=\sigma^4[1-(\xi_{12}/\sigma^2)^2]$. One can perform a large-separation expansion for $\xi_{12}/\sigma^2\ll 1$ valid in the regime of no overlaps as shown in Figure~\ref{fig:expansion_correlation_variancediff}. At leading order, one can approximate $d\simeq\sigma^4$ and expand the remaining $\xi_{12}(\theta)$ term in the exponential to obtain
    \begin{align}
    \label{eq:Gaussian_largesep}
        \frac{\mP_{\rm G}(\kappa_1,\kappa_2;\theta)}{\mP_{\rm G}(\kappa_1)\mP_{\rm G}(\kappa_2)}&= 1+\xi_{12}(\theta)\frac{\kappa_1}{\sigma^2} \frac{\kappa_2}{\sigma^2} +\mathcal O(\xi^2)
        \,,
    \end{align} 
which is a meaningful expansion in the regime $\kappa\lesssim\mathcal O(\sigma)$ and reproduces the linear `Kaiser' bias function \citep{Kaiser84,Codis16b,Uhlemann17Kaiser}
\begin{equation}
\label{eq:b1_Gauss}
b_{1,\rm G}(\kappa)=\kappa/\sigma^2\,.
\end{equation} 
We will discuss a large-separation expansion up to higher orders in $\xi(\theta)/\sigma^2$ in  Section~\ref{sec:cov_Gauss}. 

{\it The impact of mean-subtraction.} 
For small patches, the mean convergence $\kappa_s$ will significantly fluctuate from patch to patch with a sizeable average correlation $\bar\xi=\sigma_{\kappa_s}^2$ causing the {\it super-sample covariance}. To better emulate measurements of modern surveys on a significant fraction of the full sky, subtracting this mean convergence in each patch in the spirit of the peak-background split \citep{BBKS86,MoWhite96PBS,ShethTormen99PBS} can be desirable.
The two-point PDF of the mean-subtracted lensing convergence $\tilde\kappa_i=\kappa_i-\bar\kappa$ is obtained from a trivariate Gaussian PDF for $\kappa_1,\kappa_2$ and the background value in the patch $\kappa_s=\bar\kappa$, which is then integrated out
\begin{subequations}
\begin{equation}
\label{eq:jointPDF_meansub_from_marginalisation_Gauss}
\mP_{\rm G}(\tilde\kappa_1,\tilde\kappa_2)=\!\int\!\! d\kappa_s \mP_{\rm G}(\kappa_1=\tilde\kappa_1+\kappa_s,\kappa_2=\tilde\kappa_2+\kappa_s,\kappa_s)\,.
\end{equation}
The covariance matrix is of the form
\begin{equation}
\label{eq:trivariatecovariance}
    \boldsymbol{\Sigma}_{\kappa_1,\kappa_2,\kappa_s}=\begin{pmatrix} \sigma^2 & \xi_{12} & \xi_{1s} \\
    \xi_{12} & \sigma^2 & \xi_{2s}\\
    \xi_{1s} & \xi_{2s} & \sigma^2_s
    \end{pmatrix}\,,
\end{equation} 
\end{subequations}
where we assume $\bar\xi=\sigma^2_s=\xi_{1s}=\xi_{2s}$, which holds for large enough patches compared to the cell size.
The integration yields a bivariate Gaussian PDF for the mean-subtracted convergence $\mP(\tilde\kappa_1,\tilde\kappa_2)$ with reduced variance $\sigma_{\tilde\kappa}^2=\sigma^2-\bar\xi$ and reduced correlation  $\xi_{\tilde\kappa}(\theta)=\xi(\theta)-\bar\xi$, where the reduction is set by the variance at the patch scale, $\bar\xi=\sigma_s^2$. The mean-subtracted correlation $\xi-\bar\xi$ now leads to a vanishing leading order super-sample covariance term from equation~\eqref{eq:covGauss_leading} and creates a new effective leading order term that we will discuss in Section~\ref{sec:cov_Gauss_meansub}. Obviously this above described formalism has a long history which goes back to early papers on constrained Gaussian random fields e.g \citep{1987ApJ...323L.103B} and \citep{1991ApJ...380L...5H} 

{\it Covariance.} When obtaining the covariance according to equation~\eqref{eq:covfromjointPDFmodel} from the large-separation expansion of the Gaussian~\eqref{eq:Gaussian_largesep}, the leading $1$ disappears and the correlation $\xi_{12}(\theta)$ is integrated over the distance distribution to yield $\bar\xi$ which sets the amplitude of a super-sample covariance like term we will discuss further in Section~\ref{subsec:SSC}
\begin{equation}
\label{eq:covGauss_leading}
\text{cov}_{\rm G}(\mP(\kappa_i),\mP(\kappa_j))\!=\! \bar{\xi} (b_{1}\mP)_{\rm G}(\kappa_i)(b_{1}\mP)_{\rm G}(\kappa_j)+\mathcal O(\overline{\xi^2})\,.
\end{equation}

\begin{figure*}
    \centering
    \includegraphics[width=\textwidth]{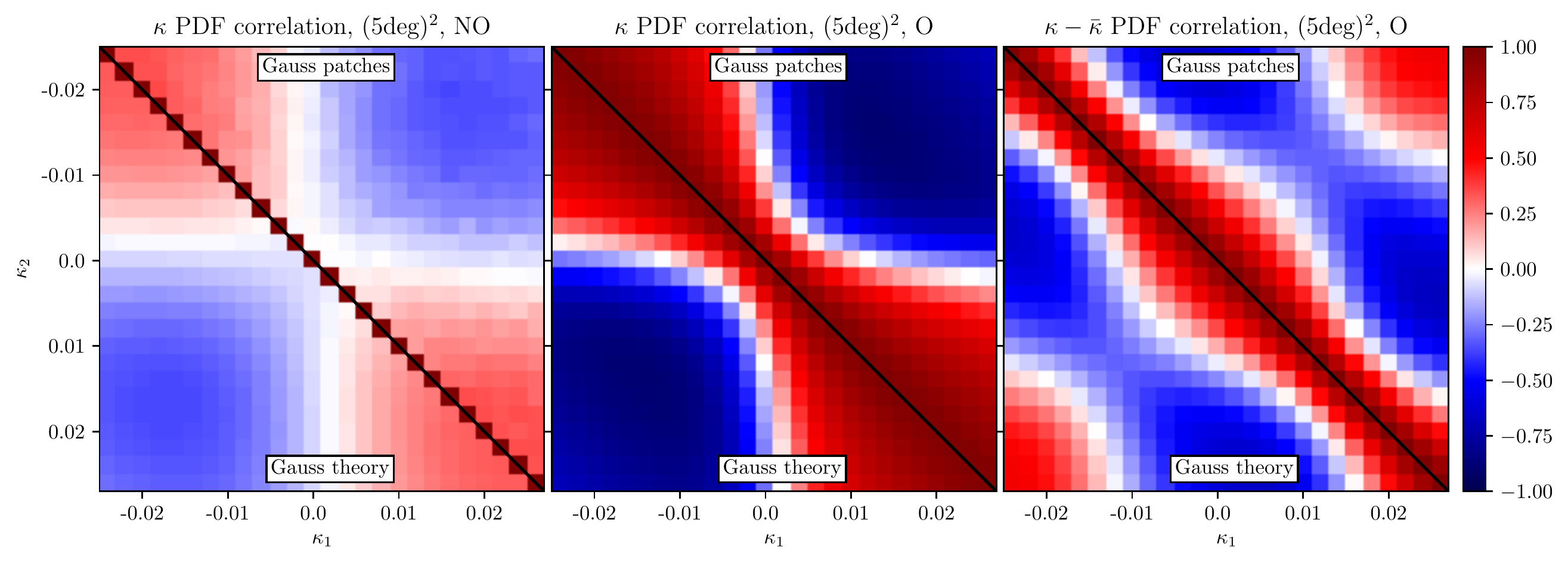}
    \caption{Correlation matrix between the bins of the Gaussian weak lensing convergence $\kappa$ PDF at a smoothing scale $\simeq 7.3'$ measured from \change{10$\times$}1200 (5 deg)$^2$ patches \change{from} 10 full sky Gaussian maps (upper triangle) and the theory prediction (lower triangle) for $\kappa$ including the nonzero mean in the patches (left and middle panel) with non-overlapping (NO) cells (left) and overlapping cells (middle) and $\kappa-\bar\kappa$ after subtracting the mean with overlapping cells (right panel). The predictions for the diagonal of the covariance matrices are shown in Figure~\ref{fig:kappa_covariance_diag_FLASK_Gauss}.}
    \label{fig:kappa_covariance_correlation_Gauss}
\end{figure*}

In Figure~\ref{fig:kappa_covariance_correlation_Gauss} we compare the correlation matrices as measured (upper triangle) and predicted (lower triangle) for Gaussian fields of the three cases discussed before (generated with FLASK, with the same initial power spectrum and smoothing as before), from non-overlapping cells with raw $\kappa$ (left) along with overlapping cells of raw $\kappa$ (middle) and mean-subtracted $\tilde\kappa=\kappa-\bar\kappa$ (right panel). In the non-overlapping case on the left-hand side, we can see how the finite sampling term~\eqref{eq:covFS} causes the diagonal to be most pronounced with a sub-dominant 4-tile pattern showing the leading order super-sample covariance like term from equation~\eqref{eq:covGauss_leading}. When making cells overlap as shown in the middle panel, the finite sampling term becomes irrelevant and the 4-tile pattern [due to the factorisation of that covariance into a product of two independent bias functions with one zero-crossing each] as seen in equation~\eqref{eq:covGauss_leading}] becomes more pronounced, while the resulting cell overlap causes a significant increase of the correlation in nearby bins as can be seen in a band along the diagonal. Considering the mean-subtracted convergence shown in the right panel will remove the leading order, super-sample covariance term from equation~\eqref{eq:covGauss_leading} and instead lead to a 9-tile pattern [due to the factorisation of that covariance contribution into a products of two bias functions with two zero-crossings each, see equation~\eqref{eq:cov_nonG}].

\subsubsection{Bivariate shifted lognormal}
\label{subsubsec:shiftedLN}
A shifted lognormal joint PDF for the lensing convergence $\kappa$ is parametrised by the variance $\sigma^2_\kappa$ and two-point correlation $\xi_\kappa=\xi_{12,\kappa}(\theta)$ along with a shift parameter $s$ related to the skewness, which we will keep fixed throughout here. We can define a Gaussian-distributed variable $g(\kappa)$ with zero-mean
\begin{equation}
\label{eq:LN_variable}
g(\kappa)=\ln\left(\frac{\kappa}{s}+1\right)-\mu_{\rm G} =\ln\left(\frac{\kappa}{s}+1\right)+\frac{\sigma_{\rm G}^2}{2}\,,
\end{equation}
where $\mu_{\rm G}=-\sigma_{\rm G}^2/2$ is fixed to ensure $\langle\kappa\rangle=0$ and the Gaussian variance is $\sigma_{\rm G}^2=\ln\left(1+\sigma^2_\kappa/s^2\right)$.
This can be used to construct the shifted lognormal univariate PDF as
\begin{align}
\label{eq:1ptPDF_LN}
    \mP_{\rm LN}(\kappa;\sigma^2_G,s) = \frac{\Theta(\kappa+s)}{\sqrt{2\pi}\sigma_{\rm G}(\kappa+s)}\exp\left[\frac{-g(\kappa)^2}{2\sigma_{\rm G}^2}\right]\,,
\end{align}
where $\Theta$ indicates the Heaviside step function, such that the PDF is zero if $\kappa\leq -s$. 
The shift parameter, $s$, can be set to replicate the desired skewness \citep[see the formula for $\lambda$ in ][]{Xavier2016}
\begin{align}\label{eq:xavier_lambda}
    s &= \frac{\sigma_\kappa}{\tilde{\mu}_3}\left(1+y(\tilde{\mu}_3)^{-1}+y(\tilde{\mu}_3)\right)-\langle \kappa\rangle \approx 3\frac{\sigma_\kappa^4}{\langle\kappa^3\rangle}, \\
    y(\tilde{\mu}_3) &= \sqrt[3]{\frac{2+\tilde{\mu}_3^2+\tilde{\mu}_3\sqrt{4+\tilde{\mu}_3^2}}{2}}\approx 1\,, \notag
\end{align}
where $\langle \kappa\rangle$ is the desired mean, $\sigma$ the target variance and $\tilde \mu_3=\langle\kappa^3\rangle/\sigma_\kappa^3$ the skewness. The shift parameter is typically positive and for our fiducial weak lensing convergence case we use $s=0.1145$ throughout the main text as done in \cite{Boyle2021}. In the executive summary, we use a different shift parameter because the \cite{Takahashi17} simulations have a different cosmology. The Gaussian limit can be reproduced by sending $s/\sigma_\kappa\rightarrow \infty$.

\begin{figure}
\centering
\includegraphics[width=\columnwidth]{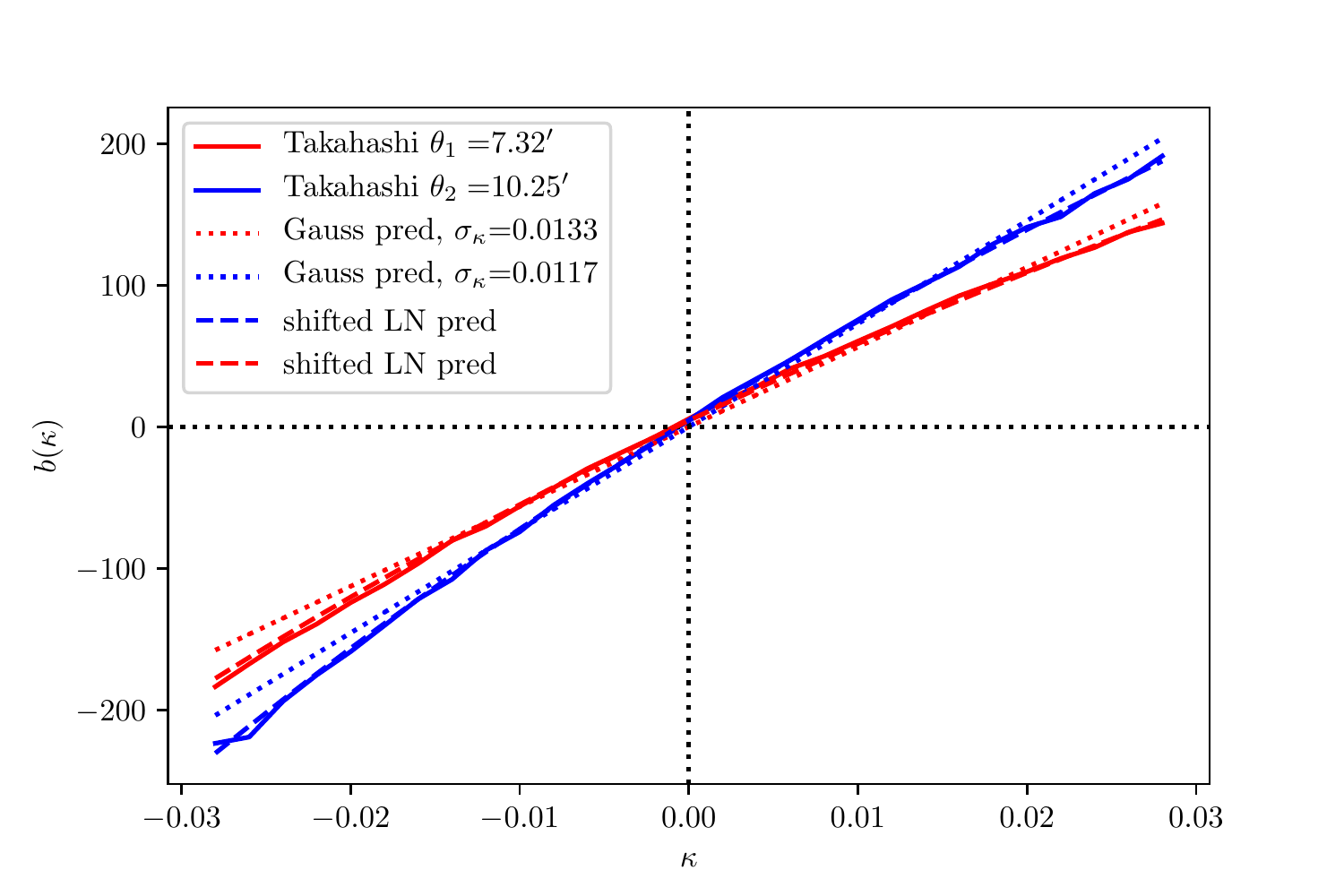}
\caption{$\kappa$-dependent clustering as measured by the cell bias from one full-sky map in the Takahashi simulation 
in comparison to the theoretical prediction for a shifted lognormal field from equation~\eqref{eq:b1_LN}.}
\label{fig:kappa_b1_LN}
\end{figure}

The bivariate shifted lognormal distribution also depends on the correlation $\xi_{\rm G}=\ln\left(1+\xi_\kappa/s^2\right)$ and the determinant $d=\sigma^4_{\rm G}-\xi_{\rm G}^2$ 
    \begin{align}
    \label{eq:joint_LN}
    \mP_{\rm LN}&(\kappa_1,\kappa_2;\sigma^2_G,\xi_G,s)=\frac{\Theta(\kappa_1+s)\Theta(\kappa_2+s)}{2\pi\sqrt{d}(\kappa_1+s)(\kappa_2+s)}\\
    \notag &\times \exp\left[-\frac{\sigma_{\rm G}^2}{2d}[g(\kappa_1)^2+g(\kappa_2)^2]+\frac{\xi_G}{d} g(\kappa_1)g(\kappa_2)\right]\,.
    \end{align}
The joint shifted lognormal PDF can be expanded at large  separations where $\xi_G(\xi_\kappa)/\sigma_G^2\ll 1$ to obtain up to leading order in $\xi_\kappa$
    \begin{align}
    \label{eq:joint_LN_largesep}
    \frac{\mP_{\rm LN}(\kappa_1,\kappa_2;\theta)}{\mP_{\rm LN}(\kappa_1)\mP_{\rm LN}(\kappa_2)}
    &=1+\frac{\xi_\kappa(\theta)}{s^2}\frac{g(\kappa_1)}{\sigma_G^2}\frac{g(\kappa_2)}{\sigma_G^2}+\mathcal O(\xi_\kappa^2)\,.
    \end{align}
In particular, it is useful to define a leading order bias function  resembling the Gaussian expression $b_{1,\rm G}=\kappa/\sigma^2$
    \begin{align}
    \label{eq:b1_LN}
    b_{1,\rm LN}(\kappa)&=\frac{g(\kappa)}{s\sigma_G^2}\simeq \frac{sg(\kappa)}{\sigma_{\kappa}^2}
    \,,
    \end{align}
where for $\kappa\ll s$ we have $sg(\kappa)\simeq \kappa+\sigma_\kappa^2/(2s)$ and hence reproduce the Gaussian result modulo a small shift. For larger $\kappa$ the shifted lognormal result deviates from the Gaussian one to closely resemble measurements from simulated $\kappa$ maps \citep{Takahashi17} using the eestimator~\eqref{eq:b1_estimator} as illustrated in Figure~\ref{fig:kappa_b1_LN}. 

{\it The impact of mean-subtraction.} \label{subsec:covariance_meansub_LN}

\begin{figure*}
\centering
    \includegraphics[width=1.15\columnwidth]{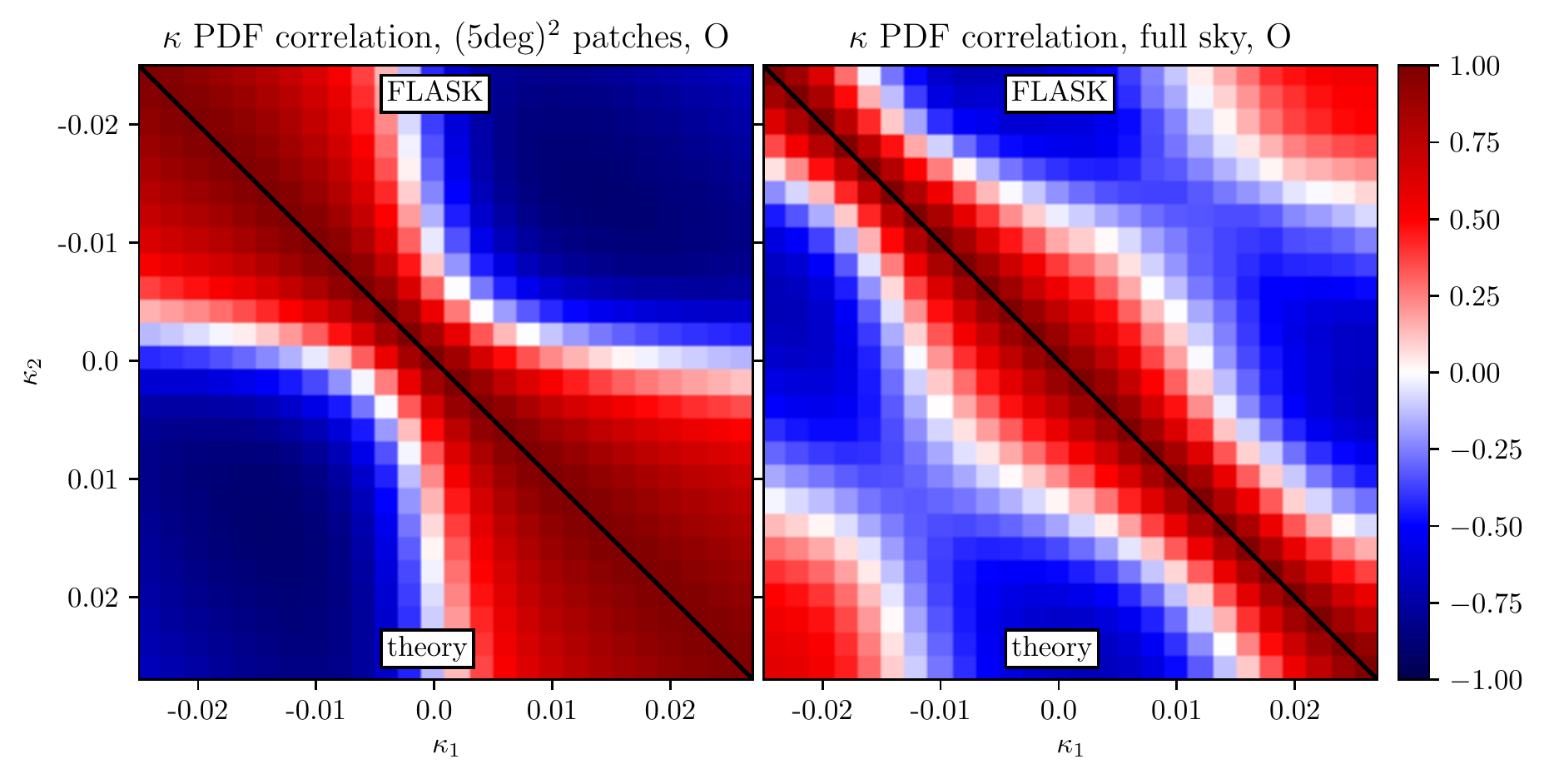}
    \includegraphics[width=0.85\columnwidth]{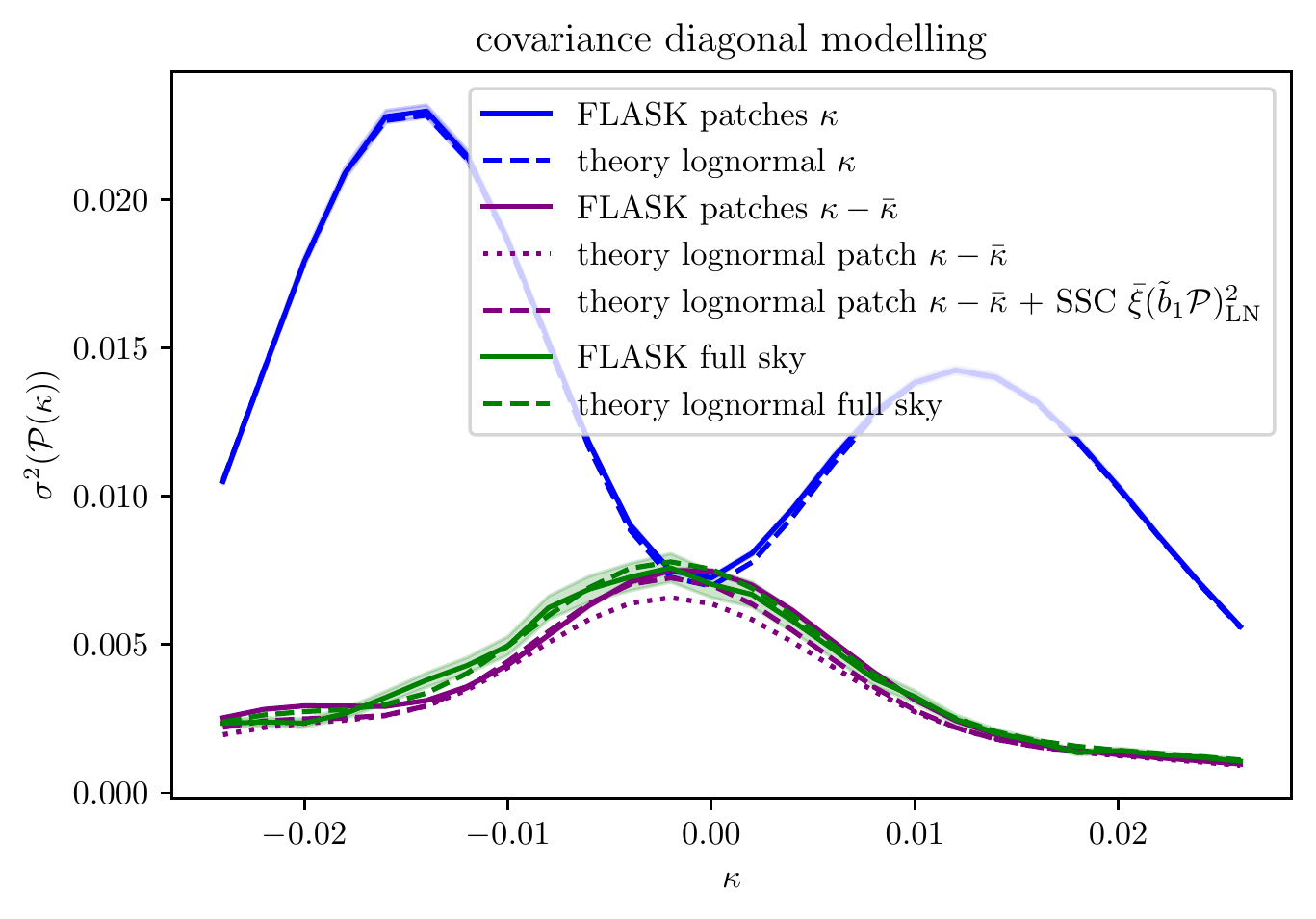}
    \caption{(\change{Left}) Correlation matrix between the bins of the shifted lognormal weak lensing convergence $\kappa$ PDF at a smoothing scale $\simeq 7.3'$ comparing FLASK measurements (upper triangles) to the theory predictions (lower triangles). 
    (left panel) Measurements from \change{20$\times$}1200 square patches of \change{(5deg)$^2$} size cut from \change{20} full sky realisations in comparison to the lognormal theory prediction in patches. While the first order bias function $b_{1,\rm LN}$~\eqref{eq:b1_LN} creates the 4-tile pattern, cell overlaps create the positive correlation band along the diagonal.
    (right panel) Measurements from 500 full sky realisations in comparison to the lognormal theory prediction on the full sky. The 9-tile pattern is set by the second order bias function $b_{2,\rm LN}$~\eqref{eq:q1b2_LN}, while cell overlaps create the strong correlation along the diagonal.
    (\change{Right}) Diagonal of the covariance matrix for the non-Gaussian FLASK PDFs measured from the full sky (green solid) and as predicted by the lognormal theory (green dotted). We also show the result for the FLASK patches without/with mean subtraction (purple/blue solid) and the lognormal theory computed for the patches without/with mean-subtraction (purple/blue dotted). The shaded bands are indicating estimated errors)}
    \label{fig:kappa_covariance_FLASK_theory}
\end{figure*}

When describing small patches which will feature different background `sample' convergence values $\kappa_s$, one needs to integrate over a trivariate PDF of the convergences $\kappa_{1/2}$ and the sample value $\kappa_s$, similarly as for the mean-subtraction in the Gaussian case \eqref{eq:jointPDF_meansub_from_marginalisation_Gauss}
\begin{align}
\label{eq:jointPDF_meansub_from_marginalisation_LN}
\mP(\tilde\kappa_1,\tilde\kappa_2)&=\!\!\int\!\! d\kappa_s \mP_{\rm LN}(\kappa_1\!=\!\tilde\kappa_1+\kappa_s,\kappa_2\!=\!\tilde\kappa_2+\kappa_s,\kappa_s)\,.
\end{align}

To evaluate this, let us assume that $\tilde\kappa$ is a shifted log-normal random variable \citep{Hilbert2011}, i.e. given in terms of a zero-mean Gaussian random variable $\tilde g=\ln(\tilde\kappa/\tilde s+1)+\tilde\sigma_G^2/2$ with variance $\tilde\sigma_{\rm G}^2=\ln\left(1+\sigma_{\tilde\kappa}^2/\tilde s^2\right)$ and a new lognormal shift $\tilde s$.
The correlation of the Gaussian variables is given by $\tilde\xi_{\rm G}=\ln(1+\xi_{\tilde\kappa}/\tilde s^2)$. To find an effective shifted lognormal approximation for $\tilde\kappa$, we must determine variance, correlation and the shift parameter $\tilde s$.
Recall that we know the variance of $\kappa \equiv \kappa_s + \tilde\kappa$ from any procedure for calculating the matter power spectrum. In the same way, we can also know the covariance $\langle \kappa_i \kappa_s\rangle=\xi_{is}$ and the variance $\langle \kappa_s^2 \rangle=\sigma^2_{s}$, which we will approximate as identical $\xi_{is}=\sigma^2_s=\bar\xi$. So we get the following variance and correlation for the mean-subtracted $\tilde\kappa$
\begin{subequations}
\label{eq:var_covar_meansub_LN}
\begin{align}
\sigma_{\tilde \kappa}^2 &= \langle \tilde\kappa^2 \rangle
= \langle \kappa^2 \rangle - 2\langle \kappa \kappa_s\rangle +\langle \kappa_s^2 \rangle 
\approx \sigma_\kappa^2-\bar\xi\,,\\
\xi_{\tilde\kappa} &= \langle \tilde\kappa_1\tilde\kappa_2 \rangle =\langle \kappa_1 \kappa_2\rangle  - \langle \kappa_1\kappa_s \rangle - \langle \kappa_2\kappa_s \rangle + \langle \kappa_s^2 \rangle\nonumber\\
&\approx\xi_\kappa-\bar\xi\,,
\end{align}
\end{subequations}
and hence the Gaussian variance $\tilde\sigma_{\rm G}^2= \ln[ 1 + (\sigma_\kappa^2-\bar\xi)/\tilde s^2]$ and correlation $\tilde\xi_{\rm G} = \ln[ 1 + (\xi_\kappa-\bar\xi)/{\tilde s^2}]$. The shift parameter can be related to the skewness as described in equation B.11 of \citet{Hilbert2011}
\begin{equation}
\label{eq:shiftkappameansub}
\langle \tilde \kappa^3 \rangle = \frac{3}{\tilde s} \langle \tilde\kappa^2 \rangle^2 + \frac{1}{\tilde s^3} \langle \tilde\kappa^2 \rangle^3 \approx\frac{3}{\tilde s} \langle \tilde\kappa^2 \rangle^2 \Rightarrow \tilde s\approx \frac{3\langle\tilde\kappa^2 \rangle^2}{\langle\tilde\kappa^3 \rangle}\,.
\end{equation}
Following \citet{Friedrich18} we use the shift parameter $\tilde s$ to match the skewness of $\tilde\kappa$, which is obtained from the third moment of the difference 
\begin{align}
\label{eq:mom3kappameansub}
\langle \tilde \kappa^3 \rangle =&\ \langle (\kappa - \kappa_s)^3 \rangle 
= \langle \kappa^3 \rangle - 3\langle \kappa^2\kappa_s \rangle + 3\langle \kappa\kappa_s^2 \rangle\,.
\end{align}
The third moment of $\kappa$ can be expressed in terms of the variance $\sigma_\kappa$ and original shift parameter $s$ using $\langle\kappa^3\rangle\approx 3\sigma_\kappa^4/s$. Additionally, we can compute the relevant mixed moments of a zero-mean normal random variable $Y$ and a lognormal  variable $Z=s_z(e^X-1)$ given in terms of a normal variable $X$. The relevant relations are $\langle Y(e^X-1)\rangle=\xi_{XY}$ and $\langle Y^2(e^X-1)\rangle=\xi_{XY}^2 \left(\sigma_X^2 + 1 - \sigma_X^{-2}\right)$ which yields $\langle Y(e^X-1)^2\rangle=0$.
This can be used to obtain $\langle \kappa^2\kappa_s \rangle=0$ and $\langle \kappa\kappa_s^2 \rangle\approx \bar\xi^2 \left(\sigma_{\rm G}^2 + 1 - \sigma_{\rm G}^{-2}\right)/s$. Inserting this into equations~\eqref{eq:shiftkappameansub}~and~\eqref{eq:mom3kappameansub} leads to a relation between the shift parameters $s$ and $\tilde s$, which we found to be practically identical in our case. 

{\it Covariance.} Again, we can integrate the large-separation expansion of the shifted lognormal~\eqref{eq:joint_LN_biasexp} over the distance distribution to obtain the covariance from equation~\eqref{eq:covfromjointPDFmodel} and eventually get at leading order
\begin{equation}
\label{eq:covlognormal_leading}
\text{cov}_{\rm LN}(\mP(\kappa_i),\mP(\kappa_j))= \bar{\xi} (b_{1}\mP)_{\rm LN}(\kappa_i)(b_{1}\mP)_{\rm LN}(\kappa_j)+\mathcal O(\overline{\xi^2})\,,
\end{equation}
which is in form analogous to the Gaussian result~\eqref{eq:covGauss_leading} despite the modified ingredients. This explains why the correlation matrix from small patches (with sizable $\bar\xi$) in the FLASK case closely resembles the Gaussian case, as seen in Figure~\ref{fig:kappa_covariance_correlation} while the diagonal is modified due to the non-Gaussian shape of the final convergence PDF as seen in Figure~\ref{fig:kappa_covariance_diag_FLASK_Gauss}.

In Figure~\ref{fig:kappa_covariance_FLASK_theory} we compare the measurements from FLASK maps to the theoretical predictions for shifted lognormal fields.
In the \change{left panel on the left side} we show the correlation matrix obtained from $1200$ FLASK patches of $5$deg each cut from 20 full sky realisations (upper triangle) along with the theoretical prediction (lower triangle) finding excellent agreement. We can see that the correlation matrix has a 4-tile structure. This is because the mean correlation $\bar\xi$ in the patches is sizeable such that the first order covariance term~\eqref{eq:covlognormal_leading} dominates the visual appearance. In the \change{right panel on the left side} we show the correlation matrix obtained from 500 full sky FLASK maps (upper triangle) along with the theoretical prediction (lower triangle) finding excellent agreement. We can see that the correlation matrix closely resembles the case of the mean-subtracted convergence measured in small patches, shown in the lower panel of Figure~\ref{fig:kappa_covariance_correlation}. This is because the mean correlation $\bar\xi$ on a full sky is so small that the first order covariance term~\eqref{eq:covlognormal_leading} becomes  subdominant compared to the second order term that we will discuss in Section~\ref{subsec:covariance_lognormal_NLO}. The latter term also appears as leading order contribution in the mean-subtracted case addressed in Section~\ref{subsec:covariance_meansub_nonG_NLO}.  
In the \change{right} panel we focus on the diagonal of the covariance matrix, comparing FLASK maps and theoretical predictions for different cases. This plot shows that while the PDF covariance of the raw $\kappa$ in small patches looks significantly different, 
the mean-subtracted $\tilde\kappa$ in small patches  resembles the full sky covariance. The theoretical prediction captures all three behaviours quite well, considering relative 1-$\sigma$ errors on the diagonal of the covariance matrix (shaded bands) following from the covariance between the numerically estimated diagonal covariance elements $\hat C_{ii}$ and $\hat C_{jj}$ as given in \cite{Taylor2013}
\begin{equation}
\label{eq:covariance_diagonal_error}
    \mathrm{Cov}\left(\hat C_{ii},\hat C_{jj}\right) = \frac{2 C_{ij}^2}{N_{\rm sim} - 1}\,.
\end{equation}
We notice that while the lognormal prediction for the full sky and the raw $\kappa$ in patches agree extremely well with the predictions, there are some minor discrepancies for the mean-subtracted $\tilde\kappa$ in small patches. As this discrepancy was absent for Gaussian fields shown in Figure~\ref{fig:kappa_covariance_diag_FLASK_Gauss}, we suspect that it is caused by a more complicated patch-to-patch correlation induced by cutting multiple patches from one full sky, which is unaccounted for in the theory describing the mean-subtracted convergence as another shifted lognormal random field.

\subsubsection{Bivariate Gaussian copula with general marginals}
In case neither the bivariate normal nor the shifted lognormal distributions can provide a good description of the target PDF as its univariate marginal, one can design a bivariate PDF by specifying the marginal and correlation structure independently.
A joint PDF can be built from a given marginal one-point PDF $\mP(\kappa)$ and a correlation structure encoded in a copula density $c_{12}$ which quantifies the correlation between ranked variables obtained from the cumulative distribution function $\mathcal C(\kappa)=\int^{\kappa}\! d\tilde\kappa\, \mathcal P(\tilde \kappa)$, such that
\begin{equation}
\label{eq:def_copula_biv}
\mP(\kappa_1,\kappa_2;\theta)=\mP(\kappa_1)\mP(\kappa_2) c_{12}(\mathcal C(\kappa_1),\mathcal C(\kappa_2);\theta)\,,
\end{equation}
where $\theta$ indicates angular separation as before.
If the $\kappa_i$ and $\kappa_2$ are independent, then the copula density is unity $c_{12,\rm indep}=1$. A Gaussian copula only depends on the distance through the pairwise correlation  $\xi_{12}(\theta)$, $c_{12,\rm G}(\mathcal C(\kappa_1),\mathcal C(\kappa_2);\xi_{12}(\theta))$ and is a known function which can be computed numerically although it has no analytical expression. 
Beyond those two cases, systematically constructing the dependence structure is possible in the large-separation regime, where the copula can be expanded in powers of the pairwise correlation $\xi_{12}(\theta)$ as we will further discuss in Section~\ref{sec:covPDF_largesep}.

\subsubsection{Inclusion of shape noise}
\label{subsec:shapenoise}
\begin{figure}
    \centering
    \includegraphics[width=\columnwidth]{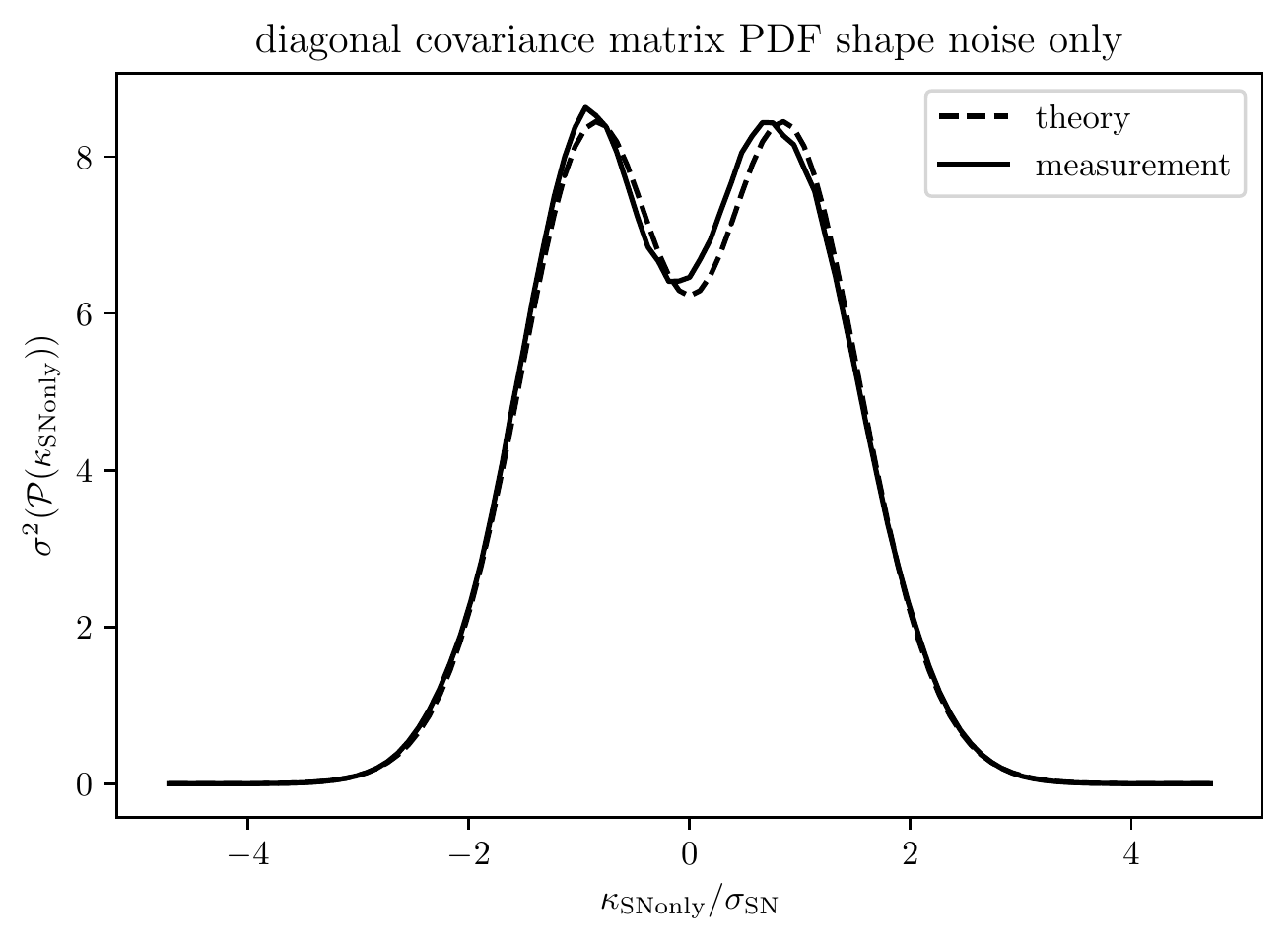}
    \includegraphics[width=\columnwidth]{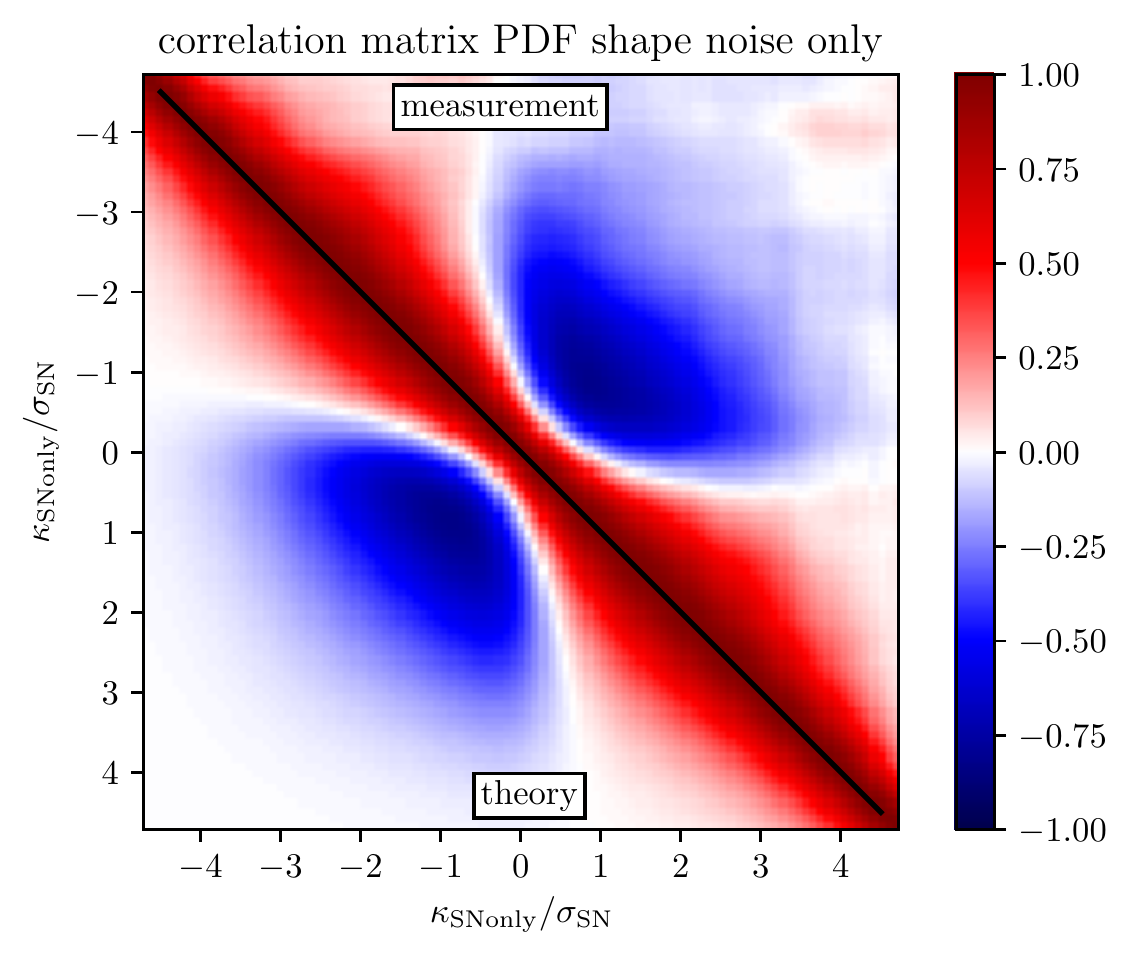}
    \caption{PDF covariance diagonal (upper panel) and correlation matrix (lower panel) for the pure shape-noise lensing convergence in circular apertures of radius $\theta_s$ comparing the prediction from equation~\eqref{eq:covPDFshapenoise_only} to the measurement from shape noise realisations. The slight apparent asymmetry between positive and negative $\kappa$ is caused by the noise due to the finite number of realisations.}
    \label{fig:PDFcovariance_shapenoise_only}
\end{figure}
Observationally, the weak lensing convergence map is obtained from cosmic shear measurements, which are subject to shape noise often constituting the dominant source of noise. Since galaxies  are intrinsically elliptical, the observed shear contains a contribution from this intrinsic signal, quantified by the variance of the intrinsic ellipticity $\sigma_\epsilon$. In simulated maps, shape noise can be included by adding a white noise -- modeled by a Gaussian random map $\kappa_{\rm SN}$ -- to the `raw' simulated convergence map $\kappa$ before smoothing. The standard deviation of this Gaussian is typically assumed to be \begin{equation}
    \label{eq:sigma_shapenoise}
    \sigma_{SN}= \frac{\sigma_\epsilon}{\sqrt{n_g \cdot \Omega_{\theta_s}}} \propto \theta_s^{-1}\,,
\end{equation}
with the shape-noise parameter $\sigma_\epsilon$ (here taken to be $\sigma_\epsilon=0.30$), the galaxy density $n_g$ (here taken to be $30$ arcmin$^{-2}$) and the solid angle $\Omega_{\theta_s}$ (here taken to correspond to circular apertures of radius $\theta_s=7.3'$). 

{\it Smoothing induced shape noise correlations and covariance.} At the raw $\kappa$ map level, the shape noise values in different pixels are uncorrelated and $\xi\equiv 0$. Top-hat smoothing the field at angular size $\theta_s$ will correlate shape noise values in overlapping cells (being at separation $\theta<2\theta_s$). Their correlation is related to the area of the cell overlap (which is a symmetric lens)
\begin{subequations}
\label{eq:correlation_shapenoise}
\begin{align}
    A(\theta<2\theta_s)=2\theta_s^2\arccos\left(\frac{\theta}{2\theta_s}\right)-\frac{\theta}{2}\sqrt{4\theta_s^2-\theta^2}\,.
\end{align}
The resulting correlation of the top-hat smoothed shot noise contribution to the weak lensing convergence is
\begin{equation}
    \xi_{\rm SN}(\theta< 2\theta_s)=\sigma_{\rm SN}^2A(\theta)/(\pi\theta_s^2)\,,\ \xi_{\rm SN}(\theta\geq 2\theta_s)=0\,.
\end{equation}
\end{subequations}
To obtain the covariance of the shot noise contribution alone, we integrate the bivariate Gaussian with variance $\sigma_{\rm SN}^2$ and covariance~\eqref{eq:correlation_shapenoise}
over distances following equation~\eqref{eq:covfromjointPDFmodel} to get
\begin{align*}
\langle\mP_{\rm SN}(\kappa_i)\mP_{\rm SN}(\kappa_j)\rangle &= \int_0^{2\theta_s}\!\!\!\!\!\! d\theta\, P_d(\theta) \mP_{\rm G}(\kappa_{i},\kappa_{j};\sigma_{\rm SN}^2,\xi_{\rm SN}(\theta)) \\
\notag&+ \mP_{\rm G} (\kappa_{i};\sigma_{\rm SN}^2)\mP_{\rm G} (\kappa_j;\sigma_{\rm SN}^2) \int_{2\theta_s}^\infty\!\!\!\!\!\! d\theta\, P_d(\theta) \,,
\end{align*}
where $P_d(\theta)$ indicates the distribution of angular distances in the survey area, which is approximately linear for small distances $P_d(\theta)\approx 2\pi\theta$. This gives the PDF covariance for the case of pure shape noise as
\begin{align}
\rm{cov}_{\rm SN}(\mP(\kappa_i),\mP(\kappa_j))&=  \int_0^{2\theta_s}\!\!\!\!\!\! d\theta\, P_d(\theta) \mP_{\rm G}(\kappa_{i},\kappa_{j};\sigma_{\rm SN}^2,\xi_{\rm SN}(\theta)) \notag\\
&-\mP_{\rm G} (\kappa_{i};\sigma_{\rm SN}^2)\mP_{\rm G} (\kappa_j;\sigma_{\rm SN}^2) \int_0^{2\theta_s}\!\!\!\!\!\! d\theta\, P_d(\theta)\,.
\label{eq:covPDFshapenoise_only}
\end{align} 
In Figure~\ref{fig:PDFcovariance_shapenoise_only} we compare this theoretical result to a measurement across 1000 shape noise realisations for square maps of length $5$deg with a top-hat smoothing of $7.3'$ finding excellent agreement.

{\it Shape noise induced convergence PDF convolution.} Shape noise impacts the one-point weak lensing convergence PDF as if it was convolved with a zero-centred Gaussian \citep{Clerkin16}
\begin{align}
\label{eq:PDFnoise}
    \notag \mathcal P_{\kappa+\rm SN}(\hat\kappa) &= (\mP_{\rm G}(\sigma_{\rm SN})*\mP)(\hat\kappa) \\
    &=\frac{1}{\sqrt{2\pi}\sigma_{SN}}\int d\kappa \exp\left(-\frac{(\hat\kappa-\kappa)^2}{2\sigma_{SN}^2}\right) \mathcal P(\kappa)\,,
\end{align}
where the shape noise standard deviation $\sigma_{SN}$ is given in equation~\eqref{eq:sigma_shapenoise}. For the two-point PDF of the smoothed weak lensing we similarly have $\hat\kappa_i=\kappa_i+\kappa_{i,SN}$ and the bivariate PDF is given in terms of a 2D-convolution
\begin{align}
\label{eq:jointPDF_incl_shapenoise}
\mP_{\kappa+\rm SN}(\hat\kappa_1,\hat\kappa_2;\theta)=\mathcal (\mP_{\rm G}(\sigma_{\rm SN},\xi_{\rm SN}(\theta))*\mP)(\hat\kappa_1,\hat\kappa_2;\theta)   \,,
\end{align}
where the shape noise contribution is described by a bivariate Gaussian PDF from equation~\eqref{eq:2ptPDFGauss} with standard  $\sigma_{\rm SN}$~\eqref{eq:sigma_shapenoise} and correlation $\xi_{\rm SN}(\theta)$~\eqref{eq:correlation_shapenoise}. If the joint PDF of raw $\kappa_i$ was Gaussian, then the noisy PDF of $\hat\kappa_i$ arising from the convolution would be another zero-mean Gaussian distribution with covariance matrix $\Sigma_{\kappa+\rm SN}=\Sigma_{\kappa}+\Sigma_{\rm SN}$. 

\begin{figure}
\centering
\includegraphics[width=\columnwidth]{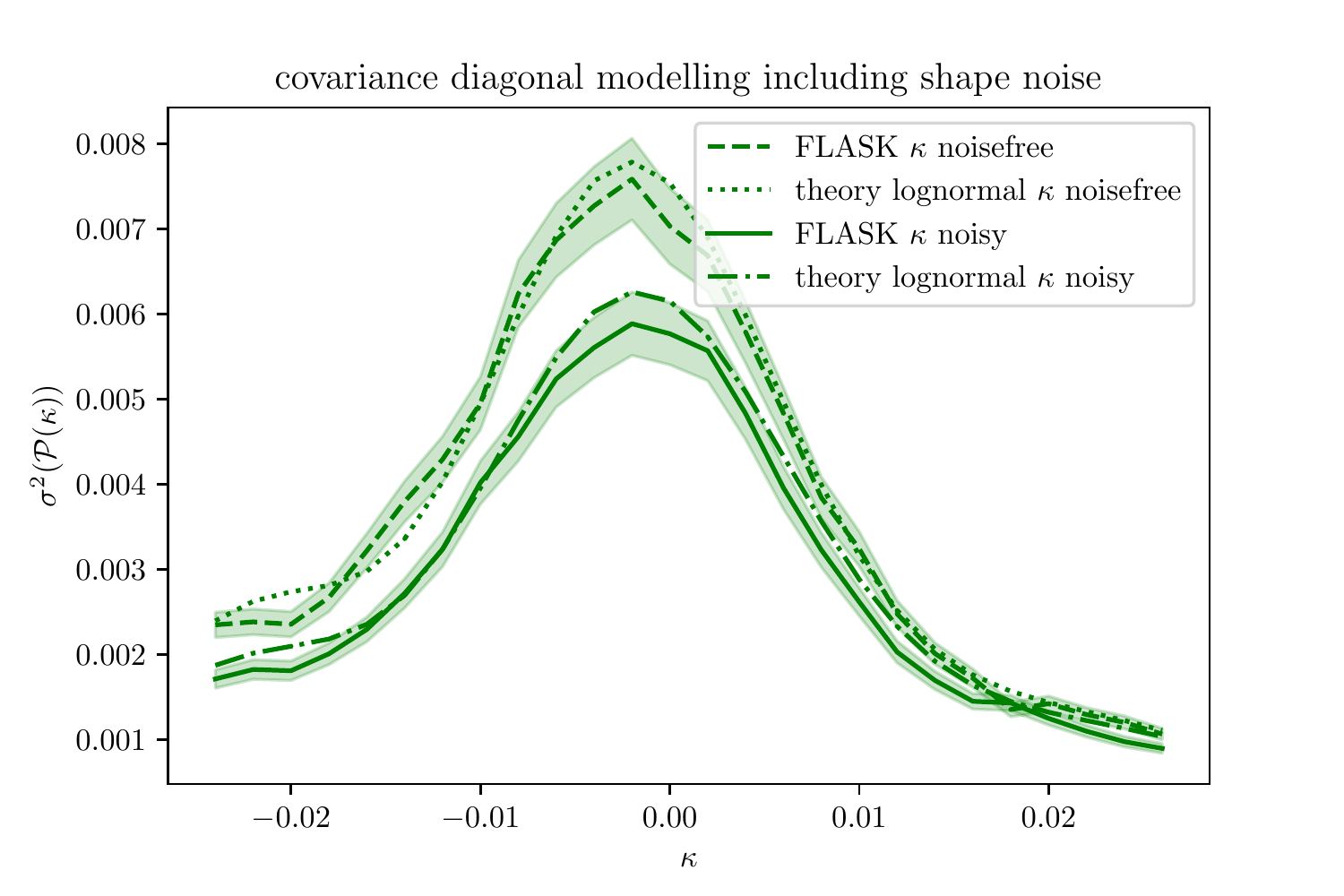}
\caption{PDF covariance diagonal for the noise-free lensing convergence as measured from 500 FLASK full sky maps (green dashed, shaded band indicating estimated errors) and predicted for the shifted lognormal (green dotted) in comparison to the noisy lensing convergence as measured from FLASK (green solid, shaded band indicating estimated errors) and predicted from equation~\eqref{eq:covfromjointPDFmodel} with with the noisy two-point PDF~\eqref{eq:jointPDF_incl_shapenoise} (green dot-dashed).}
    \label{fig:PDFcovariance_shapenoise}
\end{figure}

For the case of a shifted lognormal $\kappa$ PDF, we recompute the covariance from the convolved two-point PDF~\eqref{eq:jointPDF_incl_shapenoise} to find that the main impact of adding shape noise is to change the diagonal of the covariance matrix, shown in Figure~\ref{fig:PDFcovariance_shapenoise}, while leaving the correlation matrix largely similar. This result was obtained from $N_{\rm sim}=500$ FLASK maps of the full sky with a smoothing of $7.3'$ applied to the raw $\kappa$ and the shape-noise added $\hat\kappa$, respectively. Considering relative 1-$\sigma$ errors on the diagonal of the covariance matrix~\eqref{eq:covariance_diagonal_error} evaluated for $N_{\rm sim}$ the predictions are in good agreement with the measurements.

\section{Large separation covariance expansions}
\label{sec:covPDF_largesep}

In this Section we will perform an expansion of the joint two-point PDF in powers of the two-point cell correlation function $\xi_{12}(\theta)$, partially with the help of the cumulant generating function (CGF). We will pay particular attention to the leading and next-to-leading order contributions to the covariance matrix, which  reveals a super-sample covariance term that is controlled by the average correlation in the survey patch. We will compute the expansion to all orders in $\xi_{12}$ for a Gaussian field and the two leading order terms for the general non-Gaussian case, illustrated by results for normal and shifted lognormal fields. This will naturally lead to an expansion of the covariance matrix that resembles an eigendecomposition, which we will discuss in Section~\ref{subsec:eigendecomp}.

\subsection{Covariance expansion for raw $\kappa$}
 
\subsubsection{Leading order determined by $\kappa$-dependent clustering}
Let us first recall that at leading order in the large-separation regime, the joint two-point PDF can be approximated in terms of the one-point PDF and the `bias' functions indicating the density-dependent two-point clustering
\begin{equation}
\mP(\kappa_1,\kappa_2,\theta)\simeq \mP(\kappa_1)\mP(\kappa_2)[1+\xi(\theta)b(\kappa_1)b(\kappa_2)]\,.
\end{equation}
Hence, we have for the covariance of the binned probability
    \begin{equation}
    \label{eq:covPDFlargesep}
        \text{cov}(\mP(\kappa_i),\mP(\kappa_j)) = \bar\xi (\bar \mP b)(\kappa_i)  \bar (\mP b)(\kappa_j) + \delta_{ij} \frac{\bar\mP(\kappa_i)}{\Delta_i N_T}\,,
    \end{equation}
    where $N_T$ is the total number of cells and the mean correlation $\bar\xi$ is given by equation~\eqref{eq:meanxi}.
Note that the shot-noise term only acts on the diagonal. For the diagonal, the leading-order prediction for covariance diagonal is then 
\begin{equation}
\text{var}(\mP(\kappa)) = \bar\xi (b\bar\mP)^2(\kappa)+\frac{\bar\mP(\kappa)}{\Delta N_T}\,,
\end{equation}
where $\Delta$ is the PDF bin width. For a Gaussian field, the bias function is given by linear Kaiser bias $b(\kappa)={\kappa}/{\sigma_\kappa^2}$, which seems to be a decent approximation for the central region of $\kappa$-values, see Figure~\ref{fig:kappa_b1_LN}. Note that the cell bias $b(\kappa)$ has at least one zero crossing, for the case of raw $\kappa$, around $\kappa\simeq 0$, while for the mean-subtracted case this is modified as described in Section~\ref{sec:cov_meansub}. 

Let us also recall that for non-overlapping cells, typically the `finite sampling' effect dominates (well-described by Poisson). For overlapping cells, the number of cells is extremely large thus effectively removing the finite sampling term, but requiring the treatment of overlaps. The impact of overlapping cells can be captured by either closed-form expressions for the two-point PDF valid at small distances (like for the Gaussian and shifted-lognormal case discussed here) or a heavy overlap expansion complementary to the large separation one as described in \cite{Bernardeau2022}.

\subsubsection{Beyond leading order: Gaussian case}
\label{sec:cov_Gauss}

\begin{figure}
\centering
\includegraphics[width=\columnwidth]{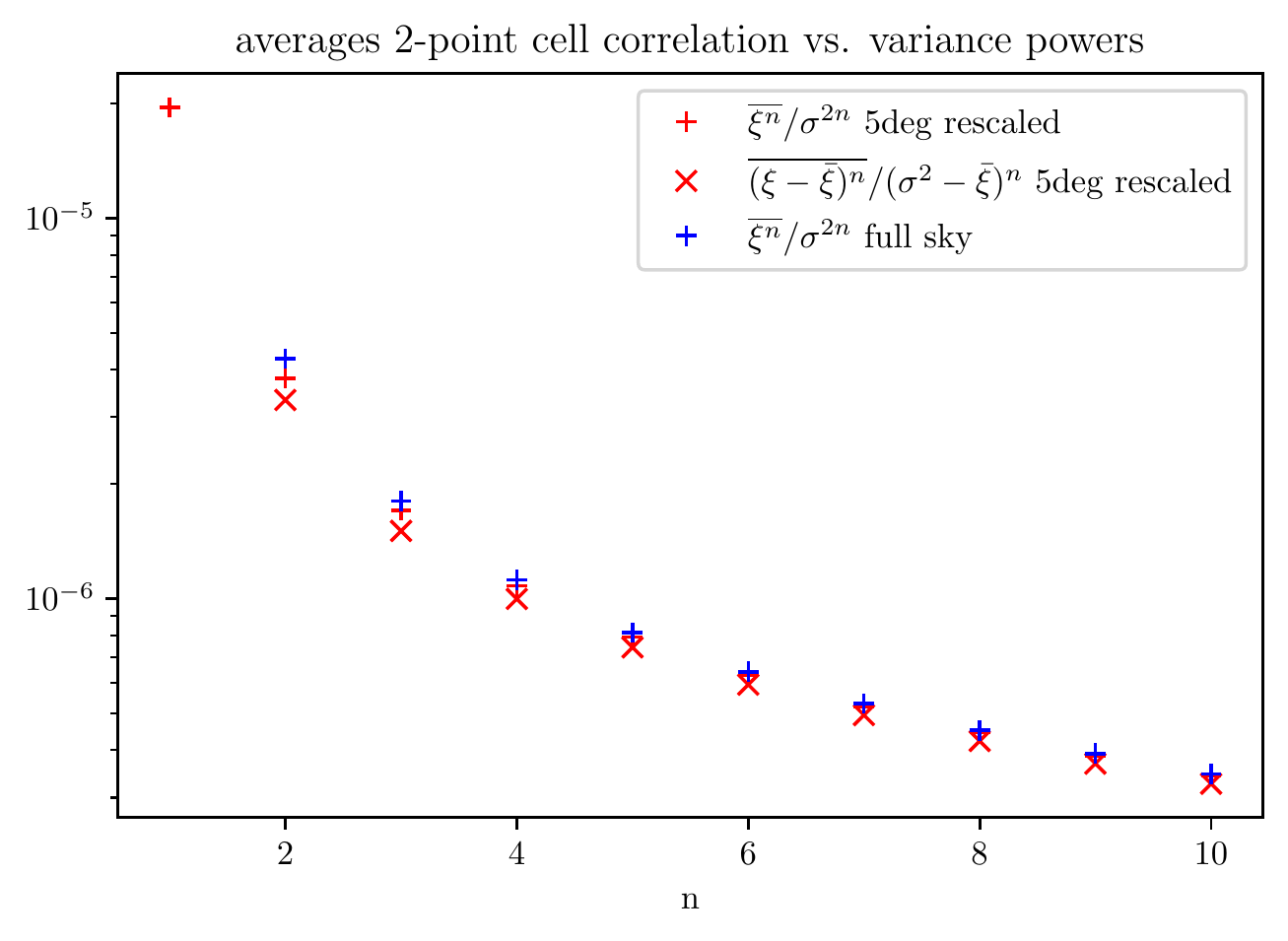}
\caption{Scaling of the average of powers of the two-point cell correlation function compared to the variance for the raw $\kappa$ maps in 5deg patches (red plusses) and the full sky result rescaled by a factor of $A_{\rm patch}/A_{\rm fullsky}$ (blue plusses). Aside from the removal of the first order term, those hierarchies closely resemble each other as well as  the mean-subtracted $\tilde\kappa$ maps in patches (red crosses).}
\label{fig:average_xi_power_scaling}
\end{figure}

For a Gaussian field of zero mean, we have the following simple structure of the joint two-point cumulant generating function in terms of the two-point correlation $\xi=\xi_{12}(\theta)$ as follows
\begin{align}
    \varphi_{\rm G}(\lambda_1,\lambda_2)&=\varphi_{0,\rm G}(\lambda_1)+\varphi_{0,\rm G}(\lambda_2) + \xi\varphi_{1,\rm G}(\lambda_1)\varphi_{1,\rm G}(\lambda_2)\,,
\end{align}
where
$\varphi_{0,\rm G}=\lambda^2\sigma^2/2$, $\varphi_{1,\rm G}(\lambda)=\lambda$ and $\varphi_{n\geq 2, \rm G}(\lambda)=0$. By expanding the last term $\exp[\xi\lambda_1\lambda_2]$ in a power series and converting $\lambda_i$ to derivatives w.r.t. $\kappa_i$, we obtain
\begin{align}
\label{eq:jointPDF_Gauss_expanded}
\frac{\mP_{\rm G}(\kappa_1,\kappa_2,\theta)}{\mP_{\rm G}(\kappa_1)\mP_{\rm G}(\kappa_2)}&= 1+\sum_{n=1}^\infty \frac{\xi^n(\theta)}{n!}b_{n,\rm G}(\kappa_1)b_{n,\rm G}(\kappa_2)\\
\label{eq:bn_Gauss_Hermite}
b_{n,\rm G}(\kappa)&=\frac{(-1)^n}{\mP_{\rm G}(\kappa)}\frac{\partial^n\mP_{\rm G}(\kappa)}{\partial\kappa^n} = \frac{1}{\sigma^n} He_n\left(\frac{\kappa}{\sigma}\right)\,,
\end{align}
which can be written in terms of the probabilist's Hermite polynomials $He_n$ 
\begin{equation}
He_n(x)=(-1)^n\exp\left[\frac{x^2}{2}\right]\frac{d}{dx^n}\exp\left[-\frac{x^2}{2}\right]\,,
\end{equation}
when using $x=\kappa/\sigma$. The first two terms read \citep[][see also]{2018PhR...733....1D}
\begin{align}
\label{eq:b1b2_Gauss}
b_{1,\rm G}(\kappa)&=\frac{\kappa}{\sigma^2}\,,\quad
b_{2,\rm G}(\kappa)=\frac{\kappa^2-\sigma^2}{\sigma^4}
=b_{1,\rm G}^2-\frac{1}{\sigma^2}\,,
\end{align}
which can be also obtained from a direct expansion of the joint PDF~\eqref{eq:2ptPDFGauss}.
This leads to the following expansion for the covariance following equation~\eqref{eq:covfromjointPDFmodel}
\begin{align}
\label{eq:cov_Gauss}
{\rm cov}&(\mP_{\rm G}(\kappa_1),\mP_{\rm G}(\kappa_2))
=\sum_{n=1}^\infty \frac{\overline{\xi^n}}{n!} \frac{\partial^n\mP_{\rm G}(\kappa_1)}{\partial{\kappa_1}^n} \frac{\partial^n\mP_{\rm G}(\kappa_2)}{\partial{\kappa_2}^n}\\
\notag &
=\sum_{n=1}^\infty \frac{\overline{\xi^n}}{n!\sigma^{2n}} He_n\left(\frac{\kappa_1}{\sigma}\right)\mP_{\rm G}(\kappa_1) He_n\left(\frac{\kappa_2}{\sigma}\right)
\mP_{\rm G}(\kappa_2)
\,,\\
\label{eq:xipowermean}
\overline{\xi^n} &:= \int d\theta P_d(\theta)\xi^n(\theta) 
\simeq \sum_{\theta\in\Theta\setminus\{0\}} \frac{N_d(\theta)}{N_d} \xi^n(\theta)
\,,
\end{align}
where in the second line we have defined the average of powers of the two-point correlation function over the survey patch in analogy to the mean from equation~\eqref{eq:meanxi}. The averages of powers of the correlation function $\overline{\xi^n}$ compared to powers of the variance $\sigma^{2n}$ control the large-separation expansion of the covariance and their hierarchy is illustrated in Figure~\ref{fig:average_xi_power_scaling}. This plot also shows that the covariance contributions scale inverse proportional to the survey area $\overline{\xi^n}\propto A^{-1}$. 

Due to the hierarchical ordering of averages of the correlation function we expect the two leading order terms in the regime $\kappa\lesssim\mathcal O(\sigma)$ to be
\begin{subequations}
\begin{align}
\label{eq:covGauss_LO}
\frac{{\rm cov}(\mP_{\rm G}(\kappa_1),\mP_{\rm G}(\kappa_2))}{\mP_{\rm G}(\kappa_1)\mP_{\rm G}(\kappa_2)} &=\frac{\bar\xi}{\sigma^2} \frac{\kappa_1}{\sigma}\frac{\kappa_2}{\sigma} \\
\label{eq:covGauss_NLO}
&+\frac{\overline{\xi^2}}{2\sigma^4} \frac{\kappa_1^2-\sigma^2}{\sigma^2} \frac{\kappa_2^2-\sigma^2}{\sigma^2}\\
\notag &+\mathcal O\left(\frac{\overline{\xi^3}}{\sigma^6}\right)\,,
\end{align}
\end{subequations}
which predicts a {\it super-sample covariance} leading order term similar to the separate universe picture~\eqref{eq:cov_SSC_SU}, being  proportional to the variance of the mean $\kappa$ in different patches, $\bar\xi=\sigma^2(\kappa_s)$, and the derivative of the PDF. This large-separation expansion of the covariance will be connected to an eigendecomposition in Section~\ref{subsec:eigendecomp}. Here, the hierarchical ordering of $\overline{\xi^n}/\sigma^{2n}$ will reflect in the eigenvalues and the bias functions multiplied by the Gaussian PDF evaluated at a given $\kappa$-value will yield the eigenvectors (after orthogonalisation and appropriate normalisation). In this context, the predicted zero-crossings of the first bias function $b_{1,\rm G}(\kappa)$ at $\kappa=0$ and of the second $b_{2,\rm G}(\kappa)$ at $\kappa=\sigma^2$ will translate into the approximate zero crossings of the first two eigenvectors as we will show later in Figure~\ref{fig:eigenvectors_vs_bn}.

\subsubsection{Beyond leading order: general non-Gaussian case}
\label{subsec:largesep_nonG}

Higher-order terms are obtained from expanding the joint two-point cumulant generating function in powers of the two-point correlation $\xi=\xi_{12}(\theta)$ as follows 
\begin{align}
    \varphi(\lambda_1,\lambda_2)&=\varphi_0(\lambda_1)+\varphi_0(\lambda_2) + \xi \varphi_1(\lambda_1)\varphi_1(\lambda_2)\\
    \notag &+\frac{\xi^2}{2} [\varphi_1^2(\lambda_1)\varphi_2(\lambda_2)+\varphi_2(\lambda_1)\varphi_1^2(\lambda_2)] + \mathcal O(\xi^3)
\end{align}
We are then in position to compute the (connected part) of the joint PDF,
\begin{subequations}
\begin{align}
&\mP(\kappa_1,\kappa_2)-\mP(\kappa_1)\mP(\kappa_2)=
\int\frac{\dd\lambda_1}{2\pi\ii}
    \int\frac{\dd\lambda_2}{2\pi\ii}
    \nonumber\\
& \exp\left[-\lambda_1\kappa_1-\lambda_2\kappa_2+\varphi(\lambda_1,-\lambda_1)+\varphi(\lambda_2,-\lambda_2)\right]
    \times \nonumber\\
& \Big( 
\xi\varphi_1(\lambda_1)\varphi_1(\lambda_2)
+\frac{\xi^2}{2}\varphi_1^2(\lambda_1)\varphi_1^2(\lambda_2)\\
&+ \frac{\xi^2}{2} [\varphi_1^2(\lambda_1)\varphi_2(\lambda_2)+\varphi_2(\lambda_1)\varphi_1^2(\lambda_2)] +\mathcal O(\xi^3)
\Big)
\end{align}
\end{subequations}
We obtain the covariance according to equation~\eqref{eq:covfromjointPDFmodel} after integrating over $\lambda_i$ and the distance distribution and symmetrising
\begin{subequations}
\label{eq:cov_nonG}
\begin{align}
&\frac{{\rm cov}(\mP(\kappa_1),\mP(\kappa_2))}{\mP(\kappa_1)\mP(\kappa_2)} =\bar\xi b_1(\kappa_1)b_{1}(\kappa_2) \\
&+\frac{\overline{\xi^2}}{2}\left[(b_{2}+q_1)(\kappa_1)(b_{2}+q_1)(\kappa_2)-q_1(\kappa_1)q_1(\kappa_2)\right]\\
\notag &+\mathcal O(\overline{\xi^3})\,,
\end{align}
where we defined
\begin{align}
\label{eq:b1_def}
(b_1\mathcal P)(\kappa)&=\int \frac{d\lambda}{2\pi\ii}\varphi_1(\lambda)\exp[-\lambda\kappa+\varphi(\lambda)] \\
\label{eq:b2_def}
(b_2\mathcal P)(\kappa)&=\int \frac{d\lambda}{2\pi\ii}\varphi_1(\lambda)^2\exp[-\lambda\kappa+\varphi(\lambda)] \\
\label{eq:q1_def}
(q_1\mathcal P)(\kappa)&=\int \frac{d\lambda}{2\pi\ii}\varphi_2(\lambda)\exp[-\lambda\kappa+\varphi(\lambda)]\,.
\end{align}
\end{subequations}
For a saddle-point approximation of the integral, we would typically expect $b_2\simeq b_1^2$, although that does not reproduce the Gaussian or shifted lognormal result.
The expression~\eqref{eq:cov_nonG} consists of dyadic products that resemble the structure of an eigendecomposition, which we will look at in Section~\ref{subsec:eigendecomp}, although eigenvectors are additionally pairwise orthogonal. 

\subsubsection{Beyond leading order: shifted lognormal case}
\label{subsec:covariance_lognormal_NLO}

\begin{figure}
\centering
\includegraphics[width=\columnwidth]{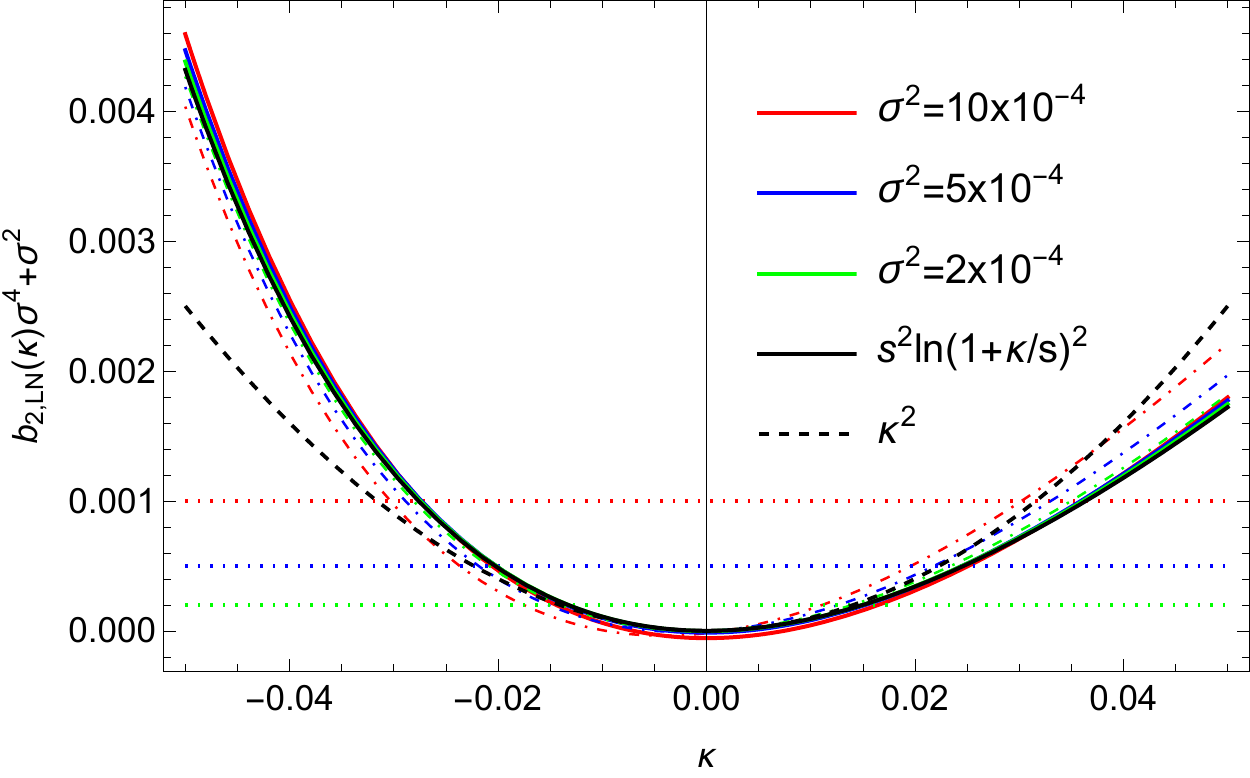}
\caption{A comparison of a rescaled version of the second order bias functions from the shifted lognormal model used here $b_{2,\rm LN}$ (solid) along with $(b_2+q_1)_{\rm LN}$ (dot-dashed) from equations~\eqref{eq:q1b2_LN} (colours for different variances) and the limiting behaviour~\eqref{eq:b2_LN_limit} (black solid) in comparison to the Gaussian result (black dashed). We also show lines of constant $\sigma^2$ (dotted) to indicate the zero-crossings of $b_{2,\rm LN}(\kappa)$ which appear in the second eigenvectors for the $\kappa$ PDF covariance in the lower panel of Figure~\ref{fig:kappa_cov_eigendecomp_fullsky}. }
\label{fig:b2_LN}
\end{figure}

The recipe described in the previous paragraph for general non-Gaussian distributions is not valid for a lognormal PDF, which is not well-described in terms of a CGF. Here we detail an approach that can be used to obtain lognormal bias functions by applying a large-separation expansion to the shifted lognormal two-point PDF~\eqref{eq:joint_LN}. This bias expansion is similar in spirit to the Gaussian case, but requires an additional expansion of the Gaussian field correlation $\xi_G$ in terms of the underlying $\kappa$ correlation $\xi_\kappa$. The joint shifted lognormal PDF can be expanded at large  separations where $\xi_G(\xi_\kappa)/\sigma_G^2\ll 1$ to obtain up to next-to-leading order in $\xi_\kappa$
    \begin{subequations}
    \begin{align}
    \label{eq:joint_LN_biasexp}
    \frac{\mP_{\rm LN}(\kappa_1,\kappa_2;\xi_\kappa)}{\mP_{\rm LN}(\kappa_1)\mP_{\rm LN}(\kappa_2)}
    &=1+\xi_G(\xi_\kappa)\frac{g(\kappa_1)}{\sigma_G^2}\frac{g(\kappa_2)}{\sigma_{\rm G}^2}\\
    \notag &+\frac{\xi_{\rm G}^2(\xi_\kappa)}{2}\frac{\left(g(\kappa_1)^2 -\sigma_{\rm G}^2\right)}{\sigma_{\rm G}^4}\frac{\left(g(\kappa_2)^2-\sigma_{\rm G}^2\right)}{\sigma_{\rm G}^4}\\
    \notag &+\mathcal O(\xi_G^3)\\
    &=1+\left(\frac{\xi_\kappa}{s^2}-\frac{\xi_\kappa^2}{2s^4}\right)\frac{g(\kappa_1)}{\sigma_G^2}\frac{g(\kappa_2)}{\sigma_G^2}\\
    \notag &+\frac{\xi_\kappa^2}{2s^4}\frac{\left(g(\kappa_1)^2 -\sigma_{\rm G}^2\right)}{\sigma_{\rm G}^4}\frac{\left(g(\kappa_2)^2-\sigma_{\rm G}^2\right)}{\sigma_{\rm G}^4}\\
    \notag &+\mathcal O(\xi_\kappa^3)\,,
    \end{align}
    \end{subequations}
    In particular, it is useful to define two additional bias functions $q_{1,\rm LN}$ and $b_{2,\rm LN}$ following the previous expansion~\eqref{eq:cov_nonG}, with their combination resembling the Gaussian expression from equation~\eqref{eq:b1b2_Gauss}
    \begin{subequations}
    \label{eq:q1b2_LN}
    \begin{align}
    \label{eq:b2_LN}
    (b_2+q_1)_{\rm LN}(\kappa)&=\frac{g(\kappa)^2-\sigma_G^2}{s^2\sigma_G^4}\simeq \frac{[sg(\kappa)]^2-\sigma^2_\kappa}{\sigma_{\kappa}^4} \\
    \label{eq:q1_LN}
    q_{1,\rm LN}(\kappa)&=\frac{1}{s} b_{1,\rm LN}(\kappa)=\frac{g(\kappa)}{s^2\sigma_G^2}\simeq \frac{g(\kappa)}{\sigma_{\kappa}^2}\,,
    \end{align}
    \end{subequations}
where we used the result for $b_{1,\rm LN}$ from equation~\eqref{eq:b1_LN}.
We see that in the limit of $s\rightarrow \infty$ we recover $sg(\kappa)\rightarrow \kappa$ and the Gaussian result with $q_{1,\rm G}=0$, while for mildly non-Gaussian fields, we typically have $q_1\ll b_2$. In a small variance expansion where $\sigma/s\ll 1$, we obtain the limiting behaviour
\begin{equation}
    \label{eq:b2_LN_limit}
    b_{2,\rm LN}(\kappa)\stackrel{\frac{\sigma}{s}\ll 1}{\longrightarrow} \frac{s^2\ln\left(1+\frac{\kappa}{s}\right)^2-\sigma^2}{\sigma^4}-\frac{1}{4s^2}
\end{equation}
Using the all-order results for $b_{n,\rm G}$~\eqref{eq:bn_Gauss_Hermite} together with the expansion of $\xi_G(\xi_\kappa)=\sum_{n=1}\frac{(-1)^{n-1}}{n} \left(\frac{\xi_\kappa}{s^2}\right)^n$ allows to compute all order expressions for the lognormal bias functions. Note that due to the difference between $\xi_G$ and $\xi_\kappa$, the higher-order terms  receive contributions from lower-order bias functions, as predicted by the CGF-based expansion from before. In Figure~\ref{fig:b2_LN} we show a comparison of the lognormal terms $b_2$ (solid) and $b_2+q_1$ (dashed) to the Gaussian result for the shift parameter adopted in this work and different variances. We can see that for our case (closely corresponding to the green line), the $q_1$ contribution is negligible. The crossing between the solid $b_2$ line and the dotted line of constant $\sigma^2$ indicates the zero-crossing of $b_2(\kappa)$, which is reflected in the second eigenvector for the $\kappa$ covariance shown in the lower panel of Figure~\ref{fig:kappa_cov_eigendecomp_fullsky}. For the case of a survey area being a significant portion of the sky, the mean correlation $\bar\xi$ will be extremely small and this second order bias function $b_2(\kappa)$ will determine the covariance structure. The two zero-crossings of $b_2$ cause the 9-tiled correlation matrix structure displayed earlier, notably in the upper right panel of Figure~\ref{fig:kappa_covariance_FLASK_theory}.

\subsection{Covariance expansion for mean-subtracted $\tilde\kappa=\kappa-\kappa_s$}
\label{sec:cov_meansub}
To determine the PDF covariance for the case of a mean-subtracted $\kappa$ field, one needs to consider the 3 variables $\kappa_1$, $\kappa_2$ and $\kappa_s$, where $\kappa_{1/2}$ are as before and $\kappa_s$ is the mean convergence in the survey patch. Like the convergence field, those variables all have zero expectation value and we are then interested in the joint PDF of the mean-subtracted $\tilde\kappa_i$ 
\begin{eqnarray}
\label{eq:tildekappa_def}
\bkap_1=\kappa_1-\kappa_s\,, \quad
\bkap_2=\kappa_2-\kappa_s\,.    
\end{eqnarray}
Let us note that this tilde variable is formally equivalent to the slope variable that was introduced in \cite{Bernardeau14,Bernardeau15}. The CGF for the three variables defined above is written here up to linear order in the two-point function at the survey scale, $\xi_s$, and  up to quadratic order in the two-point correlation at the given cell separation $\xi_{12}$
\begin{align}
\label{eq:jointCGF3lambda}
\varphi(\lambda_1,\lambda_2,\lambda_s)&=
    \frac{1}{2}\sigma^2_s\lambda_s^2+
    \varphi_0(\lambda_1)+\varphi_0(\lambda_2)+\xi_{12}\varphi_1(\lambda_1)\varphi_1(\lambda_2)
    \nonumber\\
    &
    +\lambda_s\left[\xi_{1s}\varphi_1(\lambda_1)+\xi_{2s}\varphi_1(\lambda_2)\right]\\
   \notag &+ \frac{\xi_{12}^2}{2} [\varphi_1^2(\lambda_1)\varphi_2(\lambda_2)+\varphi_2(\lambda_1)\varphi_1^2(\lambda_2)] + \mathcal O(\xi_{12}^3)\\
   \label{eq:jointCGFlambda_bg}
\varphi(\lambda_1,\lambda_s)&=
    \frac{1}{2}\xi_s\lambda_s^2+
    \varphi_0(\lambda_1)
    +\lambda_s\xi_{1s}\varphi_1(\lambda_1)\,,
\end{align}
where we assumed that the patches are big enough such that the $\kappa_s$ PDF is Gaussian.
The one- and two-point PDFs are obtained from this CGF by performing the change of variable~\eqref{eq:tildekappa_def} and then marginalising over the background $\kappa_s$, leading to
\begin{align}
\label{eq:Pbarkappa1}
    \mP(\bkap_1)&=\int\frac{\dd\lambda_1}{2\pi\ii}
    e^{-\lambda_1\bkap_1+\varphi(\lambda_1,-\lambda_1)}\\
    \mP(\bkap_1,\bkap_2)&=\!\int\!\!\!\int\!\frac{\dd\lambda_1}{2\pi\ii}\frac{\dd\lambda_2}{2\pi\ii}
    e^{-\lambda_1\bkap_1-\lambda_2\bkap_2+\varphi(\lambda_1,\lambda_2,-\lambda_1-\lambda_2)}\,.
\end{align}
When all computations are made at linear order in $\xi$, we can set $\sigma^2_s=\xi_{1s}=\xi_{2s}=\xi_{12}=\xi(\theta)$. To obtain the second order, one needs to start from $\bar\xi=\sigma^2_s=\xi_{1s}=\xi_{2s} \neq \xi_{12}=\xi(\theta)$, which we will use to illustrate the similarities and differences between the Gaussian and non-Gaussian cases.

\subsubsection{Gaussian case}
\label{sec:cov_Gauss_meansub}
For a Gaussian field, the two terms $\lambda_s^2$ and $\lambda_s[\varphi_1(\lambda_1)+\varphi_1(\lambda_2)]$ can be summarised and act to reduce the variance and correlation function by $\bar\xi$, such that joint PDF $\mP(\tilde\kappa_1,\tilde\kappa_2)$ is a Gaussian with variance $\sigma_{\tilde\kappa}^2=\sigma^2-\bar\xi$ and correlation $\xi_{\tilde\kappa}(\theta)=\xi(\theta)-\bar\xi$. Hence, the covariance can be obtained in full analogy to the previous computation that led to equation~\eqref{eq:cov_Gauss} which is modified to read
\begin{align}
\label{eq:covGauss_meansub}
{\rm cov}(\mP_{\rm G}(\tilde\kappa_1),\mP_{\rm G}(\tilde\kappa_2)) &=\sum_{n=2}^\infty \frac{\overline{(\xi-\bar\xi)^n}}{n!} \frac{\partial^n\mP_{\rm G}(\tilde\kappa_1)}{\partial\tilde\kappa_1^n} \frac{\partial^n\mP_{\rm G}(\tilde\kappa_2)}{\partial{\tilde\kappa_2}^n}\,,
\end{align}
where due to the definition of the average of powers of the correlation function~\eqref{eq:xipowermean} the $n=1$ term disappears. 
Note that $\mP_{\rm G}(\bkap_1)$ does not denote the same PDF as $\mP_{\rm G}(\kappa_1)$, for the Gaussian case they differ by the variance being $\sigma_{\tilde\kappa}^2=\sigma_{\kappa}^2-\bar\xi$ rather than $\sigma_{\kappa}^2$. Aside from this (small) difference and the removal of the $n=1$ term, we now have a different prefactor consisting of averages of powers of $\xi-\bar\xi$ rather than $\xi$. Those averages of powers of the correlation function $\overline{\xi^n}$ and $\overline{(\xi-\bar\xi)^n}$ compared to powers of the variances $\sigma^{2n}$ and $(\sigma^2-\bar\xi)^n$ control the large-separation expansion of the covariance (in the Gaussian case given in equation~\eqref{eq:cov_Gauss} for raw $\kappa$ and in equation~\eqref{eq:covGauss_meansub} for the mean-subtracted $\tilde\kappa=\kappa-\bar\kappa$) follow similar hierarchies as illustrated in Figure~\ref{fig:average_xi_power_scaling}.

In particular, the leading order term will be
\begin{align}
\notag \frac{{\rm cov}(\mP_{\rm G}(\tilde\kappa_1),\mP_{\rm G}(\tilde\kappa_2))}{\mP_{\rm G}(\tilde\kappa_1)\mP_{\rm G}(\tilde\kappa_2)} &= \frac{\overline{(\xi-\bar\xi)^2}}{2\sigma_{\tilde\kappa}^4} \underbrace{\left(\frac{\bkap_1^2}{\sigma_{\tilde\kappa}^2}-1\right)\left(
\frac{\bkap_2^2}{\sigma_{\tilde\kappa}^2}-1\right)}_{\sigma_{\tilde\kappa}^4b_{2,\rm G}(\tilde\kappa_1)b_{2,\rm G}(\tilde\kappa_2)}\\
&+ \mathcal O\left(\frac{\overline{(\xi-\bar\xi)^3}}{\sigma_{\tilde\kappa}^6}\right)
\,,
\label{eq:covGauss_meansub_LO}
\end{align}
which predicts a first eigenvalue/-vector similar to the second eigenvalue/-vector of the Gaussian case without mean-subtraction~\eqref{eq:covGauss_NLO} because $\overline{(\xi-\bar\xi)^2}/\sigma_{\tilde\kappa}^4 \simeq \overline{\xi^2}/{\sigma^4}$ as shown in Figure~\ref{fig:average_xi_power_scaling}. The  zero-crossings of the first eigenvector will be at $\tilde\kappa\approx\sigma^2-\bar\xi\approx \sigma^2$, closely following the result for full sky maps where $\bar\xi$ is negligibly small and eigenvectors are shown as green/olive lines in the middle panel of Figure~\ref{fig:kappa_cov_eigendecomp_fullsky}, as discussed in Section~\ref{subsec:eigendecomp}. Note that as before the joint two-point PDF of the mean-subtracted $\tilde\kappa$ values could be written in terms of the marginal Gaussian PDFs and a series of Hermite polynomials as in equation~\eqref{eq:bn_Gauss_Hermite} with prefactors set by averages of powers of $(\xi-\bar\xi)/\sigma_{\tilde\kappa}^2$.

\subsubsection{Leading order: general non-Gaussian case}
Similarly as for the Gaussian case, we will rewrite the joint CGF of the three variables $\lambda_1,\lambda_2,\lambda_s$ in terms of the joint CGF for just two variables $\lambda_i$ and $\lambda_s$, where $\lambda_s=-(\lambda_1+\lambda_2)$. We will first discuss the leading order result which reads
\begin{align}
    \varphi&(\lambda_1,\lambda_2,-\lambda_1-\lambda_2)-\varphi(\lambda_1,-\lambda_1)-\varphi(\lambda_2,-\lambda_2)\\
\notag =&\xi\varphi_1(\lambda_1)\varphi_1(\lambda_2)
+\bar\xi [\lambda_1\lambda_2-\lambda_1\varphi_1(\lambda_2)-\lambda_2\varphi_1(\lambda_1) ]+ \mathcal O(\xi^2)\\
\notag=&(\xi-\bar\xi)\varphi_1(\lambda_1)\varphi_1(\lambda_2)+\bar\xi [\varphi_1(\lambda_1)-\lambda_1][\varphi_1(\lambda_2)-\lambda_2]+ \mathcal O(\xi^2)\,,
\end{align}
where we used $\xi=\xi(\theta)$ for brevity and reorganised the formula such that the for a Gaussian only the first term remains, because $\varphi_1(\lambda)=\lambda$. We are then in position to compute the connected part of the joint PDF, given by
\begin{align*}
&\mP(\bkap_1,\bkap_2)-\mP(\bkap_1)\mP(\bkap_2)     \nonumber\\
&=\!\int\!\!\!\int\!\frac{\dd\lambda_1}{2\pi\ii}\frac{\dd\lambda_2}{2\pi\ii}
    e^{-\lambda_1\bkap_1-\lambda_2\bkap_2+\varphi(\lambda_1,-\lambda_1)+\varphi(\lambda_2,-\lambda_2)} \nonumber\\
& \times\Big[
(\xi-\bar\xi)\varphi_1(\lambda_1)\varphi_1(\lambda_2)
+\bar\xi [\varphi_1(\lambda_1)-\lambda_1][\varphi_1(\lambda_2)-\lambda_2]\\
&\quad \nonumber +\mathcal O(\{\overline{\xi^2},\bar\xi^2,\overline{(\xi-\bar\xi)^2}\})\Big]\,.
\end{align*}
When integrated over the distance distribution, the term linear in $\xi-\bar\xi$ will vanish and the term involving products of $\varphi_1(\lambda_i)-\lambda_i$ will create a new bias function $\tilde b_{1,\rm NG}(\tilde\kappa)$ for the mean-subtracted convergence $\tilde\kappa$ only present for non-Gaussian fields. In terms of this, the PDF covariance can be written as
\begin{subequations}
\label{eq:cov_nonG_meansub_LO}
\begin{align}
{\rm cov}(\mP(\tilde\kappa_1),\mP(\tilde\kappa_2)) =&\bar\xi (\tilde b_{1,\rm NG}\mP)(\tilde\kappa_1)(\tilde b_{1,\rm NG}\mP)(\tilde\kappa_2) \\
\notag &+ \mathcal O(\{\overline{\xi^2},\bar\xi^2,\overline{(\xi-\bar\xi)^2}\})\,,
\end{align}
where we introduced the leading order composite bias 
\begin{equation}
\label{eq:eff_b1_kappa}
\tilde b_{1,\rm NG}(\bkap)=b_1(\bkap)+\frac{\partial}{\partial\bkap}\log \mP(\bkap)\,.
\end{equation}
\begin{figure}
\centering
\includegraphics[width=\columnwidth]{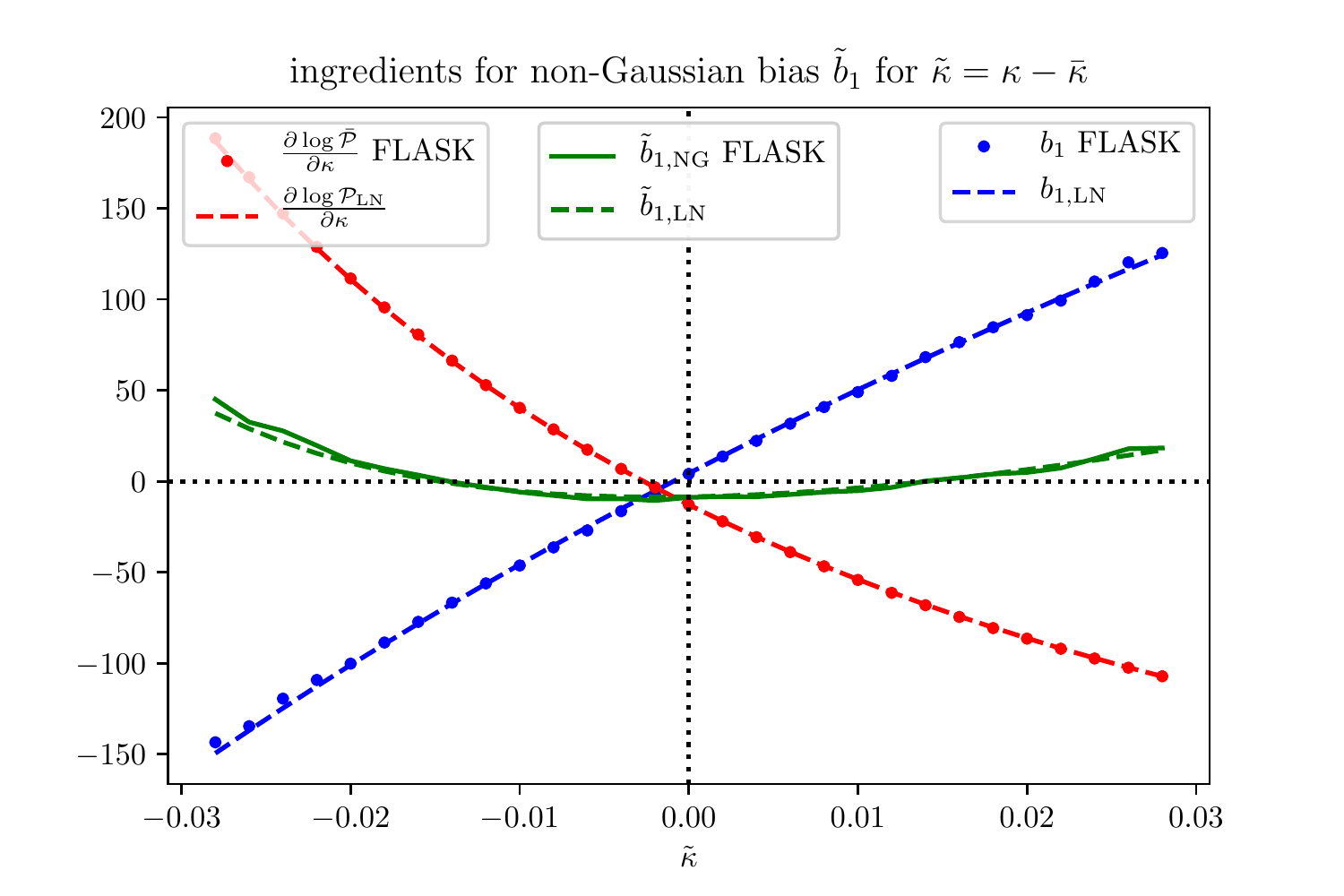}
\caption{Additional first-order bias $\tilde b_{1,\rm NG}(\tilde\kappa)$ (green) of the mean-subtracted $\tilde\kappa=\kappa-\bar\kappa$  obtained from the sum of the `raw' bias $b_1$ (green) and the response of the PDF $\partial\log\bar \mP/\partial\kappa$ (red) as given in equation~\eqref{eq:eff_b1_kappa}. We also show predictions from the shifted lognormal case (dashed) relying on the PDF from equation~\eqref{eq:1ptPDF_LN} and the bias $b_{1,\rm LN}$ from equation~\eqref{eq:b1_LN}}
\label{fig:kappa_bias_meansub}
\end{figure}
In Figure~\ref{fig:kappa_bias_meansub} we illustrate the first composite non-Gaussian bias function (green) along with its ingredients, the first order bias $b_1$ (blue) and the logarithmic derivative of the PDF (red) as measured in one FLASK realisation (data points) and predicted for a shifted lognormal (dashed lines). We can see that for a mildly non-Gaussian field, this composite non-Gaussian bias function is more than an order of magnitude smaller than the raw bias $b_1$, such that there is no guarantee that our result at leading order in $\bar\xi$ constitutes the most relevant correction. This can be established analytically for a shifted lognormal distribution using equations~\eqref{eq:1ptPDF_LN}~and~\eqref{eq:b1_LN}, which gives
\begin{equation}
\label{eq:eff_b1_kappa_LN}
\tilde b_{1,\rm LN}(\bkap)=-\frac{1}{\bkap+s}+\frac{x(\bkap)}{s\sigma^2_{\rm G}(\sigma_{\bkap}^2)}\left(1-\frac{1}{1+\bkap/s}\right)\,.
\end{equation}
\end{subequations}
The minimum value is $\tilde b_{1,\rm LN}(\tilde\kappa\simeq 0)\simeq -1/s$ and in the limit of $s\rightarrow \infty$ we recover the Gaussian result with vanishing $\tilde b_{1}$.

\subsubsection{Next-to-leading order: general non-Gaussian case}
\label{subsec:covariance_meansub_nonG_NLO}

We now generalise the previous result up to next-to-leading order, where we find
\begin{align}
    \varphi&(\lambda_1,\lambda_2,-\lambda_1-\lambda_2)-\varphi(\lambda_1,-\lambda_1)-\varphi(\lambda_2,-\lambda_2)\\
\notag=&(\xi-\bar\xi)\varphi_1(\lambda_1)\varphi_1(\lambda_2)+\bar\xi [\varphi_1(\lambda_1)-\lambda_1][\varphi_1(\lambda_2)-\lambda_2]\\
\notag&+ \frac{\xi^2}{2} [\varphi_1^2(\lambda_1)\varphi_2(\lambda_2)+\varphi_2(\lambda_1)\varphi_1^2(\lambda_2)]+ \mathcal O(\xi^3)\,.
\end{align}
Again, we compute the connected part of the joint PDF, given by
\begin{subequations}
\begin{align}
&\mP(\bkap_1,\bkap_2)-\mP(\bkap_1)\mP(\bkap_2)    \nonumber\\
&=\!\int\!\!\!\int\!\frac{\dd\lambda_1}{2\pi\ii}\frac{\dd\lambda_2}{2\pi\ii}
e^{-\lambda_1\bkap_1-\lambda_2\bkap_2+\varphi(\lambda_1,-\lambda_1)+\varphi(\lambda_2,-\lambda_2)}
    \nonumber\\
\nonumber &\times \Big[ 
(\xi-\bar\xi)\varphi_1(\lambda_1)\varphi_1(\lambda_2)
+\bar\xi [\varphi_1(\lambda_1)-\lambda_1][\varphi_1(\lambda_2)-\lambda_2]\\
&\quad+\frac{\bar\xi^2}{2} [\varphi_1(\lambda_1)-\lambda_1]^2[\varphi_1(\lambda_2)-\lambda_2]^2\\
&\quad+ \bar\xi (\xi-\bar\xi) \varphi_1(\lambda_1)\varphi_1(\lambda_2)  [\varphi_1(\lambda_1)-\lambda_1][\varphi_1(\lambda_2)-\lambda_2]\nonumber \\
&\quad+ \frac{\xi^2}{2} [\varphi_1^2(\lambda_1)\varphi_2(\lambda_2)+\varphi_2(\lambda_1)\varphi_1^2(\lambda_2)]\\
&\quad+ \frac{(\xi-\bar\xi)^2}{2}\varphi_1^2(\lambda_1)\varphi_1^2(\lambda_2) +\mathcal O(\xi^3)\Big] \,.\nonumber 
\end{align}
\end{subequations}
When integrated over the distance distribution, all terms linear in $\xi-\bar\xi$ will vanish. The remaining contributions are
\begin{enumerate}[label=\alph*)]
    \item terms in the first two lines, involving powers of $\varphi_1(\lambda)-\lambda$ creating new bias functions, $\tilde b_{n,\rm NG}$, only present for non-Gaussian fields.\footnote{At higher orders, this will generalise to a class of terms with products of positive powers of $\varphi_1(\lambda-\lambda)$ and powers of $\varphi_1(\lambda)^2$ and/or $\varphi_{n\geq 2}(\lambda)$.} 
    \item terms in the last two lines containing products of powers of $\varphi_n$ that yield the bias functions encountered for the covariance of the raw $\kappa$ PDF, at second order $b_2$ from equation~\eqref{eq:b2_def} and $q_1$ from equation~\eqref{eq:q1_def}.
\end{enumerate}
This yields
\begin{subequations}
\label{eq:cov_nonG_meansub}
\begin{align}
\nonumber &{\rm cov}(\mP(\tilde\kappa_1),\mP(\tilde\kappa_2)) =\bar\xi (\tilde b_{1,\rm NG}\mP)(\tilde\kappa_1)(\tilde b_{1,\rm NG}\mP)(\tilde\kappa_2)\\
&+\frac{\bar\xi^2}{2}(\tilde b_{2,\rm NG}\mP)(\tilde\kappa_1)(\tilde b_{2,\rm NG}\mP)(\tilde\kappa_2)\\
\nonumber &+\frac{\overline{\xi^2}}{2} [(b_2\mP)(\tilde\kappa_1)(q_1\mP)(\tilde\kappa_2)+(q_1\mP)(\tilde\kappa_1)(b_2\mP)(\tilde\kappa_2)]\\
&+\frac{\overline{(\xi-\bar\xi)^2}}{2}(b_{2}\mP)(\tilde\kappa_1)(b_{2}\mP)(\tilde\kappa_2)\\
\notag &+ \mathcal O(\{\overline{\xi^3},\bar\xi\,\overline{\xi^2},\bar\xi^3,\bar\xi\, \overline{(\xi-\bar\xi)^2},\overline{(\xi-\bar\xi)^3}\})\,,
\end{align}
where we defined the following NLO composite bias function only present for non-Gaussian fields
\begin{align}
\label{eq:eff_b2_kappa}
\tilde b_{2,\rm NG}(\bkap)&=b_2(\bkap)+2\frac{1}{\mP(\bkap)}\frac{\partial(b_1\mP)}{\partial\bkap}+\frac{1}{\mP(\bkap)}\frac{\partial^2\mP}{\partial\bkap^2}
\,.
\end{align}
\end{subequations}
For a Gaussian field, the first three lines in equation~\eqref{eq:cov_nonG_meansub} do not contribute because $\varphi_1(\lambda)=\lambda$ and $\varphi_{n\geq 2}=0$, so we recover the leading order of equation~\eqref{eq:covGauss_meansub}. Note that while we ordered terms according to powers of $\xi$ here, there is no guarantee that the leading order term proportional to $\bar\xi$ will be larger than the next-to-leading order terms proportional to $\overline{\xi^2}$ and $\overline{(\xi-\bar\xi)^2}$, as highlighted by the Gaussian case where the leading order term vanishes and the shifted lognormal case where we found the term to be subdominant. Typically, power of averages of the mean correlation are much smaller than averages of powers, so $\bar\xi^2\ll \overline{\xi^2}\approx \overline{(\xi-\bar\xi)^2}$ as evident from Figure~\ref{fig:average_xi_power_scaling} and for a weakly non-Gaussian field the $\tilde b_{2,\rm NG}$ term will only give a small contribution. For the mildly non-Gaussian $\kappa$ PDF we are considering here, we observe that the first order term is effectively removed by the mean-subtraction, as evident from the visual inspection of the covariance matrix shown in Figure~\ref{fig:kappa_covariance_correlation} and the dominant eigenvectors, which resemble the full sky case shown in green/olive in the lower panel of Figure~\ref{fig:kappa_cov_eigendecomp_fullsky}. Given the similarity between the Gaussian and FLASK correlation matrices, we can deduce that the $b_2$ term given by equation~\eqref{eq:q1b2_LN} for the shifted lognormal case is in fact the dominant contribution.

Our results from this section can largely be carried over to the covariance modelling for the 3-dimensional matter PDF on mildly nonlinear scales (roughly $10$Mpc$/h$ at low redshifts), as we will discuss in Section~\ref{sec:covPDF_3D} below. Typically, the matter PDF features stronger non-Gaussianities, such that the lognormal shift parameter will be close to unity $s_{\rm 3D}\lesssim 1$. In Appendix~\ref{app:MinimalTreeModel} we additionally discuss tree models as a useful class of non-Gaussian models for which all order covariance expansion results can be obtained.

\section{Quantitative covariance model tests}
\label{sec:quant_tests}

Beyond performing a visual comparison between the covariances obtained from simulated maps and the theoretical models, we quantitatively compare their accuracy using 
\begin{enumerate}[label=\arabic*.]
    \item an eigendecomposition of the covariance validating the set of eigenvalues and leading order eigenvectors,
    \item a comparison of the dominant bands in the almost band-diagonal precision matrix as inverse of the covariance,
    \item a $\chi^2$-test \citep[advocated in][]{Friedrich2021cov} to establish the Gaussianity of the PDF data vector and accuracy of the model covariance, and 
    \item a Fisher forecast for two cosmological parameters $\{\Omega_m,\sigma_8\}$.
\end{enumerate}

When comparing our measurements to the predictions in tests involving the inverse covariance or precision matrix, we account for that fact that matrix inversion is a non-linear operation, so the noise in the estimate of the covariance elements will lead to a bias in the elements of the precision matrix. This bias from covariance estimation noise turns out to be just a factor multiplying the entire inverse covariance matrix, the so called 
Kaufman-Hartlap factor \citep{Kaufman67,Hartlap06},
\begin{equation}
\label{eq:hartlap}
    h=(N_{\rm sim} - N_d- 2 )/(N_{\rm sim} - 1) \ ,
\end{equation}
where $N_{\rm sim}$ is the number of samples used and $N_d$ is the length of the data vector. This effect closely resembles the impact of the non-Gaussian distribution of a data vector that is a consequence of a finite number of realisations \citep{SellentinHeavens2017} and of size $(N_{\rm sim}-N_d + N_p-1)/(N_{\rm sim}-1)$, where $N_p$ is the number of parameters.
Another effect is due to the nonlinear relationship between the precision matrix and the resulting parameter covariance matrix \citep{Percival2014cov}, which can partially counteract the previous effect, but was found to be negligible in our setting. Beyond direct effects for the width of the likelihood contours, noise also leads to an uncertainty for their location in parameter space \citep{DodelsonSchneider2013}, which is only negligible if $N_{\rm sim}-N_d \gg N_d-N_p$ as is the case for our simplified setting with a short data vector. 

\subsection{Eigendecomposition of the covariance matrix}
\label{subsec:eigendecomp}

The eigendecomposition of the covariance matrix of dimension $n\times n$ (set by the length $n$ of the data vector) can be written in terms of a complete set of eigenvalues $\lambda_i$ (for convenience ordered in descending order, so $\lambda_1>\ldots >\lambda_n$) and associated normalised eigenvectors $\mathbf{e_i}$ arranged in an orthogonal matrix $O$
\begin{align}
\notag{\rm cov}(\mP,\mP)&=O\cdot {\rm diag}(\lambda_1,\ldots,\lambda_n)\cdot O^T\,,\ O=\begin{pmatrix} \mathbf{e_1}, \ldots ,\mathbf{e_n}\end{pmatrix}\\
&=\sum_{k=1}^n \lambda_k \mathbf{e_k}\mathbf{e_k}^T\,.
\label{eq:cov_eigendecomp}
\end{align}
In terms of components, we can write this as
\begin{align}
{\rm cov}(\mP_i,\mP_j)&=\sum_{k=1}^n \lambda_k \left(\mathbf{e_k}\mathbf{e_k}^T\right)_{ij}= 
\sum_{k=1}^n \lambda_k \left(\mathbf{e_k}\right)_i\left(\mathbf{e_k}\right)_j\,,
\end{align}
where $\left(\mathbf{e_k}\right)_i$ denotes the $i$-th component of the eigenvector $\mathbf{e_k}$.
The eigendecomposition can be understood in the context of the analytical results from the large-separation expansion in Section~\ref{sec:covPDF_largesep}. For the Gaussian case, we obtained an expansion consisting of dyadic products with building blocks of the form $He_n(x)\exp(-x^2/2)$ with $x=\kappa/\sigma$ corresponding to $(b_n\mP)(\kappa)$. While the Hermite polynomials are orthogonal with respect to the Gaussian weight function $\exp(-x^2/2)$ corresponding to the PDF $\mP_G(\kappa)$, the dyadic product building blocks  are not generally orthogonal as they carry a combined weight of $\exp(-x^2)$. However, the orthogonality of neighbouring bias contributions $b_n\mP$ and $b_{n+1}\mP$ is ensured by the alternating symmetry of the Hermite functions. 

\begin{figure}
\centering
\includegraphics[width=\columnwidth]{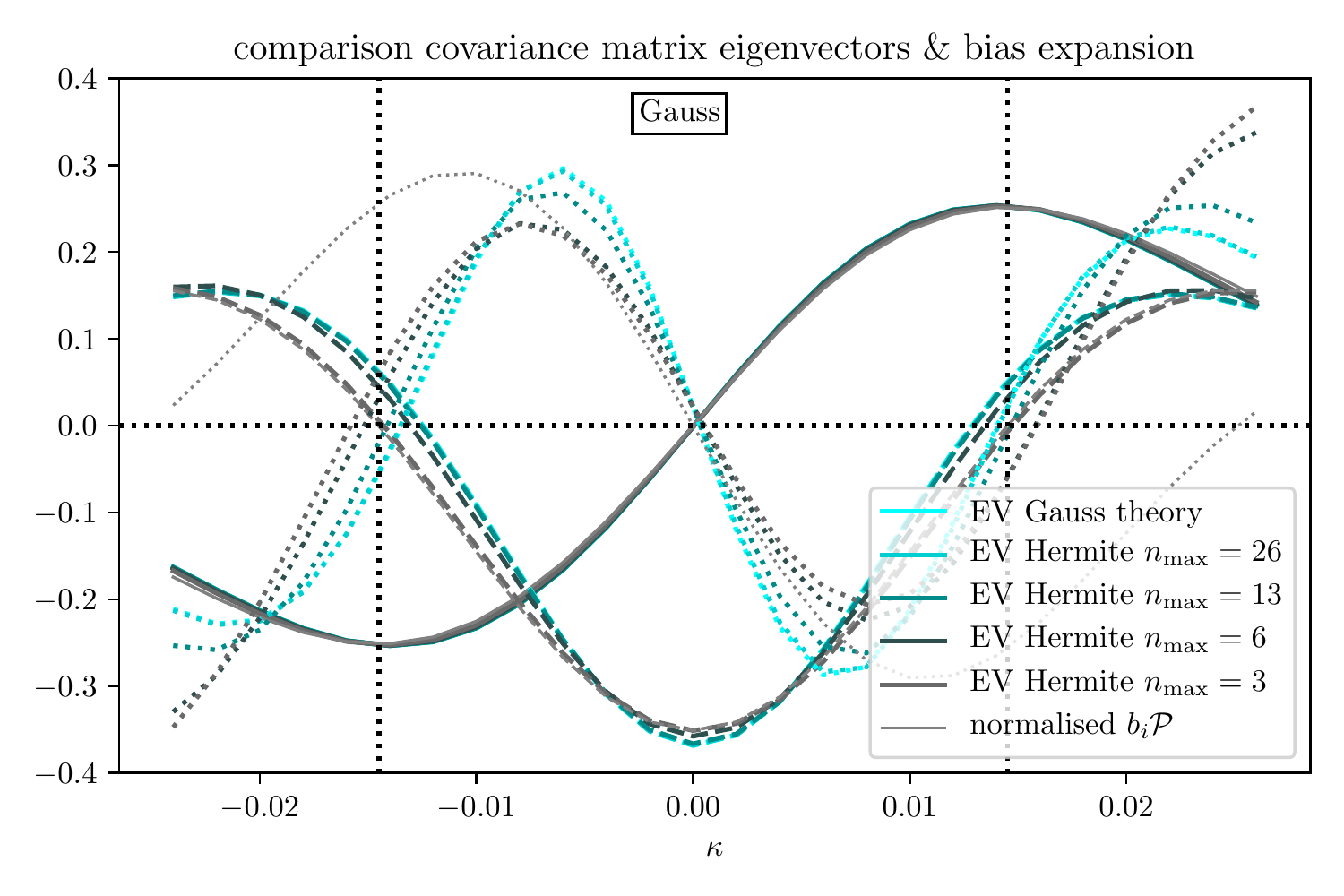}
\caption{First (solid), second (dashed) and third (dotted) eigenvectors of the Gaussian $\kappa$ PDF covariance predicted for small patches (cyan) and the normalised bias functions $(b_n\mathcal P)_{\rm G}(\kappa)$ (light grey). We also show the eigenvectors obtained from a Hermite expansion~\eqref{eq:cov_Gauss} up to order $n_{\rm max}$ (shaded from dark grey to cyan). The vertical black dotted lines indicate where $\kappa=\pm\sigma$.}
\label{fig:eigenvectors_vs_bn}
\end{figure}

\begin{figure}
\centering
\includegraphics[width=\columnwidth]{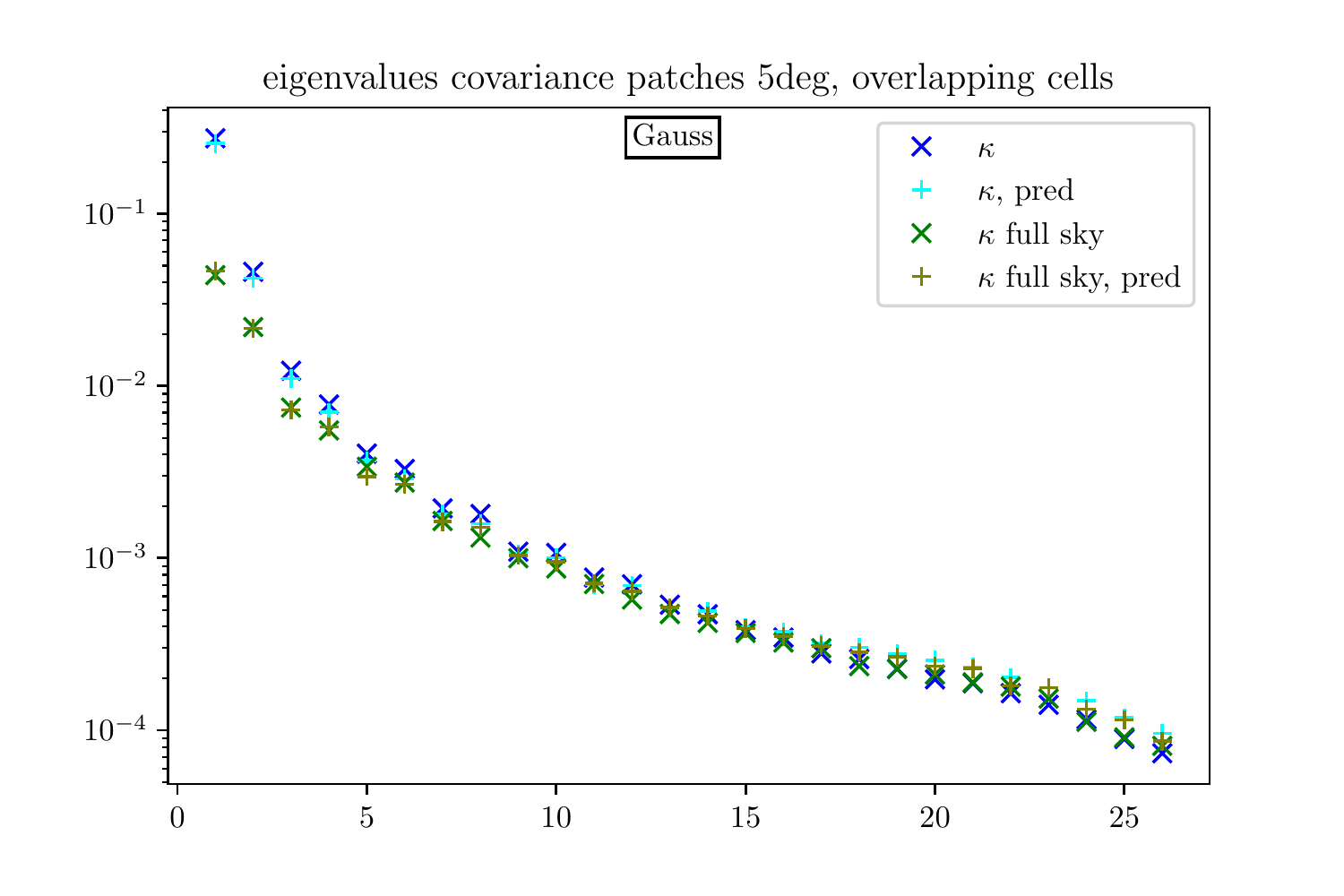}
\includegraphics[width=\columnwidth]{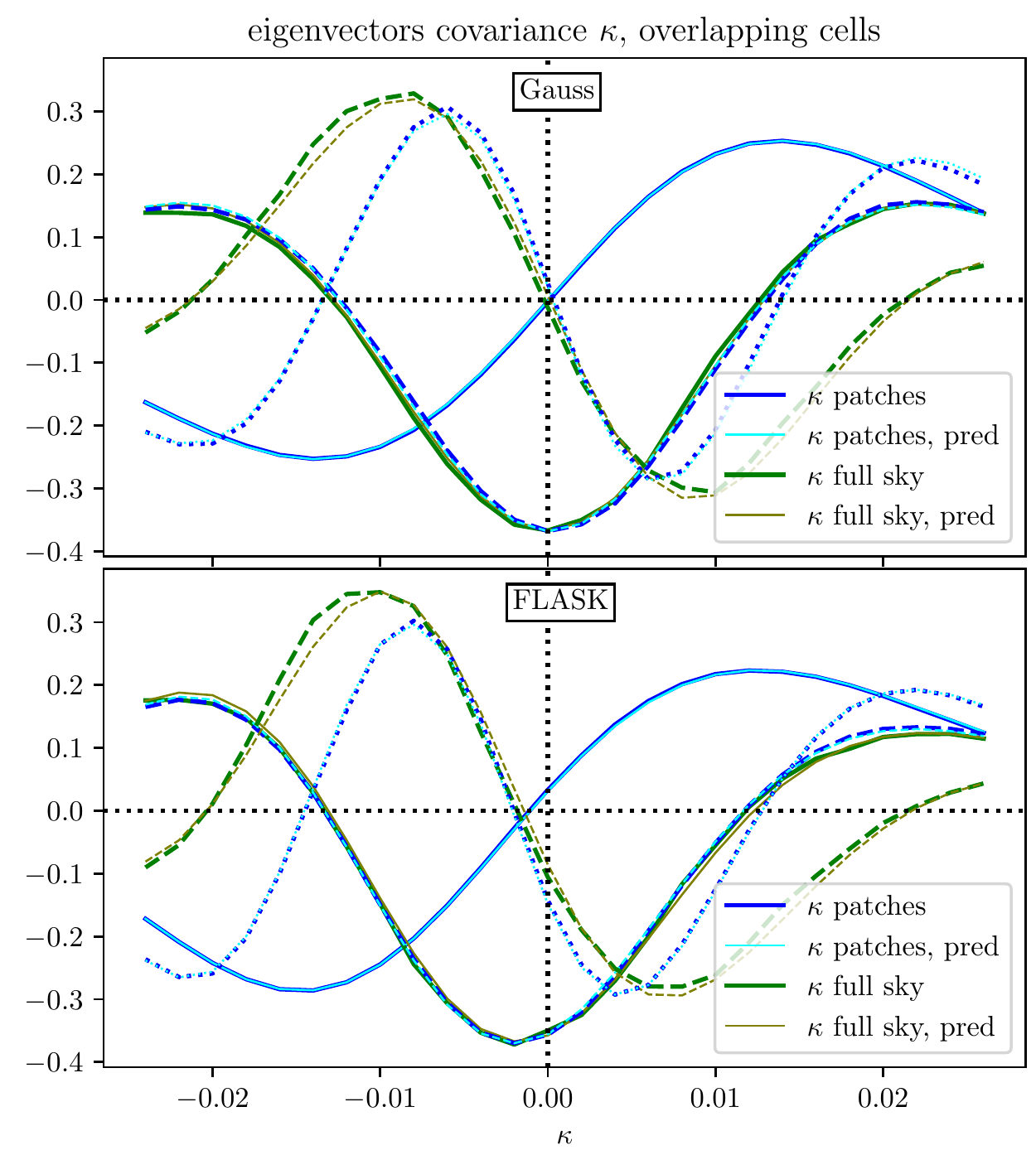}
\caption{(Upper panel) Eigenvalues of the Gaussian $\kappa$ PDF covariance for $\kappa$ from small patches (measurement in blue, prediction in cyan) and the full sky  (measurement in green, prediction in olive). 
(Middle and lower panel) The dominant eigenvectors of the Gaussian (middle) and FLASK (lower) $\kappa$ PDF covariance, including the 1st eigenvector (solid lines), 2nd eigenvector (dashed lines) and 3rd eigenvector for raw $\kappa$ from small patches (dotted lines). The move from small patches to the full sky (solid and dashed lines) essentially removes the first term in the eigendecomposition and replaces it with the second term as evident from the agreement of the first green cross with the second blue cross, and the agreement of the solid green and the dashed blue lines.
The FLASK results are qualitatively similar to the Gaussian case except for subtle shifts in the location of the zero-crossings and the asymmetry between positive and negative $\kappa$, both related to the non-Gaussianity of the underlying field.}
\label{fig:kappa_cov_eigendecomp_fullsky}
\end{figure}

Considering the binned PDF means that there is just a finite set of eigenvectors, which receive contributions from a whole series of $b_n\mP$ terms. We illustrate that in Figure~\ref{fig:eigenvectors_vs_bn} by comparing the first three eigenvectors (solid/dashed/dotted) of the PDF covariance for a Gaussian $\kappa$ field obtained from the full theory expression (cyan) with the normalised bias functions (light grey). The leading order bias contribution $b_1\mP$ agrees with the first eigenvector almost perfectly, while $b_2\mP$ resembles the overall shape of the second eigenvector with slightly modified zero crossings. This agreement reflects that there is a hierarchy in the contributions from different powers of the correlation function compared to the variance $\overline{\xi^n}/\sigma^{2n}$ as shown in Figure~\ref{fig:average_xi_power_scaling}. The slight difference between $b_2\mP$ and the second eigenvector is due to contributions from higher-order terms as we illustrate by showing the eigenvectors of the Hermite expansions~\eqref{eq:cov_Gauss} up to a finite $n_{\rm max}$ (shaded from dark grey to cyan with increasing $n_{\rm max}$). The large difference between $b_3\mP$ and the third eigenvector comes from its non-orthogonality with $b_1\mP$ as can be seen from the difference between the dotted lines in light and darker grey.
For the case of the full sky (or the mean-subtracted $\kappa$ in patches), the smallness (or vanishing) of the mean correlation $\bar\xi$ will practically remove the contribution of $b_1\mP$ such that the first and second eigenvectors will be resembling $b_2\mP$ and $b_3\mP$, respectively. This qualitative statement remains true for mildly non-Gaussian fields such as our shifted lognormal (FLASK) case.
We show the eigenvalues and dominant eigenvectors for the Gaussian and FLASK maps in Figure~\ref{fig:kappa_cov_eigendecomp_fullsky}. 
The eigenvalues in the upper panel of  Figure~\ref{fig:kappa_cov_eigendecomp_fullsky} show a good agreement between measurements and theory and suggest a clear ordering between contributions from different bias functions arising from a large-separation expansion, qualitatively resembling the behaviour seen in Figure~\ref{fig:average_xi_power_scaling} for the averages of powers of the cell correlation in comparison with the powers of the variance. The middle and lower panels of Figure~\ref{fig:kappa_cov_eigendecomp_fullsky} show the leading three eigenvectors for Gaussian (middle) and lognormal (bottom) $\kappa$ fields in small patches (blue/cyan) and the full sky (green/olive). The first eigenvector (solid) is determined by $b_1\mP$ and hence shows one zero-crossing close to $\kappa\simeq 0$ as predicted by $b_{1,\rm G}$~\eqref{eq:b1_Gauss}~and~$b_{1,\rm LN}$\eqref{eq:b1_LN}. The second eigenvector (dashed) is driven by $b_2\mP$ with two zero-crossings around $\kappa\simeq\sigma_\kappa$ as predicted by $b_{2,\rm G}$~\eqref{eq:b1b2_Gauss}~and~$b_{2,\rm G}$~\eqref{eq:q1b2_LN}. The third eigenvector (dotted) involves $b_3\mP$, which is predicted to have three zero-crossings for the Gaussian case from equation~\eqref{eq:bn_Gauss_Hermite}, but is also sensitive to higher-order terms as we illustrated in Figure~\ref{fig:eigenvectors_vs_bn}. 
For the covariance of the $\kappa$ PDF on the full sky we show the first two eigenvectors (measurements in green, predictions in olive). We observe that the first eigenvector closely resembles the second eigenvector of the $\kappa$ PDF from small patches (measurements in blue, predictions in cyan), as predicted for Gaussian fields in equation~\eqref{eq:covGauss_meansub} and for mildly non-Gaussian fields in equation~\eqref{eq:cov_nonG_meansub}. Overall, the theoretical predictions are in good agreement with the measurements from Gaussian and FLASK maps. 
The Gaussian and the lognormal cases are qualitatively similar with a few important differences appearing mostly as shifts towards negative $\kappa$ (reflecting the PDF peak location) and an asymmetry between positive and negative $\kappa$ reflecting the skewness of the PDF. Due to the nature of the eigendecomposition, the contribution of the finite sampling term on the diagonal of the covariance matrix~\eqref{eq:covFS} to individual eigenvectors is typically small, such that the leading order results for the non-overlapping case (not shown) resemble the overlapping case aside from additional noise caused by the small number of sampling cells. 

In principle, one can reconstruct the covariance matrix from the terms in the eigendecomposition, although in the presence of an incomplete set of eigenvectors (a number less than the length of the data vector), this matrix will not be invertible and hence unsuitable for statistical analysis. For the case of non-overlapping cells, there is a large contribution to the diagonal from shot noise, so one can attempt a covariance reconstruction using only the first few eigenvectors along with a matched diagonal mostly arising from shot noise. In practice, we opt for the approach to use an invertible covariance baseline model, such as the ones obtained from the Gaussian or shifted lognormal model. If this was found to be insufficiently accurate, it could be successively improved using the precision matrix expansion \citep{Friedrich2017precisionmatrixexp} or iteratively corrected by adjusting the dominant eigenvectors to their desired values as predicted by a large-scale expansion discussed in Section~\ref{sec:covPDF_largesep}.

\subsection{Bands of the precision matrix}
\label{subsec:precision_matrix}
\begin{figure}
    \centering
    \includegraphics[width=\columnwidth]{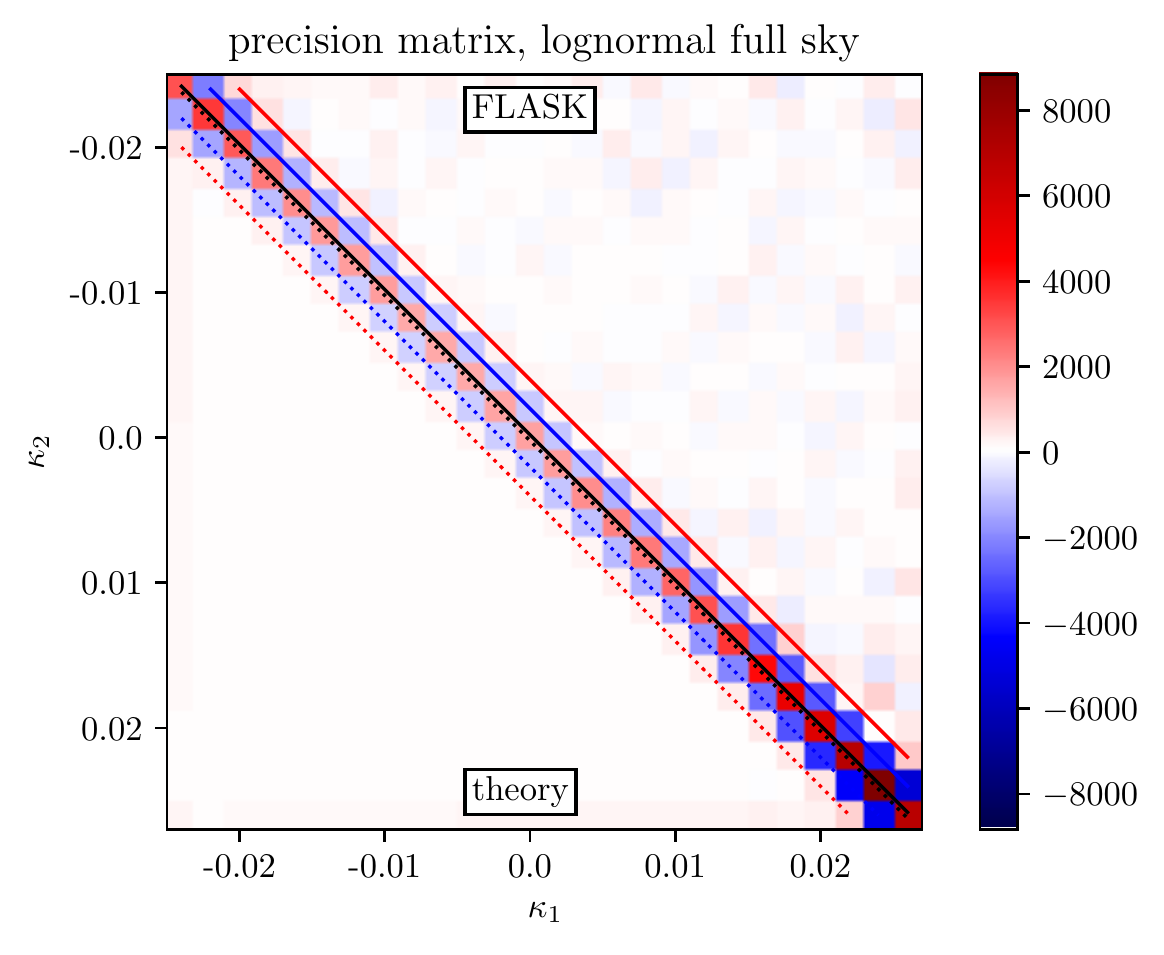}
    \includegraphics[width=\columnwidth]{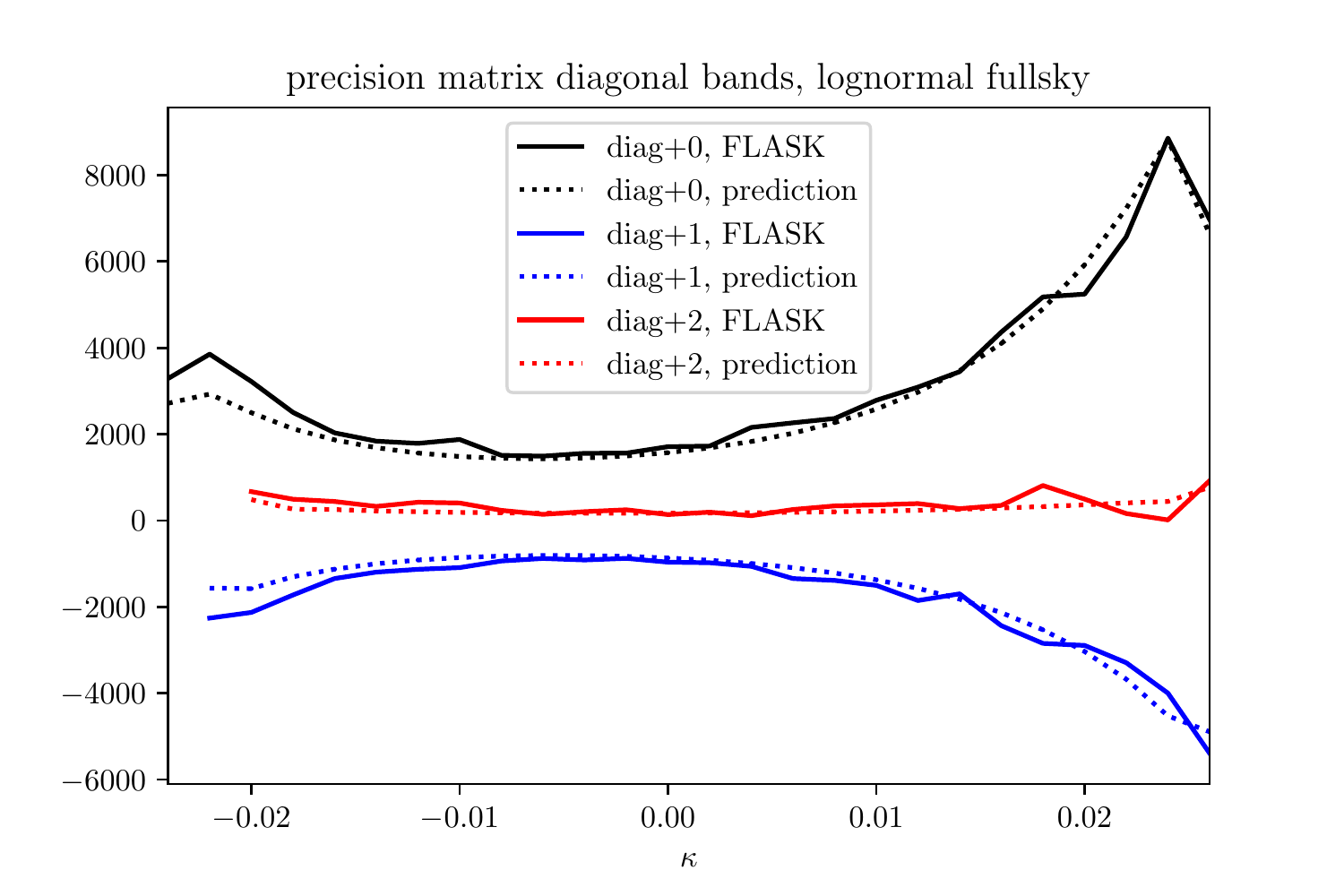}
    \caption{(Upper panel) Impression of the approximately band-diagonal precision matrix obtained as inverse of the $\kappa$ covariance matrix for overlapping cells for shifted lognormal fields as predicted (lower triangle) and measured in 500 FLASK full sky realisations corrected by the Hartlap factor~\eqref{eq:hartlap} (upper triangle).  This matrix is close to penta-diagonal, featuring 5 central bands, one diagonal band (black) and 2 neighbouring bands above/below (blue, red). We highlight those bands with with solid/dotted lines for the measured and predicted case. On the diagonal we show the mean of the FLASK measurements and the prediction. The measurement from a finite number of realisations in the upper panel exhibits additional noise off the diagonal. 
    (Lower panel) Comparison of the bands in the precision matrix between the the shifted lognormal theory prediction (dotted lines) and the measurement from FLASK realisations (solid lines). }
    \label{fig:inverse_covariance}
\end{figure}

The accuracy of cosmological parameter estimation relies on an accurate modelling of the precision (inverse covariance) matrix which enter the $\chi^2$ tests and Fisher forecasts \citep{Taylor2013}. We show this precision matrix in the case of shifted lognormal fields from the full sky (rescaled to the survey area) in the upper panel of Figure~\ref{fig:inverse_covariance} as measured in FLASK maps (upper triangle) and predicted (lower triangle). We notice that the inverse covariance is close to band-diagonal with alternating signs. In the lower panel we focus on the three dominant bands (black, blue, red) comparing the prediction (solid) to the measurements (dotted). We find similar results for the precision matrix shape obtained from FLASK patches, and for both the raw $\kappa$ and the mean-subtracted $\tilde\kappa=\kappa-\bar\kappa$ from overlapping cells. We can qualitatively understand the agreement between $\kappa$ and $\tilde\kappa$ in terms of the previously discussed eigendecomposition. In principle, the precision matrix can be obtained from the eigenvectors and eigenvalues of the covariance matrix as
\begin{align}
\notag{\rm cov}^{-1}(\mP,\mP)&=O\cdot {\rm diag}(\lambda_1^{-1},\ldots,\lambda_n^{-1})\cdot O^T\,,
\end{align}
which means eigenvectors are retained but their order is reversed. From the upper panel of Figure~\ref{fig:kappa_cov_eigendecomp_fullsky} we expect a flat hierarchy of the highest eigenvalues that determine the inverse, with the lowest order eigenvectors (associated with the largest eigenvalues) playing a minor role for the shape of the precision matrix.

The band-diagonal structure of the precision matrix is mainly driven by the strong correlations of neighbouring bins induced by the overlap of cells arising at small separations. This can be inferred from a comparison with the non-overlapping case and could be formalised by evaluating the small separation contributions described in \cite{Bernardeau2022} for the analogous case of densities.
We can illustrate this effect in a simpler setting by computing the covariance between 1D self-centered cells for a Poisson (constant) power spectrum with no correlation between non-overlapping cells. Thus for two cells of half-length $x_i$ and $x_j$, their covariance is given by the integral over the (constant) power spectrum multiplied by the 1D top-hat window function of each cell. We thus have
\begin{subequations}
\label{eq:cov1D}
\begin{equation}
    {\rm cov}_{\rm 1D}(x_i,x_j) \propto\!\!\!\int\! {\rm d}k \frac{\sin(kx_i)}{kx_i}\frac{\sin(kx_j)}{kx_j}\propto \left[\max(x_i,x_j)\right]^{-1}\,.
\end{equation}
This special covariance shape is called a hook matrix
\begin{equation}
    \rm{cov}_{\rm 1D} \propto \begin{pmatrix} 
    1/x_1 & 1/x_2 & 1/x_3 & \dots & 1/x_n \\
    1/x_2 & 1/x_2 & 1/x_3 & \dots & 1/x_n \\
    1/x_3 & 1/x_3 & 1/x_3 & \dots & 1/x_n \\
    \vdots & & & & \vdots \\
    1/x_n &  & \dots & & 1/x_n 
    \end{pmatrix}.
\end{equation}
\end{subequations}
This matrix can be inverted by obtaining the cofactor matrix $C$ which determines the inverse as $\rm{cov}_{\rm 1D}^{-1}=C^T/\det \rm{cov}_{\rm 1D}$. The elements of the cofactor matrix are given by Cramer's rule in terms of determinants of submatrices obtained from $\rm{cov}_{\rm 1D}$ by deleting the $i$th row and $j$th column
\begin{equation}
C_{i,j}=(-1)^{i+j}\det (\rm{cov}_{\rm 1D}|_{i,j})\,.
\end{equation}

It can then be shown that indeed the cofactor elements vanish everywhere except on the diagonal and the directly adjacent bands to the main diagonal. This is because the removal of a row $i$ and column $j$ in the covariance~\eqref{eq:cov1D} leads to linearly dependent rows/columns and hence $C_{i,j}=0$ for $|i-j|<2$. As the covariance is symmetric, so is the cofactor matrix and hence the inverse of the covariance will have a band-diagonal structure. When the values of the radii $x_i$ (resembling $\kappa_i$) are an increasing sequence, the diagonal will have a positive sign and the band adjacent to the diagonal will have a negative sign.

\subsection{$\chi^2$ tests}
\label{subsec:chi2}

We will use a $\chi^2$-test \citep[advocated in][]{Friedrich2021cov} to establish the Gaussianity of the data vector and assess the accuracy of our predicted covariances. Let us consider the $\chi^2$ associated with the original $N$ samples of the data vector. For every sample, labelled by $i$, of the data vector, $D_{\rm org}[i]$, of length $n$ (encoding the number of degrees of freedom), one computes the $\chi^2$
\begin{equation}
\label{eq:chi2}
    \chi^2_{\rm org}[i]=(D_{\rm org}[i]-\mu)^T \cdot C^{-1} \cdot (D_{\rm org}[i]-\mu)\,,
\end{equation}
using the mean data vector, $\mu$ (also of length $n$) and data covariance $C$ (of dimension $n\times n$). The distribution of $\chi^2$ follows a $\chi^2$ distribution with the appropriate number of degrees of freedom $n$ (here the number of PDF bins). To test the Gaussianity of the data vector distribution, one can create
another sample of data vectors, $D_{\rm G}$,  by
drawing $N$ times from a multivariate Gaussian using the mean data vector, $\mu$ and covariance $C$. Then one recomputes the $\chi^2$ for every sample using the covariance to check whether it follows the same distribution
\begin{equation}
\label{eq:chi2_Gauss}
    \chi^2_{\rm G}[i]=(D_{\rm G}[i]-\mu)^T \cdot C^{-1} \cdot (D_{\rm G}[i]-\mu)\,.
\end{equation}
If so, one can assert that the data vector is `close enough' to Gaussian distributed around the mean with the measured covariance. Both of those quantities will have the mean $\overline{\chi^2}=\overline{\chi^2_{\rm org}}=n$, since the covariance matrix is identical and is reproduced in the expectation value of the fluctuations of the data vector around the mean. 
\begin{figure}
    \centering
    \includegraphics[width=\columnwidth]{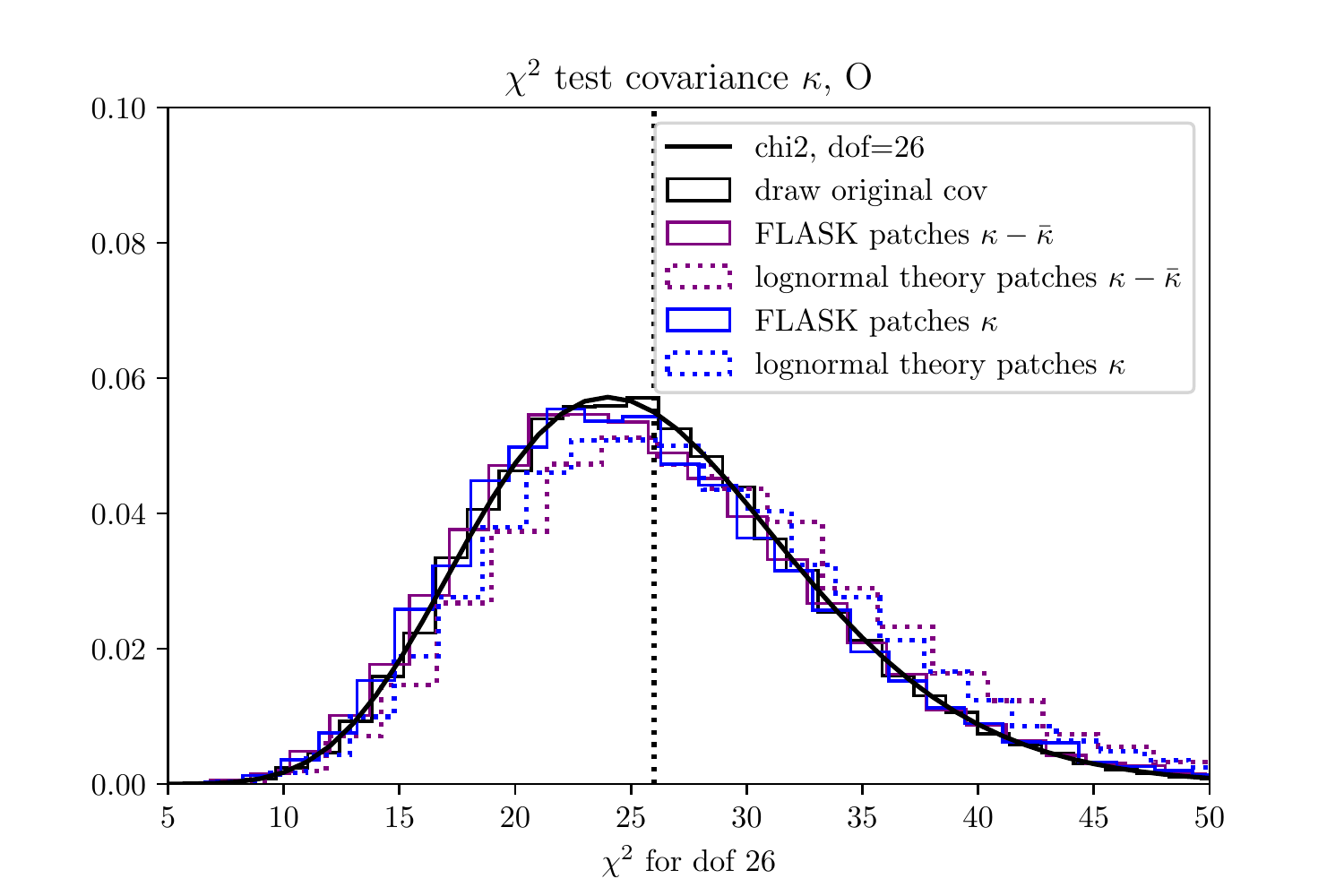}
    \includegraphics[width=\columnwidth]{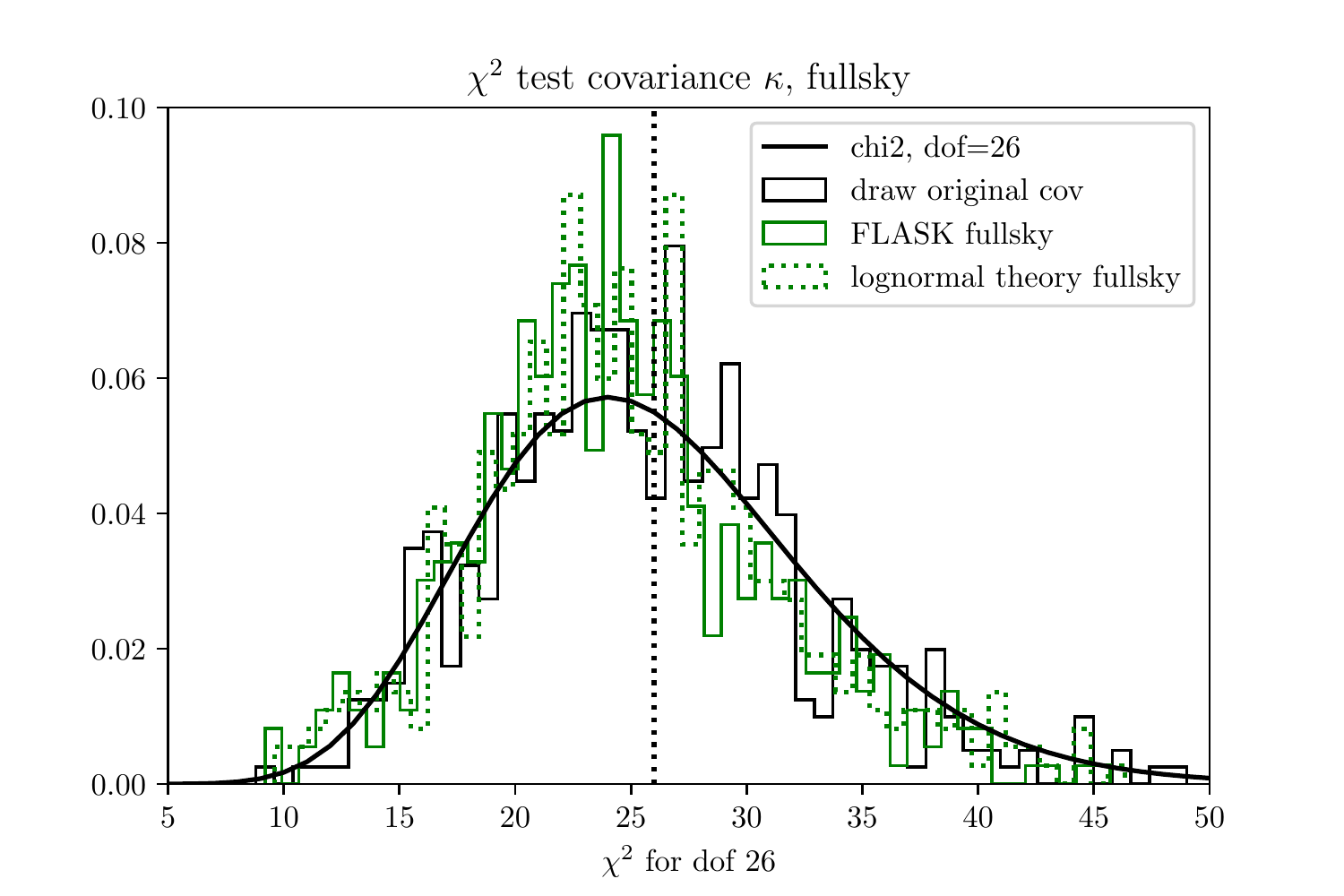}
    \caption{$\chi^2$-tests for the validation of our predicted covariances and the Gaussianity of our data vector showing the $\chi^2$ distribution (black line) and the result for a data vector drawn from a Gaussian with the measured covariance (black histogram).
    (Upper panel) Covariance matrix test for the PDF of the convergence $\kappa$ (blue) or mean-subtracted convergence $\kappa-\bar\kappa$ (purple) measured from overlapping cells computed in small patches comparing the FLASK measurements (solid) to the predictions from a shifted lognormal (dotted).
    (Lower panel) Covariance matrix test for overlapping cells computed for the full sky comparing the measurements from 500 FLASK realisations (solid) to the lognormal prediction (dotted).}
    \label{fig:chi2_test_FLASK_theory}
\end{figure}

One can also use a $\chi^2$ test to compare the closeness of two covariances, where $C$ denotes the original `true' (or target) covariance and $C_{\rm mod}$ the model (or approximated) covariance. By recomputing the $\chi^2$ using the model covariance 
\begin{equation}
\label{eq:chi2_model}
    \chi^2_{\rm mod}[i]=(D_{\rm org}[i]-\mu)^T \cdot (C_{\rm mod})^{-1} \cdot (D_{\rm org}[i]-\mu)\,
\end{equation}
and checking its distribution compared to the original one $\chi^2_{\rm org}$ mentioned above, one can assess how well the model and true covariance agree.
We found that the covariance of non-overlapping cells from the FLASK patches is well captured in both the Gaussian and shifted lognormal theory, as it is dominated by the finite sampling term and thus not shown here. The covariance of heavily overlapping cells from FLASK patches was visually different from the Gaussian case, mainly driven by the asymmetry in the diagonal of the covariance matrix shown in Figure~\ref{fig:kappa_covariance_diag_FLASK_Gauss}. In Figure~\ref{fig:chi2_test_FLASK_theory} we show the $\chi^2$ test results for the overlapping case with FLASK covariances for patches (upper panel) and the full sky (lower panel). The agreement of the $\chi^2$ obtained from the measured covariances (coloured solid) and a draw from a Gaussian with the same variance (black) shows that the data vectors are close to Gaussian distributed. We find good agreement between the FLASK covariance measurements (coloured solid) and the shifted lognormal predictions (coloured dotted). The results for patches (upper panel) show a slight (6-9\%) bias of the theory towards higher $\chi^2$, which could be caused by the patch-to-patch correlation arising from cutting multiple patches from each FLASK full sky that is unaccounted for in the theory. Interestingly, we find no evidence for a corresponding discrepancy in the Fisher forecasts for those two cases, demonstrating the complementarity of the two tests adopted in \cite{Friedrich2021cov}.

\subsection{Fisher forecast validation} 
\label{subsec:Fisher}
To forecast the errors on a set of cosmological parameters, $\vec{p}$, we use the Fisher matrix formalism. The Fisher matrix given a (set of) summary statistics in the data vector $\vec{D}$ is defined as
\begin{equation}
F_{ij}= \sum_{\alpha,\beta}\frac{\partial D_\alpha}{\partial p_i}(C^{-1})_{\alpha \beta}\frac{\partial D_\beta}{\partial p_j}~, \quad 
\label{eq:Fisher}
\end{equation}
where $D_\alpha$ is the $\alpha$-th element of the data vector $\vec{D}$, $C^{-1}$ denotes the precision matrix introduced in Section~\ref{subsec:precision_matrix}, which is the matrix-inverse of the data covariance $C$ given in equation~\eqref{eq:covariance} and is was assumed that the covariance matrix is independent of cosmology. The Fisher formalism rests on the assumption that the summary data vector is Gaussian distributed which we have explicitly checked in the last subsection for the central region of the weak lensing PDF considered here. The parameter covariance matrix $\mathbf{C}(\vec{p})$ is then obtained as inverse of the Fisher matrix, $C^{\rm P}_{ij}=(F^{-1})_{ij}$.
In the Fisher formalism, marginalisation over a subset of parameters is achieved by simply selecting the appropriate sub-elements of the parameter covariance. 

To further validate our theoretical covariances, we perform simple Fisher forecasts for two cosmological parameters $\{\Omega_{\rm cdm},\sigma_8\}$ with the measured and predicted covariance matrices, while keeping the data vector and its dependence on cosmological parameters in \eqref{eq:Fisher} as predicted from large-deviation statistics \citep[see][]{Boyle2021}.\footnote{The predicted derivatives rely on large-deviations theory for the convergence PDF \citep{Barthelemy2020}. Using CLASS \citep{CLASS}, linear and non-linear \citep[Halofit,][]{halofit} matter density variances are calculated to rescale the matter density CGF in each redshift slice. The convergence CGF is obtained by integration over the redshift slices, and translated to the final PDF by an inverse Laplace transform.}
\begin{figure}
    \centering
    \includegraphics[width=\columnwidth]{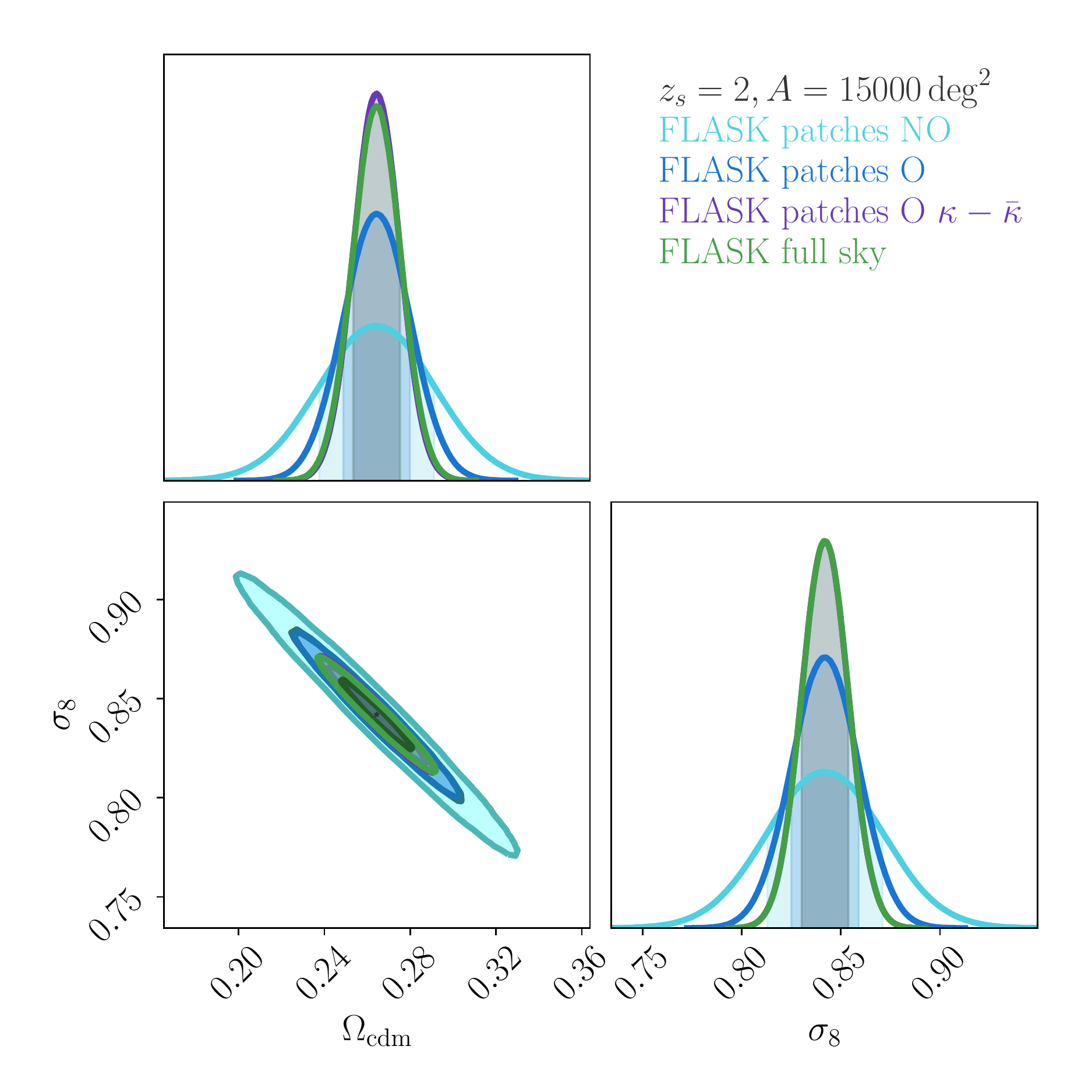}
    \caption{Fisher forecasts comparing different FLASK-generated covariances using non-overlapping cells (cyan), overlapping cells (blue), overlapping cells with mean-subtraction (purple) and overlapping cells on full sky maps (green). All covariances were rescaled to mimic the same survey area.}
    \label{fig:fisher_FLASK_patches_fullsky}
\end{figure}
The Fisher forecasts in Figure~\ref{fig:fisher_FLASK_patches_fullsky} demonstrate that measuring non-overlapping cells only (cyan) leads to much wider parameter errors than including cell overlaps (blue), such that capturing cell overlaps is imperative to avoid artificially inflating the error budgets \citep[as already pointed out in][]{SzapudiColombi96}. Additionally we show that when measuring covariances from small patches obtained from FLASK and subtracting the mean in each small patch (purple) this reproduces the result obtained from full sky maps rescaled to the survey area (green). 
\begin{figure}
    \centering
    \includegraphics[width=0.99\columnwidth]{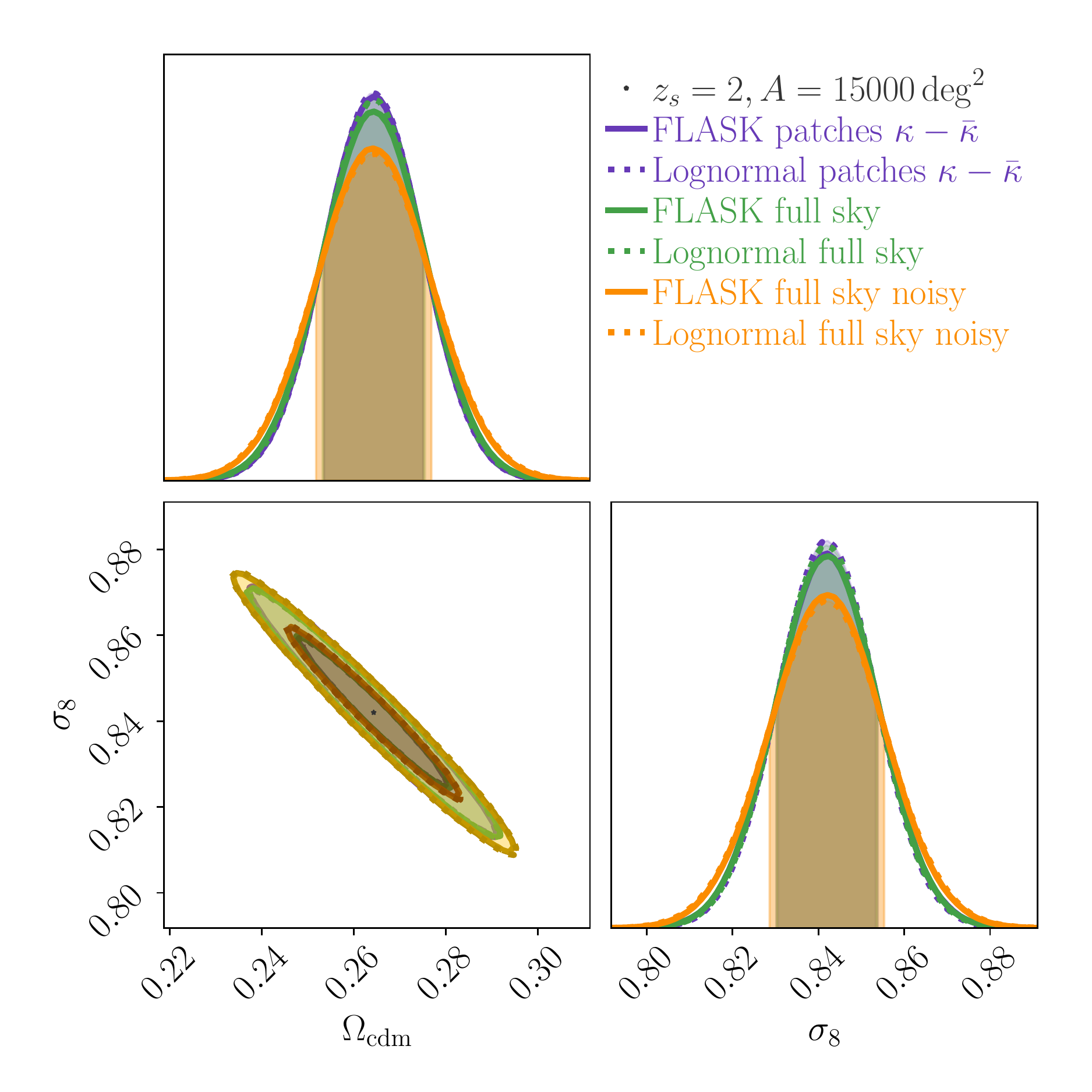}
    \caption{Fisher forecasts validating the predicted covariances (dotted) against FLASK measurements (solid) for shifted lognormal fields on small patches (purple), the full sky (green) and including shape noise (orange).}
    \label{fig:fisher_lognormal_FLASK}
\end{figure}
In Figure~\ref{fig:fisher_lognormal_FLASK} we show that our shifted lognormal predicted covariances (dotted) for small patches with mean-subtraction (purple) and the full sky (green) agree well with the FLASK measurements (solid). We also illustrate the impact of shape noise (orange) and validate our convolution approach from equation~\eqref{eq:jointPDF_incl_shapenoise} against FLASK measurements with added shape noise.

\section{Extension to 3D clustering PDF covariance}
\label{sec:covPDF_3D}

In this Section, we describe the essential steps needed to adapt our covariance model to 3-dimensional fields using the 3D matter PDF as an example. While the line-of-sight projection involved in computing the weak lensing convergence and photometric galaxy density on mildly nonlinear scales tends to Gaussianise the underlying field, 3D matter and spectroscopic galaxy densities typically show a greater degree of non-Gaussianity on mildly nonlinear scales. This might complicate the two-point PDF modelling, but we will show that the previously discussed shifted lognormal model from Section~\ref{subsec:jointPDFmodels} has flexibility to match the skewness of the one-point distribution and capture the clustering properties. 
Finally, we will focus on the theoretical modelling of the super-sample covariance, which can be used to complement simulated covariances from  periodic boxes missing this effect. 

\subsection{Applicability of the shifted lognormal model}

The three-dimensional matter PDF is typically extracted in terms of the normalised cell density $\hat\rho=\rho/\rho_s$ as ratio of the physical cell density $\rho$ and the mean sample density $\rho_s$, or the associated zero-mean density contrast $\delta=\hat\rho-1$. For illustration purposes, we use previously obtained measurements for the matter PDF at redshift $z=0$ in spheres of radius $R=10$ Mpc$/h$ \citep[][]{Uhlemann2020Fisher} in  the Quijote simulation suite \citep{Quijote}. The one-point PDF of the normalised density\footnote{ensuring the compatibility of measurements in simulations with different particle numbers (changing the mean density of particles in a sphere).} $\hat\rho$ was extracted in logarithmically spaced bins designed to equally capture under- and overdensities in the strongly non-Gaussian distribution and means and covariances obtained from 15,000 realisations at the fiducial cosmology.

\begin{figure}
\centering
\includegraphics[width=\columnwidth]{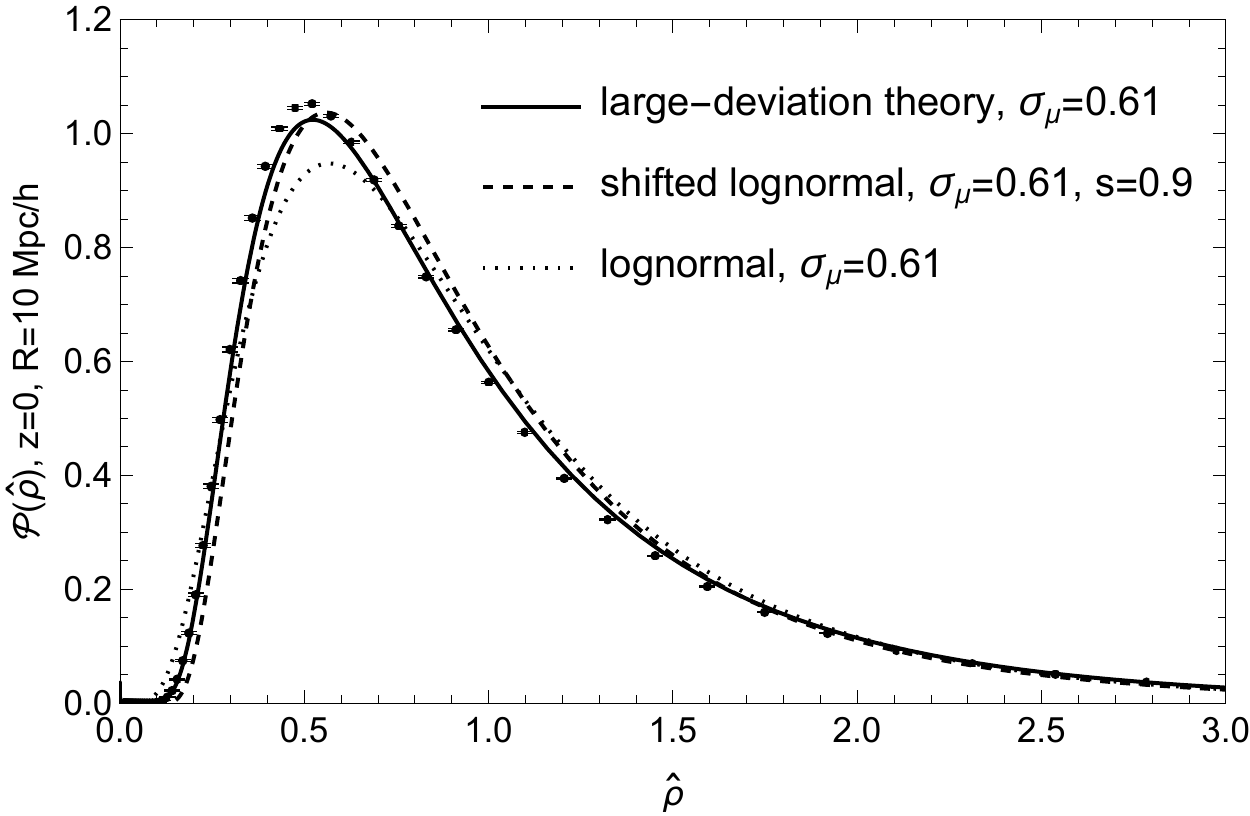}
\caption{3D matter density PDF at redshift $z=0$ and smoothing radius $R=10$ Mpc$/h$ as measured from the Quijote simulations (data points, with error bars indicating the sample variance), predicted by large-deviation theory (black line) and approximated with a lognormal distribution (black dotted) and a shifted lognormal distribution with tuned shift parameter (black dashed).}
\label{fig:Quijote_PDF}
\end{figure}

In Figure~\ref{fig:Quijote_PDF} we give a visual impression of the mean PDF for the  normalised matter density $\hat\rho$ as measured from the simulations (data points) and large-deviation theory prediction \citep[black line, following][]{Uhlemann16,Uhlemann2020Fisher}.
To describe clustering observables with the shifted lognormal model, we can simply replace the weak lensing convergence $\kappa$ with the density contrast $\delta$ in equations~\eqref{eq:1ptPDF_LN}~and~\eqref{eq:joint_LN}. Choosing a unit shift parameter $s=1$ corresponds to the usual lognormal approximation \citep[dotted black line,][]{ColesJones91}. This already provides a decent phenomenological fit to first principle predictions for the matter PDF from large deviation statistics \citep{Bernardeau14,Uhlemann16,Uhlemann2020Fisher}, and can be improved by including a shift factor below unity (dashed black line).\footnote{The lognormal model is however insufficient to accurately predict the response to cosmological parameters \citep[see][]{Uhlemann2020Fisher}.} The good match for the fiducial PDF in Figure~\ref{fig:Quijote_PDF} and the leading order bias in Figure~\ref{fig:sphere_bias_DCmode} suggests that the shifted lognormal model can be used to obtain accurate analytical PDF covariance estimates. This can be done by computing the joint PDF~\eqref{eq:jointPDF_meandivided} following a similar lognormal matching procedure as sketched for the mean-subtracted weak lensing convergence $\tilde\kappa$. For 3D observables for which the shifted lognormal model would not yield a good match of the PDF and leading order bias, the class
of tree models discussed in Appendix~\ref{app:MinimalTreeModel} and \cite{Bernardeau2022} offer an alternative with the capability of skewness matching.

\subsection{Relevance and prediction of super-sample covariance}

{\it Measurements.} The covariance measured from simulated boxes with fixed mean density (due to a fixed particle number) $\bar\rho$ cannot capture the super-sample effect created by having different sample densities $\rho_s$ \change{as would arise naturally from computing the covariance of subboxes within a much larger simulation box}. 
To estimate this \change{background density driven} super-sample covariance effect for the Quijote simulation suite, \change{we make use of their publicly available} separate-universe style `DC' runs emulating a background density contrast $\delta_b=\pm 0.035$ \change{through} changed cosmological parameters and simulation snapshot times from the separate universe approach \citep{Sirko2005}.
The super-sample covariance between two data vector entries $D_i$ and $D_j$ can be estimated by 
\begin{equation}
\label{eq:cov_SSC_SU}
  \text{cov}^{\rm SSC}_{\rm SU}(D_i,D_j)
  = \sigma_b^2 \frac{\partial D_i}{\partial \delta_b}
               \frac{\partial D_j}{\partial \delta_b},
\end{equation}
where $\sigma_b^2$ is the variance of $\delta_b$ (here just $\delta_b^2$), and the other two
terms encode the linear response of the data vector, which can be determined from the simulations using finite differences. \change{The impact of super-sample covariance on the 3D matter and halo power spectra, bispectra, void size distributions and the halo mass function has been recently studied in \cite{Bayer2022SSC}. As mentioned therein, in addition to the background density effect captured by the separate universe approach, there can be an additional super-sample effect due by tidal fields affecting higher-order statistics in a non-trivial way. For the PDF, we expect those tidal effects to be negligible due to its intrinsic symmetry and averaging.}

\begin{figure}
\centering
\includegraphics[width=\columnwidth]{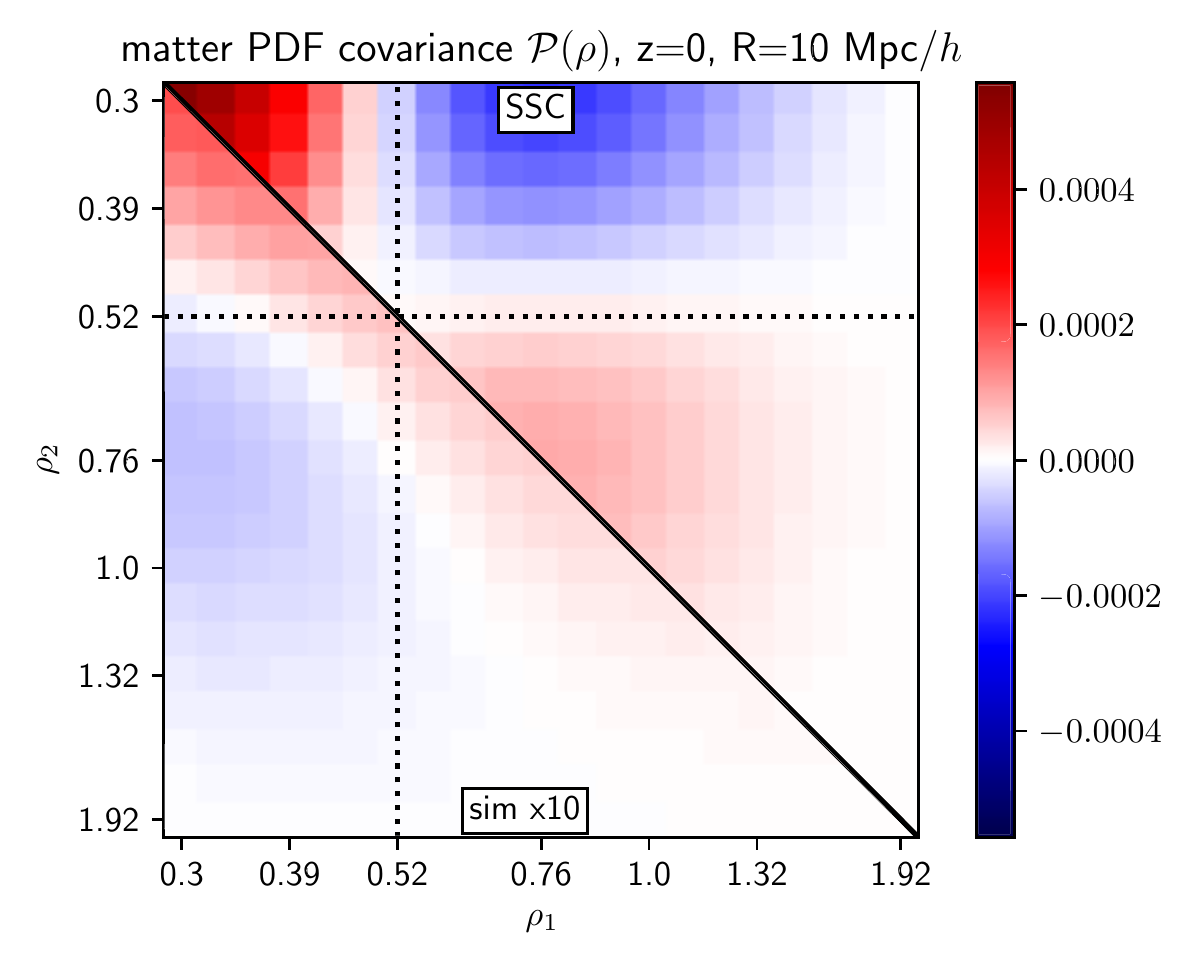}
\caption{Comparison of the PDF covariance matrix contributions for $z=0$, $R=10$ Mpc$/h$ as measured from the $1($Gpc$/h)^3$ Quijote simulations and the super-sample covariance contribution constructed from the DC mode following equation~\eqref{eq:cov_SSC_SU}.
The upper triangle shows the super-sample covariance contribution while the lower triangle shows the measured covariance from the simulations enhanced by a factor of 10, while the diagonal from the upper panel is masked. \change{The dotted lines indicate the PDF peak location.}}
\label{fig:Quijote_SSC}
\end{figure}

In the lower triangle of the lower panel of Figure~\ref{fig:Quijote_SSC} we show the covariance matrix of the matter density PDF  showing the correlation between different density bins in the central region of the PDF. We see that, as expected, neighbouring bins are positively correlated, while intermediate underdense and overdense bins are anticorrelated with each other. The strong positive correlation in the bands along the diagonal is induced by cell overlaps, similarly as discussed previously for the analogous weak lensing case. 
We illustrate the separate universe super-sample  effect on the PDF covariance in the upper triangle of Figure~\ref{fig:Quijote_SSC}, showing that the super-sample covariance term completely dominates the overall covariance for a volume of $1($Gpc$/h)^3$ and will still lead to a significant effect for a volume of $\mathcal O(10)$(Gpc$/h)^3$. 

{\it Predictions.}
\label{subsec:PDFcovariance_3D}
Predicting the covariance of the PDF of the normalised matter density $\hat\rho$ in a three-dimensional volume is conceptually analogous to the previous two-dimensional case relevant for weak lensing, so we can write the covariance in terms of the two-point PDF as in equation~\eqref{eq:covfromjointPDFmodel} by replacing the weak lensing convergence with the density $\kappa\rightarrow\rho$ and angular by 3D-distances $\theta\rightarrow r$, where $P_d(r)$ indicates the distribution of distances $r$ in a given survey volume. The distance distribution in a cubic box is easily obtained from the one for a 3D unit cube that corresponds to the problem of cube line picking for which a closed-form expression is available \citep{Mathai1999cubelinepicking,WeissteinCubeLinePicking}.
Similarly to the weak lensing case, where the mean-subtraction leads to a modification of the relevant bias functions determining the covariance, the relative density definition will result in modified bias functions. This can be predicted by considering the matter PDF in different sample densities $\rho_s$, where the joint PDF of normalised densities $\hat\rho_i$ is obtained from the trivariate PDF of physical densities $\rho_i=\rho_s\hat\rho_i$ and the background density $\rho_s$ by integrating out the latter
\begin{equation}
\label{eq:jointPDF_meandivided}
    \mP(\hat\rho_1,\hat\rho_2)=\int d\rho_s\, \rho_s^2 \mP(\rho_s,\rho_1=\rho_s \hat\rho_1,\rho_2=\rho_s\hat\rho_2)\,.
\end{equation}
The super-sample covariance effect~\eqref{eq:cov_SSC} for the PDF of the normalised matter density $\hat\rho$ is obtained from the leading order effective bias function in the resulting covariance, given by equation~(28) in \cite{Bernardeau2022}
\begin{equation}
\label{eq:bias_rho_meansub}
\hat b_1(\hat\rho)=1+\frac{\partial\log \mP(\hat\rho)}{\partial\log \hat\rho} + b_1(\hat\rho) \,.
\end{equation} 
Note that while the expression looks similar to the one for the mean-subtracted weak lensing case $\tilde b_{1,\rm NG}(\tilde\kappa)$ from equation~\eqref{eq:eff_b1_kappa}, this contribution does not vanish even for a Gaussian field. Predictions for the ingredients, the one-point PDF $\mP(\rho)$ and the sphere bias $b_1(\rho)$, can be obtained from large-deviation statistics \citep{Codis16a,Uhlemann17Kaiser}, shifted lognormal models (Section~\ref{subsec:jointPDFmodels}) and tree models \citep[Appendix~\ref{app:MinimalTreeModel} and][]{Bernardeau2022}.

\begin{figure}
\includegraphics[width=1\columnwidth]{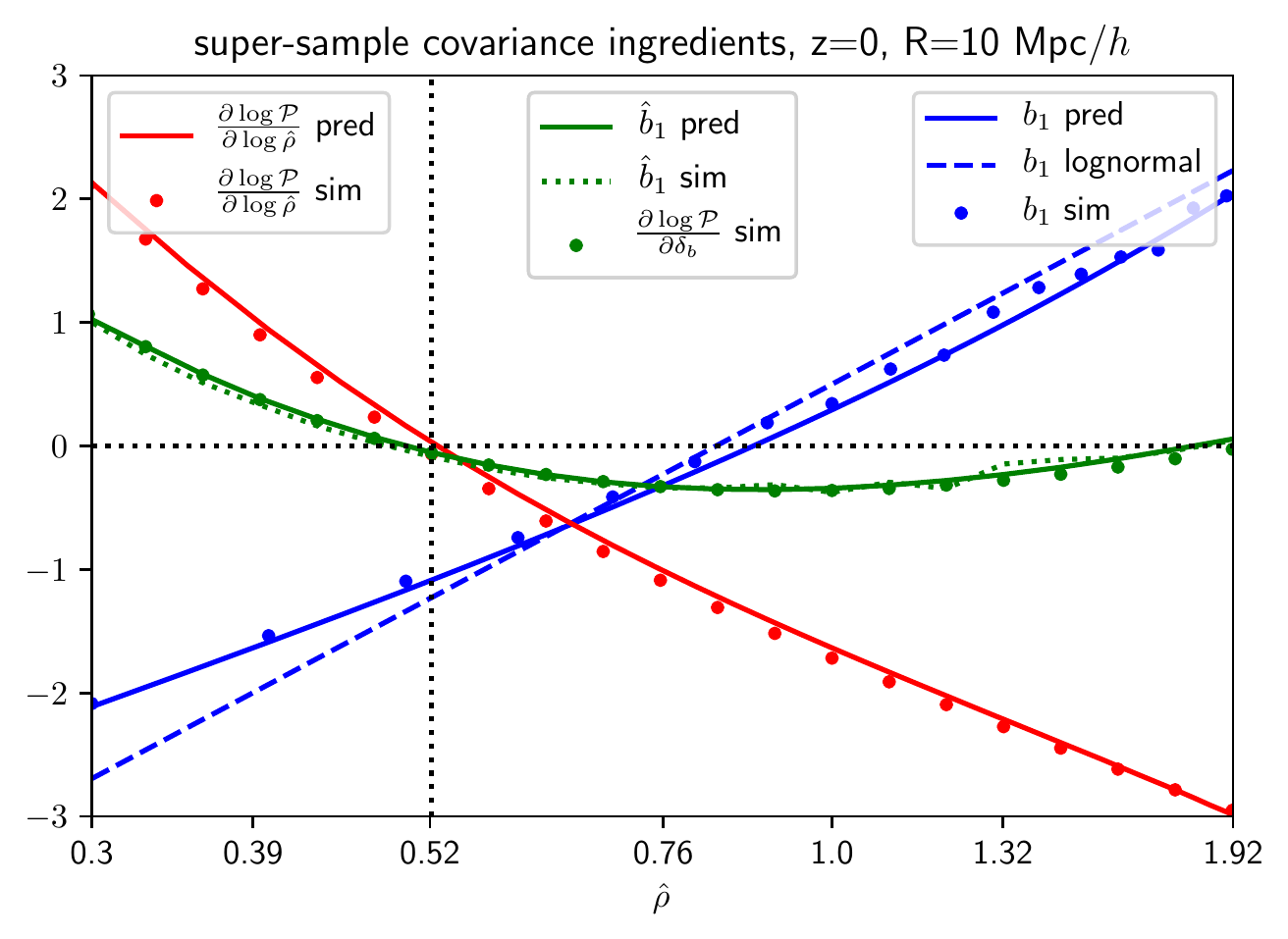}
\caption{Comparison of the theoretically predicted sphere bias (blue solid), the lognormal approximation (blue dashed) and the measurement in one realisation of the fiducial cosmology of Quijote (blue data points) at radius $R=10$ Mpc$/h$, distance $d=20$ Mpc$/h$ and redshift $z=0$. 
Additionally shown is the logarithmic derivative of the PDF (red) as predicted (solid) and measured (data points) and the resulting prediction from the effective bias~\eqref{eq:bias_rho_meansub} using the theoretical ingredients (green solid) or the simulated ones (green dotted). The predictions agrees very well with the measurements from separate universe simulations (green data points). 
The vertical dotted line indicates the peak location of the PDF. The $\rho$-range shown equals the one used in the covariance plots in Figure~\ref{fig:Quijote_SSC}.
}
\label{fig:sphere_bias_DCmode}
\end{figure}

{\it Comparison.} In Figure~\ref{fig:sphere_bias_DCmode} we demonstrate that the derivative of the PDF with respect to the background density (green data points) determining the super-sample covariance~\eqref{eq:cov_SSC_SU} is well predicted by the effective bias $\hat b_1$ from equation~\eqref{eq:bias_rho_meansub} using theoretical ingredients (green solid line) or simulated ones (green dotted line). We also show the $\hat b_1$ ingredients consisting of
the first order sphere bias function $b_1$ as measured from the correlation between neighbouring spheres in one realisation (blue data points) in comparison to the large-deviation theory prediction \citep[blue solid line, following][]{Uhlemann17Kaiser} and the lognormal approximation (blue dashed line), and the logarithmic derivative of the one-point PDF as measured from finite differences (red data points) and predicted from large-deviation theory (red solid line). We conclude that the super-sample covariance effect can be robustly predicted by the theory presented here and that the shifted lognormal model holds promise to compute analytical covariances for the PDF of 3D clustering observables.

{\it Applications.}  The total covariance matrix is a sum of the simulated covariance, $C_{\rm sim}$ and the super-sample covariance term consisting of a dyadic product, $C_{\rm SSC}=v_{\rm SSC}v_{\rm SSC}^\textsf{T}$ with $v_{\rm SSC}=\sqrt{\bar\xi}(\hat b_1\mP)(\hat\rho)$. This sum can be inverted using the Sherman-Morrison formula \citep{ShermanMorrison1950} to obtain the precision matrix
\begin{equation}
\left(C_{\rm sim} + C_{\rm SSC}\right)^{-1} = C_{\rm sim}^{-1} - \frac{C^{-1}_{\rm sim} C_{\rm SSC}C_{\rm sim}^{-1}}{1 + v_{\rm SSC}^\textsf{T}C_{\rm sim}^{-1}v_{\rm SSC}}\,.
\end{equation}
The result can be used to predict the super-sample covariance impact on statistical tests like $\chi^2$ and Fisher contours, which generally leads to a degradation of parameter constraints as detailed in \citep{Lacasa2019cov} for the case of Euclid-like photometric galaxy clustering observables.

\section{Conclusion \& Outlook}
\label{sec:conclusion}
{\it Conclusion.} In this work, we have shown how to compute covariances for the one-point PDF of the weak lensing convergence on mildly nonlinear scales. Our covariance predictions~\eqref{eq:covfromjointPDFmodel} rely on the two-point PDF, which is integrated over all separations according to the distribution of distances in the survey area. Our formalism includes predictions for the key effects including {\it finite sampling}~\eqref{eq:covFS} and {\it super-sample covariance}~\eqref{eq:cov_SSC} which were shown to scale with the inverse survey area.
We presented effective models for this joint PDF for a Gaussian field~\eqref{eq:2ptPDFGauss}, a shifted lognormal field~\eqref{eq:joint_LN} matching the PDF skewness and  density-dependence of the two-point clustering (Figure~\ref{fig:kappa_b1_LN}), and a field affected by {\it shape noise}~\eqref{eq:jointPDF_incl_shapenoise} from galaxy shape measurements. Going beyond (log-)normal models, we laid out a large-separation expansion for the two-point PDF that captures the key effects of the PDF covariance for general non-Gaussian fields.
The validity of our predictions has been established through a set of tests including visual inspection of the covariance, leading order eigenvalues and eigenvectors (Figure~\ref{fig:kappa_cov_eigendecomp_fullsky}), the precision matrix (Figure~\ref{fig:inverse_covariance}), $\chi^2$ tests (Figure~\ref{fig:chi2_test_FLASK_theory}) and a two-parameter Fisher forecast (Figure~\ref{fig:fisher_lognormal_FLASK}). With this suite of tests we have demonstrated that our PDF covariance model not only accurately describes large sets of simulated FLASK \citep{Xavier2016} maps, but also measurements of realistic full sky convergence maps obtained from full $N$-body simulations \citep{Takahashi17}.

Finally, we have sketched how the formalism described here for mildly non-Gaussian weak lensing observables is also applicable to more strongly non-Gaussian 3D clustering quantities. Most importantly, we showed how this can yield predictions for the {\it super-sample covariance effect} (Figures~\ref{fig:Quijote_SSC}~and~\ref{fig:sphere_bias_DCmode}) to complement covariance measurements from finite-box simulations.

{\it Outlook.} We expect our formalism can be straightforwardly adapted to describe the one-point PDF of related weak-lensing observables such as the CMB weak lensing convergence \citep{Barthelemy2020postBorn} and aperture mass \citep{Barthelemy2021aperturemass}, as well as photometric galaxy clustering. Our model can be extended to describe the cross-correlations between weak lensing PDFs at different smoothing scales and/or tomographic redshift bins. Based on our PDF estimator, one can further attempt to model cross-covariances between the one-point PDF and 2-point statistics, which in the mildly non-linear regime are dominated by the correlation between the PDF and the variance \citep[see Figure~12 in][]{Uhlemann2020Fisher}. Note that our covariance model and its possible extensions can serve as a starting point for improving covariance estimates from simulations, even in regimes where the model itself becomes inaccurate. The Gaussian and shifted lognormal covariance models can be used as a starting point for a {\it precision matrix expansion} \citep{Friedrich2017precisionmatrixexp} in which the leading order eigenvectors could iteratively refine this baseline covariance model. Additionally, analytical models can be used to construct priors for a Bayesian covariance estimation with the help of cheap surrogates \citep{Chartier2022}.

\section*{Acknowledgements}
This work has made use of the Infinity Cluster hosted by Institut d'Astrophysique de Paris. We thank St\'ephane Rouberol for smoothly running this cluster for us. We thank the authors of \cite{Takahashi17} for their publicly available lensing maps, \cite{Quijote} for their publicly available simulation suite products, and \cite{Xavier2016} for their publicly available FLASK code. The figures in this work were created with \textsc{matplotlib} \citep{matplotlib} making use of the \textsc{numpy} \citep{numpy}, \textsc{scipy} \citep{2020SciPy-NMeth}, \textsc{sci-kit image} \citep{scikit-image} and \href{https://samreay.github.io/ChainConsumer/}{ChainConsumer} Python packages.

CU is supported by the STFC Astronomy Theory Consolidated Grant ST/W001020/1 from UK Research \& Innovation. CU's work was performed in part at Aspen Center for Physics, which is supported by National Science Foundation grant PHY-1607611 and partially supported by a grant from the Simons Foundation. AG is supported by an EPSRC studentship under Project 2441314 from UK Research \& Innovation. SC thanks Fondation Merac and the French {\sl Agence Nationale de la Recherche} (grant ANR-18-CE31-0009) for funding. The authors thank the Euclid work packages for additional galaxy clustering probes and higher-order weak lensing statistics, as well as the Rubin LSST:DESC higher-order statistics group for useful discussions.

\bibliographystyle{mnras}
\bibliography{LSStructure}

\appendix

\section{Large-separation expansion in the minimal tree model}
\label{app:MinimalTreeModel}
\begin{figure}
\centering
\includegraphics[width=\columnwidth]{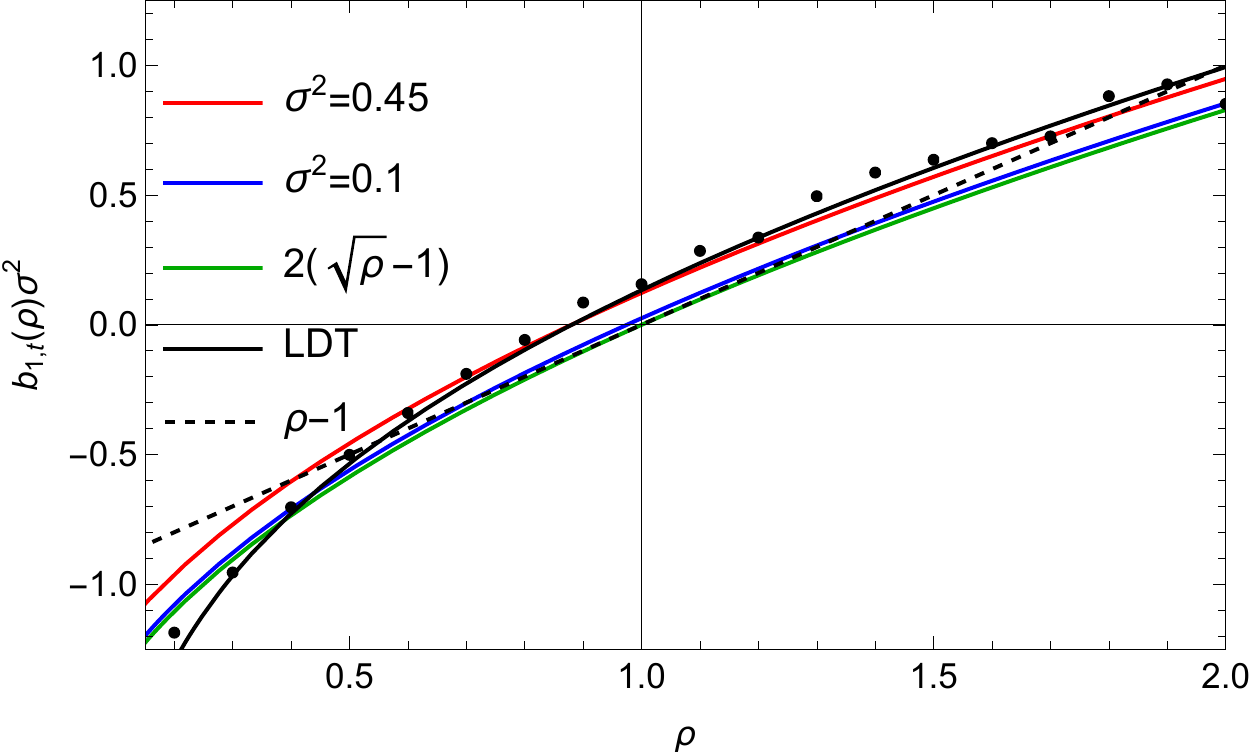}
\includegraphics[width=\columnwidth]{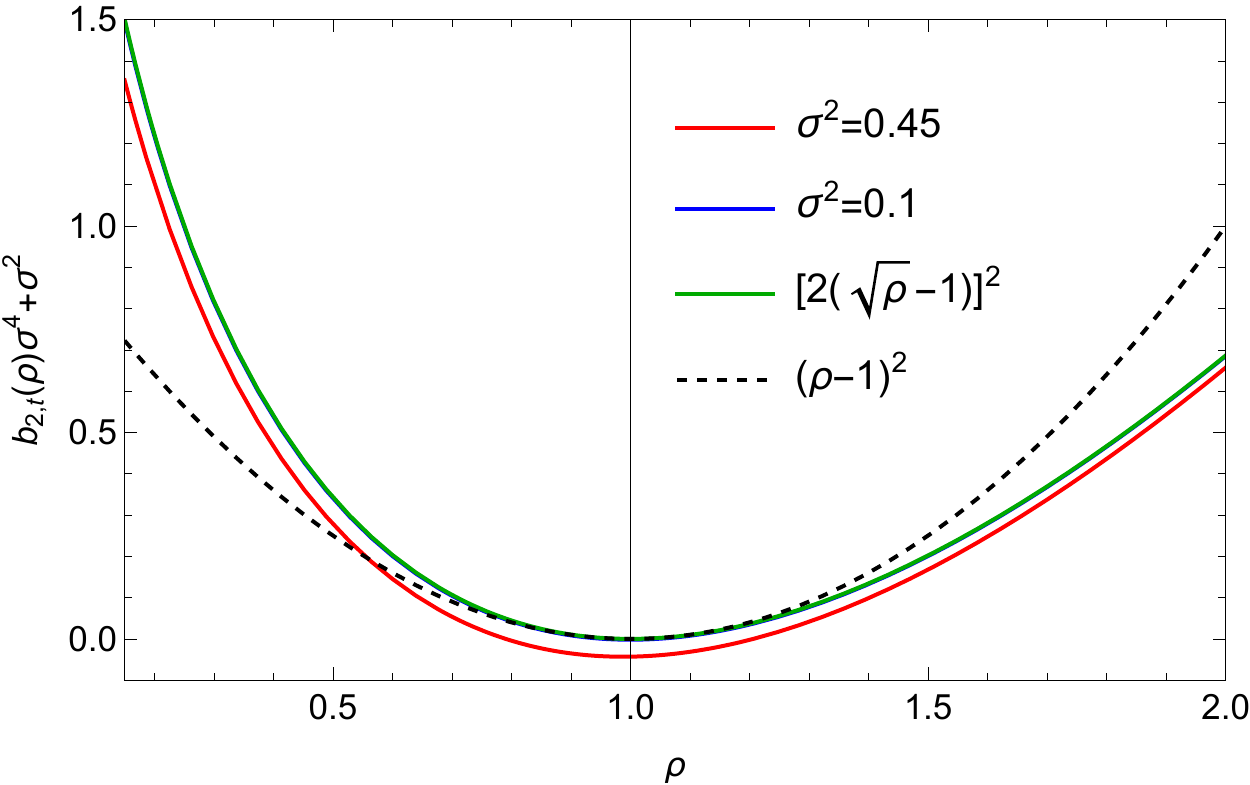}
\caption{A comparison of the first two leading order bias functions from the tree model~\eqref{eq:bias_tree}, $b_{1,\rm t}$ (upper panel) and $b_{2,\rm t}$ (lower panel), for different variances (coloured lines) approaching the small-variance limit~\eqref{eq:bias_tree_lim} (green) and the Gaussian expectation (dashed line). In the upper panel we also show the measurements for $b_1$ from the Quijote simulations (data points) and the large-deviation theory prediction (black line) as in Figure~\ref{fig:sphere_bias_DCmode}, but in linear rather than logarithmic scale.}
\label{fig:b1b2_treemodel}
\end{figure}

In addition to shifted lognormal models used heavily in the main text, let us discuss one representative from the class of so-called tree models considered in \cite{Bernardeau2022} in the context of PDF covariance modelling. The minimal tree model is the simplest example for which analytical results can be obtained at all orders. Here, we show how the phenomenology of tree models resembles predictions from large-deviation theory for the matter PDF (see the upper panel of Figure~\ref{fig:b1b2_treemodel}). Additionally, we 
present a closed-form expression for the $n$-th order bias parameter $b_{n,\rm t}$~\eqref{eq:bntree} that can be used to construct the full large separation series expansion yielding an invertible analytical covariance model.

The one-point cumulant generating function of the minimal tree model is given by
\begin{equation}
\label{eq:CGF_minimaltree}
    \varphi_{\rm t}(\lambda)=\frac{\lambda}{1-\lambda \sigma^2/2} \simeq \lambda+\frac{\sigma^2}{2}\lambda^2 + \underbrace{\frac{\sigma^4}{4}}_{\kappa_3/3!}\lambda^3 + \mathcal O(\sigma^6)\,,
\end{equation}
where the expansion shows that the mean is one, $\sigma^2$ is the variance\footnote{Note that \cite{Bernardeau2022} uses $\bar\xi$ to denote the variance, which should not be confused with our average correlation $\bar\xi$ from equation~\eqref{eq:meanxi}.} and the reduced skewness is $S_3=\kappa_3/\sigma^4=3/2$.
The one-point PDF of the matter density in the minimal tree model can be computed in closed form and is given by a hypergeometric function
\begin{equation}
    \label{eq:PDF_tree}
\mP_{\rm t}(\rho)=\frac{4}{\sigma^4} \exp\left[-
\frac{2}{\sigma^2}(1+\rho)\right]\ _0F_1\left(2,\frac{4\rho}{\sigma^4}\right)\,.
\end{equation}
The two-point PDF can be obtained from the two-point CGF which  can be written in terms of the individual one-point CGFs
\begin{subequations}
\label{eq:jointCGFclose-full}
\begin{align}
\varphi_{\rm t}(\lambda_1,\lambda_2)
&=\frac{\varphi(\lambda_1)+\varphi(\lambda_2)+\xi_{12}\varphi(\lambda_1)\varphi(\lambda_2)}{1-\xi_{12}^2\varphi(\lambda_1)\varphi(\lambda_2)/4}
\,.
\end{align}
\end{subequations}
By expanding that expression in the correlation $\xi_{12}$, one can obtain the two leading order bias functions and the additional $q_1$ term in the large-separation covariance expansion~\eqref{eq:cov_nonG} 
\begin{subequations}
\label{eq:bias_tree}
\begin{align}
    \label{eq:b1_tree}
    b_{1,\rm t}(\rho)&=\frac{_0F_1\left(1,\frac{4\rho}{\sigma^4}\right)}{_0F_1\left(2,\frac{4\rho}{\sigma^4}\right)}-\frac{2}{\sigma^2}\,,\\
\notag    b_{2,\rm t}(\rho)&=\frac{4[\rho-1-\sigma^2 b_1(\rho)]}{\sigma^4}= \frac{4}{\sigma^2}\left[\frac{\rho-1}{\sigma^2}- b_1(\rho)\right] \\
\label{eq:b2_tree}
&= \frac{4}{\sigma^2}\left[\frac{\rho+1}{\sigma^2}-\frac{_0F_1\left(1,\frac{4\rho}{\sigma^4}\right)}{_0F_1\left(2,\frac{4\rho}{\sigma^4}\right)}\right]\\
\label{eq:q1_tree}
q_{1,\rm t}(\rho)&=b_{1,\rm t}(\rho)/2\,.
\end{align}
\end{subequations}
In the limit where $\sigma^2\rightarrow 0$, those expressions tend to the following results, which can be further simplified for small density contrasts $\delta=\rho-1$ to reproduce the Gaussian result
\begin{subequations}
\label{eq:bias_tree_lim}
\begin{align}
    \label{eq:b1_tree_lim}
    b_{1,\rm t}(\rho)& \stackrel{\sigma\rightarrow 0}{\longrightarrow} \frac{2(\sqrt{\rho}-1)}{\sigma^2}\stackrel{\delta\rightarrow 0}{\longrightarrow} \frac{\delta}{\sigma^2}=b_{1,\rm G}(\delta)
    \,,\\
\label{eq:b2_tree_lim}
    b_{2,\rm t}(\rho)& \stackrel{\sigma\rightarrow 0}{\longrightarrow} \frac{[2(\sqrt{\rho}-1)]^2-\sigma^2}{\sigma^4}
    \stackrel{\delta\rightarrow 0}{\longrightarrow} \frac{\delta^2-\sigma^2}{\sigma^4}=b_{2,\rm G}(\delta)\,.
\end{align}
\end{subequations}
The two stages of this limiting behaviour are illustrated in Figure~\ref{fig:b1b2_treemodel} by a sequence of decreasing $\sigma^2$ (from red to green) showing fast approach to the small variance limit (black solid) which reproduces the Gaussian result (black dashed) for small density contrast. 
The quadratic $q_1$ term in the covariance~\eqref{eq:cov_nonG}
can be combined with the $b_1$ term (and hence be captured by the first eigenvector), while the $q_1$ term adding to $b_{2,\rm t}$ in \eqref{eq:cov_nonG} does not change much in the case of a small variance $\sigma^2\ll 1$ and an 
intermediate density regime.

For the minimal tree model, one can deduce that the covariance at any order is given in terms of the set of bias functions $b_{n,\rm t}(\rho)$. A recursion relation can be obtained as shown in \cite{Bernardeau2022}, which can be converted to an explicit closed-form expression as 
we show here. 

\label{app:bntree}
{\it Derivation of closed form bias functions.}
Define an auxilliary version of the Laplace transform of the CGF as the integral $I_j(\rho)$:
\begin{equation}
    I_j(\rho) = \int \frac{\dd{\lambda}}{2\pi i}\exp\left[{-\lambda\rho + j\varphi(\lambda)}\right].
\end{equation}
The PDF is then the value $I_1(\rho)$. In the minimal tree model, the bias functions are predicted via
\begin{equation}
    b_{n,\rm t}(\rho)\mP_{\rm t}(\rho) =  \int \frac{\dd{\lambda}}{2\pi i}\varphi_{0,\rm t}(\lambda)^n \exp\left[{-\lambda\rho + \varphi_{0,\rm t}(\lambda)}\right]
\end{equation}
where the one-point CGF $\varphi_{0,\rm t}(\lambda)$ is given by equation~\eqref{eq:CGF_minimaltree}.
The $n$th bias function equation can also be written as 
\begin{equation}
    b_{n,\rm t}(\rho)\mP_{\rm t}(\rho) = \dv[n]{I_j(\rho)}{j}\eval_{j=1}.
\end{equation}
and for the minimal tree model the $I_j(\rho)$ can be obtained as
\begin{equation}
\label{eqn:Ij}
    I_j(\rho) = \frac{4j}{\sigma^4}\exp\left[{-\frac{2}{\sigma^2}(j+\rho)}\right]\hypgeo{2, \frac{4j\rho}{\sigma^4}}\,.
\end{equation}
We can therefore obtain a closed form equation for $b_n(\rho)$ by differentiation of equation \ref{eqn:Ij}.
The $n$th derivative of $I_j$ is given by
\begin{equation}
    \dv[n]{I_j}{j} = \sum_{k=0}^n \sum_{\ell = 0}^k \binom{n}{k}\binom{k}{\ell} f^{(n-k)}(j)g^{(k-\ell)}(j)h^{(\ell)}(j)\,,
\end{equation}
where we've split $I_j$ into three functions, $f,g,h$, which we can differentiate separately as follows
\begin{subequations}
\begin{align}
f(j) &= \frac{4j}{\sigma^4}\,,\qquad
g(j) = \exp\left[-\frac{2}{\sigma^2}(j+\rho)\right]\,,\\
\dv[n]{f}{j} &= \frac{4}{\sigma^4}(j\delta_{n,0}+\delta_{n,1})\,,\\
\dv[n]{g}{j} &= \left(\frac{-2}{\sigma^2}\right)^n \exp\left[-\frac{2}{\sigma^2}(j+\rho)\right]
\end{align}
\end{subequations}
For the hypergeometric function we make use of the fact that $d/dz\, \hypgeo{n, z} = \hypgeo{n+1,z}/n$, where we defined $z = 4\rho j /\sigma^4$ so that $d/dj = \frac{4\rho}{\sigma^4}d/dz$ and
\begin{subequations}
\begin{align}
    h(j) &= \hypgeo{2, \frac{4\rho j }{\sigma^4}}\,, \\
    \dv[n]{h}{j} &= \left(\frac{4\rho}{\sigma^4}\right)^n \frac{1}{(n+1)!} \hypgeo{2+n, \frac{4\rho j}{\sigma^4}}\,.
\end{align}
\end{subequations}
Putting this into the full sum we obtain (using the Kronecker $\delta$s to collapse the $k$ sum, as the only surviving terms are when $n=k$ and when $n-k=1$).
\begin{align}
    \notag\dv[n]{I_j}{j} &= \frac{4}{\sigma^4} e^{-\frac{2}{\sigma^2}(j+\rho)}\sum_{k=0}^n \sum_{\ell = 0}^k \binom{n}{k}\binom{k}{\ell}(j\delta_{n-k,0} + \delta_{n-k,1})\\
    &\times\left(\frac{-2}{\sigma^2}\right)^{k-\ell} \left(\frac{4\rho}{\sigma^4}\right)^k \frac{\hypgeo{k+2,\frac{4j\rho}{\sigma^4}}}{(k+1)!} \,.
\end{align}
Now, we separate this sum into the case where $k=n$ and when $k=n-1$,
factor out common terms only depending on $n$ and use the binomial theorem to do these sums as simply $(1-2/\sigma^2)^{n}$ and $(1-2/\sigma^2)^{n-1}$ respectively. This equation is then the form of $b_{n,\rm t}(\rho)\mP_{\rm t}(\rho)$ when $j=1$. 
Dividing through by $\mP_{\rm t}(\rho)$ from equation~\eqref{eq:PDF_tree} everywhere and setting $j=1$ gives us the functional form of the $n$-th order bias function for the minimal tree model as
\begin{align}
\label{eq:bntree}
    b_{n,\rm t}(\rho) &= \frac{1}{n!}\left[\left(\frac{4\rho}{\sigma^4}\right)\left(1-\frac{2}{\sigma^2}\right)\right]^{n-1}\\
\notag &\left[ \frac{4\rho\left(1-\frac{2}{\sigma^2}\right)}{\sigma^4(n+1)}\frac{\hypgeo{n+2,\frac{4\rho}{\sigma^4}}}{\hypgeo{2, \frac{4\rho}{\sigma^4}}}+ \frac{\hypgeo{n+1,\frac{4\rho}{\sigma^4}}}{\hypgeo{2,\frac{4\rho}{\sigma^4}}}\right]\,.
\end{align}
Note that, the hypergeometrics can always be re-expanded down in terms of $\hypgeo{1, z}$ and $\hypgeo{2,z}$ using the identity
\begin{equation}
    \hypgeo{n-1,z} - \hypgeo{n,z} = \frac{z}{n(n-1)}\hypgeo{n+1,z}\,.
\end{equation}

\section{Refined derivation of covariance for cells in large-separation limit}
\label{app:large_sep_refine}
In Appendix C of \cite{Codis16a} the result of the covariance of finding $N_i$ cells with density $\rho_i$  and $N_j$ spheres of density $\rho_j$ is given as (with no summation over repeated indices implied)
    \begin{equation}
        \text{cov}(N_i,N_j) = \bar N_i \bar N_j \bar \xi b(\rho_i) b(\rho_j) +\delta_{ij} \bar N_i \,,
    \end{equation}
    where $b(\rho)$ is the cell bias and $\bar\xi$ is the mean correlation of spheres defined in equation~\eqref{eq:averagexi}. We can rewrite the ratio of the number of cells in the bin to the total number in terms of the measured PDF value in the bin and associated density bin size $\Delta_i$ as $N_i \simeq N_T \mP(\rho_i) \Delta_i$ and similarly $\bar N_i \simeq N_T\bar \mP(\rho_i) \Delta_i$.
    Hence, we have for the covariance of the histogram probability $\mP_i$:
    \begin{equation}
        \text{cov}(\mP(\rho_i),\mP(\rho_j)) = \bar \mP(\rho_i) \bar \mP(\rho_j) \bar\xi b(\rho_i) b(\rho_j) + \delta_{ij} \frac{\bar\mP(\rho_i)}{\Delta_i N_T}\,,
    \end{equation}
    Note that the shot-noise term only acts on the diagonal. Let us first focus on the diagonal, where the prediction for the noise to signal ratio is then is
        \begin{equation}
        \frac{\text{var}(\mP(\rho_i))}{\bar\mP^2(\rho_i)} = \bar\xi b^2(\rho_i)+\frac{1}{\bar\mP(\rho_i)\Delta_i N_T}\,,
    \end{equation}
The correlation matrix can be obtained by dividing the covariance by $\sqrt{\text{var}(\mP(\rho_1) \text{var}(\mP(\rho_2)}$. In the limit of infinite volume, where one can place $N_T\rightarrow \infty$ well-separated cells, the correlation matrix would read
\begin{align}
\text{corr}(\mP(\rho_i),\mP(\rho_j)) &\stackrel{N_t\rightarrow \infty}{\longrightarrow} \text{sgn} \left(b(\rho_i) b(\rho_j)\right) =\pm 1\,,
\end{align}
which predicts a characteristic 4-tiled pattern.

\subsection{Derivation for autocorrelation of PDF bins}
Let us assume that the joint PDF of the density in $N_T$ cells (for simplicity with regular spacing) at large separation\footnote{A generalisation can be obtained by restoring the density-dependent cell two-point correlation $\xi_R(\hat\rho_I,\hat \rho_J;r_{IJ})$ instead of the factorised version that is valid at large separations.} reads 
\begin{align}
\mP&(\hrho_1,\dots,\hrho_{N_T};\{r_{IJ}\}) \nonumber\\
&=\prod_{\rm I=1}^{N_T} \mP(\hrho_{\rm I})
\left[1+\
\sum_\mathrm{\rm I<J}
b(\hrho_{\rm I}) b(\hrho_{\rm J}) \xi(r_{IJ})
\right]\,.
\label{eq:defbiasPDF-C}
\end{align}
To derive the result for the biased distribution of finding exactly $N$ cells in a given density bin $\mP^{\rm b}(N)$, we further expand the product in \eqref{eq:defbiasPDF-C} assuming $b^2\xi \ll 1$ as follows
\begin{align}
\nonumber &\mP(\hrho_1,\dots,\hrho_{N_T};\{r_{IJ}\}) \\
=&\left[\prod_{\rm I=1}^{N_T} \mP(\hrho_{\rm I})\right]
\left[1+\sum_{I=1}^{N_T}\
\sum_\mathrm{\rm I<J}
b(\hrho_{\rm I}) b(\hrho_{\rm J}) \xi(r_{IJ})
\right]\,.
\label{eq:defbiasPDF-Cexp}
\end{align}
The probability to find exactly $N$ cells in a density bin $\hat \Delta=[\hat \rho-\Delta\rho/2,\hat \rho+\Delta\rho/2]$ is given by a) integrating over the probabilities of finding $N$ of the values of $\rho_1$, \ldots, $\rho_{N_T}$ within $\hat \Delta$ and $N_T-N$ outside $\neg\hat\Delta=\mathbb{R}^+\setminus\hat\Delta$, and b) summing over all different possibilities for the spatial distribution of the $N$ cells. Let us assume that the value $\hat\rho_I$ is fixed to a grid point $I$, such that the distance between the densities $\hat\rho_I$ and $\hat\rho_J$ are given by the distance of grid point $J$ to grid point $I$. Hence, we will drop $\{r_{IJ}\}$ from the argument of the joint PDF in the following.
Then considering different spatial distributions of the cells corresponds to selecting a subset $S$ of length $N$ from the set of indices $\{1,\ldots,N_T\}$
\begin{align}
\mP^{\rm b}(N)&=
\sum_{S}
\prod_{I\in S} \left(\int_{\hat\Delta}\!\!\!\dd\hrho_{I}\right)  \prod_{J\notin S} \left(\int_{\neg\hat\Delta}\!\!\!\dd\hrho_{J}\right)
\mP(\hrho_1,\dots,\hrho_{N_T})\,.
\label{eq:Pbiased}
\end{align}
Now we can insert equation~\eqref{eq:defbiasPDF-Cexp} and simplify using
\begin{align}
p&=\int_{\hat \Delta}\dd\hrho\,\mP(\hrho),\quad
1-p=\int_{\neg\hat\Delta}\dd\hrho\,\mP(\hrho)\\
pb&=\int_{\hat \Delta}\dd\hrho\,\mP(\hrho)b(\hrho),\quad
-pb=\int_{\neg\hat\Delta}\dd\hrho\,\mP(\hrho)b(\rho)\,,
\end{align}
where the normalisation of $\mP$ enforces $\int_{\mathbb{R}^{+}}\dd\hrho\,\mP(\hrho)b(\hrho)=0$ and we will further use $\bar b:=pb/p$. This leads to
\begin{align*}
\mP^{\rm b}(N)&=\sum_{S} \Bigg( p^{N}(1-p)^{N_T-N}\nonumber\\
&\qquad +\sum_{S\ni I<J\in S}\xi(r_{IJ})
(pb)^{2}p^{N-2}(1-p)^{N_T-N}\nonumber\\
&\qquad +2\!\!\!\!\!\!\sum_{S\ni I<J\notin S}\xi(r_{IJ})
p^{N-1} pb (-pb)(1-p)^{N_T-N-1}\\
&\qquad+\sum_{S\niton I<J\notin S}\xi(r_{IJ})
p^{N}(-pb)^{2}(1-p)^{N_T-N-2}\Bigg)\nonumber\\
&=p^{N}(1-p)^{N_T-N}\times\left\{\binom{N_T}{N} \right.\\
&\qquad\qquad+\bar b^2 \sum_{S}\Bigg(
\sum_{S\ni I<J\in S} \xi(r_{IJ})\\
&\qquad\qquad\qquad
+\frac{\bar N^2}{(N_T-\bar N)^2} \sum_{S\niton I<J\notin S}\xi(r_{IJ})\\
&\qquad\qquad\qquad-\frac{2\bar N}{N_T-\bar N}\sum_{S\ni I<J\notin S}\xi(r_{IJ})
\left.
\Bigg) \right\}\nonumber\,,
\end{align*}
where in the second step we have converted $p=\bar N/N_T$. 

Note that one can rewrite the sums over all $N$-component subsets $S$ (of which there are $\binom{N_T}{N}$) in terms of the mean spatial correlation (which agrees with the definition used in \cite{Codis16a})
\begin{equation}
    \label{eq:averagexi}
   \sum \xi := \sum_{I<J=1}^{N_T}\!\!\!\xi(r_{IJ})\ , \quad \bar \xi = \frac{\sum_{I<J=1}^{N_T} \xi(r_{IJ})}{N_T(N_T-1)/2}
\end{equation}
as follows
\begin{align}
    \sum_{S}\sum_{S\ni I<J\in S} \xi(r_{IJ})
    &= \binom{N_T-2}{N-2} \sum\xi\nonumber\\
    &= \binom{N_T}{N} \frac{N(N-1)}{2} \bar\xi\,,\\
2\sum_{S}\sum_{S\ni I<J\notin S}\!\!\!\xi(r_{IJ})
&=2\binom{N_T-2}{N-1} \sum\xi\nonumber\\
&=\binom{N_T}{N} N(N_T-N) \bar\xi\,,\\
\sum_{S} \sum_{S\niton I<J\notin S}\!\!\!\xi(r_{IJ}) 
&=\binom{N_T-2}{N} \sum\xi\nonumber\\
&=\binom{N_T}{N} \frac{(N_T-N)(N_T-N-1)}{2} \bar\xi\,.
\end{align}
Where the full array of $r_{IJ}$ with $I<J$ is split into three subsets of separations a) among cells in the density bin (length $N(N-1)/2$), b) between cells in the density bin and outside (length $N(N_T-N)$), and c) among cells outside the density bin (length $(N_T-N)(N_T-N-1)/2$). The combinatorical prefactors come from fixing two indices which corresponds to selecting from a set of $N_T-2$ remaining indices, $N-2$ when both indices are in S, $N-1$ when one index is in $S$ and the other in $\mathcal I\setminus S$, and $N$ when both indices are in $\mathcal I\setminus S$.
With this, we obtain
\begin{align*}
\mP^{\rm b}(N)
&=p^{N}(1-p)^{N_T-N} \binom{N_T}{N}\times\\
& \Bigg[ 1
+\bar b^2 \bar\xi \Bigg(\frac{N(N-1)}{2}
-\frac{\bar N N(N_T-N)}{N_T-\bar N} \\
&\qquad\qquad+ \frac{\bar N^2(N_T-N)(N_T-N-1)}{2(N_T-\bar N)^2}
\Bigg) \Bigg]\nonumber\,,
\end{align*}
In the Poisson limit this gives\footnote{Note that without the Poisson limit, the variance of the binomial distribution becomes $\bar N(1-p)$.}
\begin{align}
\mP^{\rm b}(N)
&=P^{\textrm{Poiss}}(N) \Bigg[ 1
+\bar b^2 \bar\xi \Bigg(\frac{N(N-1)}{2}
-\bar N N+ \frac{\bar N^2}{2}
\Bigg) \Bigg]\,,
\end{align}
where $P^{\textrm{Poiss}}(N)=\bar N^N\exp(-\bar N)/N!$.
In \cite{Codis16b} it was assumed that correlations are the same at all separations, $\xi(r_{IJ})=\bar\xi$ with $\bar\xi$ defined in equation~\eqref{eq:averagexi}. In this case, one can replace all sums, the sum over $r_{IJ}$ by a combinatoric factor for choosing $N$ cells out of $N_T$, and the sums over $I$ and $J$ by the number of pairs. As we showed here, it turns out that this result seems holds true even for the more general case of correlations that depend on separation. From this expression we can verify that the biased PDF is normalised and has mean $\langle N \rangle=\bar N$ and variance
\begin{equation}
    \text{var}(N)=\langle N^2\rangle - \langle N \rangle^2 = \bar N + \bar N^2 \bar \xi\, \bar b^2
\end{equation}
In the continuous limit, one can write the mean correlation as
$\bar\xi=\int_{r_{\rm min}}^{r_{\rm max}} dr\, P_d(r) \xi(r)\,,$
where $P_d(r)$ is the probability distribution of cell distances $r$. The maximum distance is set by the size of the survey patch/volume, for a $d$-dim box of side length $L$, we have $r_{\rm max}\simeq \sqrt{d} L$. The minimum distance is set by the cell separation under consideration, so $r_{\rm min}=2R$ for non-overlapping cells and $r_{\rm min}\rightarrow 0$ for heavily overlapping cells.
Note that this is not the same as attempting to measure the average correlation in a periodic simulation box (with fixed particle number) via the following definition
\begin{subequations}
 \label{eq:meanxi_wrong}
\begin{align}
\tilde\xi 
&= \frac{\sum_{i\neq j=1}^{N_T}  \delta_i\delta_j}{N_T(N_T-1)} = \frac{\sum_{i=1}^N \delta_i \left(\sum_{j=1}^N \delta_j - \delta_i\right)}{N_T(N_T-1)}\\
&\stackrel{\langle\delta\rangle=0}{=} - \frac{\sum_{i=1}^N \delta_i^2}{N_T(N_T-1)} <0\,,
\end{align}
\end{subequations}
which tends to be small but negative, because by construction the mean density contrast in a box vanishes.

\subsection{Derivation of cross-correlation between bins}
Considering different spatial distributions of the cells where a given number are in two density bins corresponds to first selecting a subset of length $N_1$, $S_{N_1}$ from the set of indices $\mathcal I=\{1,\ldots,N_T\}$ and then selecting another subset of length $N_2$, $S_{N_2}$ out of the remaining indices $\mathcal I \setminus S_{N_1}$. For brevity, let us denote $S_{N_{12}}=S_{N_1}\cup S_{N_2}$
\begin{align}
\nonumber \mP^{\rm b}(N_1,N_2)&=
\sum_{S_{N_1},S_{N_2}}
 \prod_{I\in S_{N_1}}\left(\int_{\hat\Delta_1}\!\!\!\dd\hrho_{I}\right) \prod_{J\in S_{N_2}} \left(\int_{\hat\Delta_2}\!\!\!\dd\hrho_{J}\right) \\
&\prod_{K\notin S_{N_{12}}} \left(\int_{\neg\hat\Delta}\!\!\!\dd\hrho_{K}\right)
\mP(\hrho_1,\dots,\hrho_{N_T})\,.
\label{eq:Pbiasedcross}
\end{align}
As before, the integrals can be simplified 
\begin{align*}
&\nonumber \mP^{\rm b}(N_1,N_2)\\
&=p_1^{N_1}p_2^{N_2}(1-p)^{\Delta N_{T,12}}\times\Bigg\{\binom{N_T}{N_1} \binom{N_T-N_1}{N_2} \\
&+\sum_{S_{N_1},S_{N_2}}\Bigg(
\bar b_1^2\!\!\!\!\!
\sum_{S_{N_1}\ni I<J\in S_{N_1}} \!\!\!\!\!\xi(r_{IJ})
+\bar b_2^2 \!\!\!\!\!
\sum_{S_{N_2}\ni I<J\in S_{N_2}} \!\!\!\!\!\xi(r_{IJ})\\
&\qquad\qquad
+2\bar b_1\bar b_2 
\sum_{S_{N_1}\ni I<J\in S_{N_2}} \xi(r_{IJ}) \\
&\qquad\qquad -2\frac{p_1\bar b_1+p_2\bar b_2}{1-p_1-p_2} 
\Bigg[\bar b_1\!\!\!\!\!
\sum_{S_{N_1}\ni I<J\notin S_{N_{12}}} \!\!\!\!\!\xi(r_{IJ})
\\
&\qquad\qquad\qquad\qquad +\bar b_2\!\!\!\!\!
\sum_{ S_{N_2}\ni I<J\notin S_{N_{12}}} \!\!\!\!\!\xi(r_{IJ})
\Bigg]\\
&\qquad\qquad +\left(\frac{p_1\bar b_1+p_2\bar b_2}{1-p_1-p_2}\right)^2\!\!\!\!\!
\sum_{S_{N_{12}}\niton I<J\notin S_{N_{12}}}\!\!\!\!\! \xi(r_{IJ})
\Bigg)\Bigg\}\,,
\end{align*}The sums over the different subsets $S_{N_1}$ and $S_{N_2}$ can be simplified using combinatorical properties. We use the average correlation $\bar\xi$ and its numerator $\sum\xi$ as defined in equation~\eqref{eq:averagexi}, $\neg i$ to indicate the opposite index $\neg i=2$ if $i=1$ and $\neg i=1$ if $i=2$, respectively. Introducing the  short hand notation $N_{T,12}:=N_T-N_1-N_2$,
we obtain
\begin{align}
\notag &    \sum_{S_{N_1},S_{N_2}}
\sum_{S_{N_i}\ni I<J\in S_{N_i}} \xi(r_{IJ}) \\
\notag &= \binom{N_T-2}{N_i-2}\binom{N_T-N_i}{N_T-N_{\neg i}}\sum \xi  \\
&= \binom{N_T}{N_1}\binom{N_T-N_1}{N_2} \bar\xi \frac{N_i(N_i-1)}{2} 
\,,\\
\notag &  2\sum_{S_{N_1},S_{N_2}}
\sum_{S_{N_1}\ni I<J\in S_{N_2}} \xi(r_{IJ}) \\
\notag &= 2\binom{N_T-2}{N_1-1}\binom{N_T-N_1-1}{N_2-1} \sum \xi\\
&=  \binom{N_T}{N_1}\binom{N_T-N_1}{N_2} \bar\xi N_1N_2
\,,
\end{align}
\begin{align}
\notag &    2\sum_{S_{N_1},S_{N_2}}
\sum_{S_{N_i}\ni I<J\notin S_{N_{12}}} \xi(r_{IJ})\\
\notag &= 2\binom{N_T-2}{N_i-1}\binom{N_T-N_i-1}{N_{\neg i}} \sum \xi\\
&=\binom{N_T}{N_1}\binom{N_T-N_1}{N_2} \bar \xi N_i\Delta N_{T,12}
\,,\\
\notag & \sum_{S_{N_1},S_{N_2}}\sum_{S_{N_{12}}\niton I<J\notin S_{N_{12}}} \xi(r_{IJ}) \\
\notag &= \binom{N_T-2}{N_1}\binom{N_T-N_1-2}{N_2} \sum \xi\\
 &= \binom{N_T}{N_1}\binom{N_T-N_1}{N_2} \bar\xi\frac{\Delta N_{T,12}(\Delta N_{T,12}-1)}{2}\,.
\end{align}
Additionally, one can rewrite
\begin{equation}
    \frac{p_1\bar b_1+p_2\bar b_2}{1-p_1-p_2} =
    \frac{\bar N_1\bar b_1+\bar N_2\bar b_2}{N_T-\bar N_1- \bar N_2}\,.
\end{equation}
With this one obtains
\begin{align*}
&\nonumber \mP^{\rm b}(N_1,N_2)\\
&=p_1^{N_1}p_2^{N_2}(1-p)^{\Delta N_{T,12}} \binom{N_T}{N_1} \binom{N_T-N_1}{N_2}\times\Bigg\{ 1+ \\
&+\bar\xi \Bigg[\frac{N_1(N_1-1)}{2} \bar b_1^2 
+ \frac{N_2(N_2-1)}{2} \bar b_2^2
+ N_1N_2 \bar b_1\bar b_2\\
&\quad - ( N_1 \bar b_1 + N_2\bar b_2 ) \Delta N_{T,12}\frac{\bar N_1\bar b_1+ \bar N_2\bar b_2}{N_T-\bar N_1- \bar N_2}
\\
&\quad + \frac{\Delta N_{T,12}(\Delta N_{T,12}-1)}{2}\left(\frac{\bar N_1\bar b_1+\bar N_2\bar b_2}{N_T-\bar N_1-\bar N_2}\right)^2
\Bigg] \Bigg\}\,.
\end{align*}
In the Poisson limit this becomes
\footnote{Note that without the Poisson limit, the covariance of the multinomial distribution becomes $-N_Tp_1p_2=-\bar N_1\bar N_2/N_T$.}
\begin{align*}
\nonumber \mP^{\rm b}(N_1,N_2)&=P^{\textrm{Poiss}}(N_1) P^{\textrm{Poiss}}(N_2) \times\Bigg[ 1+ \\
&+\bar\xi\, \bar b_1^2 \Bigg(\frac{N_1(N_1-1)}{2} +\frac{\bar N_1^2}{2} - N_1\bar N_1 \Bigg)\\
&+\bar\xi\, \bar b_2^2\Bigg(\frac{N_2(N_2-1)}{2} +\frac{\bar N_2^2}{2} - N_2\bar N_2 \Bigg)\\
&+\bar\xi\, \bar b_1\bar b_2 \Bigg( N_1N_2 + \bar N_1\bar N_2 - N_1\bar N_2 - \bar N_1 N_2 \Bigg) 
\Bigg]\,.
\end{align*}
With this we reproduce the result from \cite{Codis16b} for the Poisson limit and the associated covariance
\begin{equation}
    \langle N_1N_2\rangle= \bar N_1\bar N_2(1+\bar \xi\, \bar b_1\bar b_2)\,.
\end{equation}

\end{document}